\documentclass[a4paper,11pt]{article}
\usepackage[latin1]{inputenc}
\usepackage{indentfirst}
\usepackage[left]{lineno}
\usepackage{amsfonts}
\usepackage{float}

\usepackage{latexsym,amssymb,amsmath}
\usepackage[dvips]{graphics,graphicx,epsfig,color}
\usepackage[lofdepth,lotdepth]{subfig}
\usepackage[lofdepth,lotdepth]{subfig}
\begin{document}
\title{\bf The spurious resonance disease and how to cure it: application to the seismic response of a canyon}
\author{Armand Wirgin\thanks{LMA, CNRS, UPR 7051, Aix-Marseille Univ, Centrale Marseille, F-13453 Marseille Cedex 13, France, ({\tt wirgin@lma.cnrs-mrs.fr})} }
\date{\today}
\maketitle
\begin{abstract}
Three types of boundary integral equation (BIE) methods are employed to obtain closed-form solutions of a wave-scattering problem which are compared to the exact, closed-form (reference), solution deriving from the separation-of-variables technique. The problem involves either Dirichlet (D) or Neumann (N) boundary conditions (BC) for a scatterer that is a circular cylinder submitted to one or two incident waves. The three BIE methods lead to different expressions  for the traction (for D-BC) or boundary displacement (for N-BC) by which numerous resonances are predicted whose frequency of occurrence differs from one method to another. This is interpreted as being the sign that the three methods are generally-defective and the resonances are 'spurious'. This 'disease' is cured by combining two BIE into one in such a way that the resulting BIE gives rise to a closed-form solution identical to the exact reference solution devoid of spurious resonances.
\end{abstract}
\newline
\newline
Abbreviated title: Spurious resonances in canyon seismic response
\newline
\newline
Corresponding author: Armand Wirgin, \\ e-mail: wirgin@lma.cnrs-mrs.fr
\newpage
\tableofcontents
\newpage
\newpage
\section{Introduction}\label{intro}
In a recent contribution \cite{wi20} I suggested that the seismic response of both above- (e.g., hill) and below (e.g. valley)-(otherwise-flat) ground features are dominated by so-called surface shape resonances. I showed in  \cite{wi20} that both filled (with a softer material than the underlying rock)  and unfilled (i.e. the material in the above-ground feature is the same as in that of the underground) hills indeed exhibit this pronounced resonant behavior. Previously, I showed \cite{wi95} that resonant behavior dominates the seismic response of a particular below-ground feature (i.e., basin) filled with a material that is softer than the underlying rock (see also \cite{sl96}). There remained the question as to whether an unfilled basin or valley (e.g., canyon) exhibits the same sort of resonant response (this question also applies to  trenches (for screening seismic waves and other types of vibrations \cite{bd86,cd95}), surface-breaking cracks \cite{ts10} and subsurface tunnels and cavities; see \cite{ll19} for a very complete bibliography on this latter subject).

\begin{figure}[ptb]
\begin{center}
\includegraphics[width=0.55\textwidth]{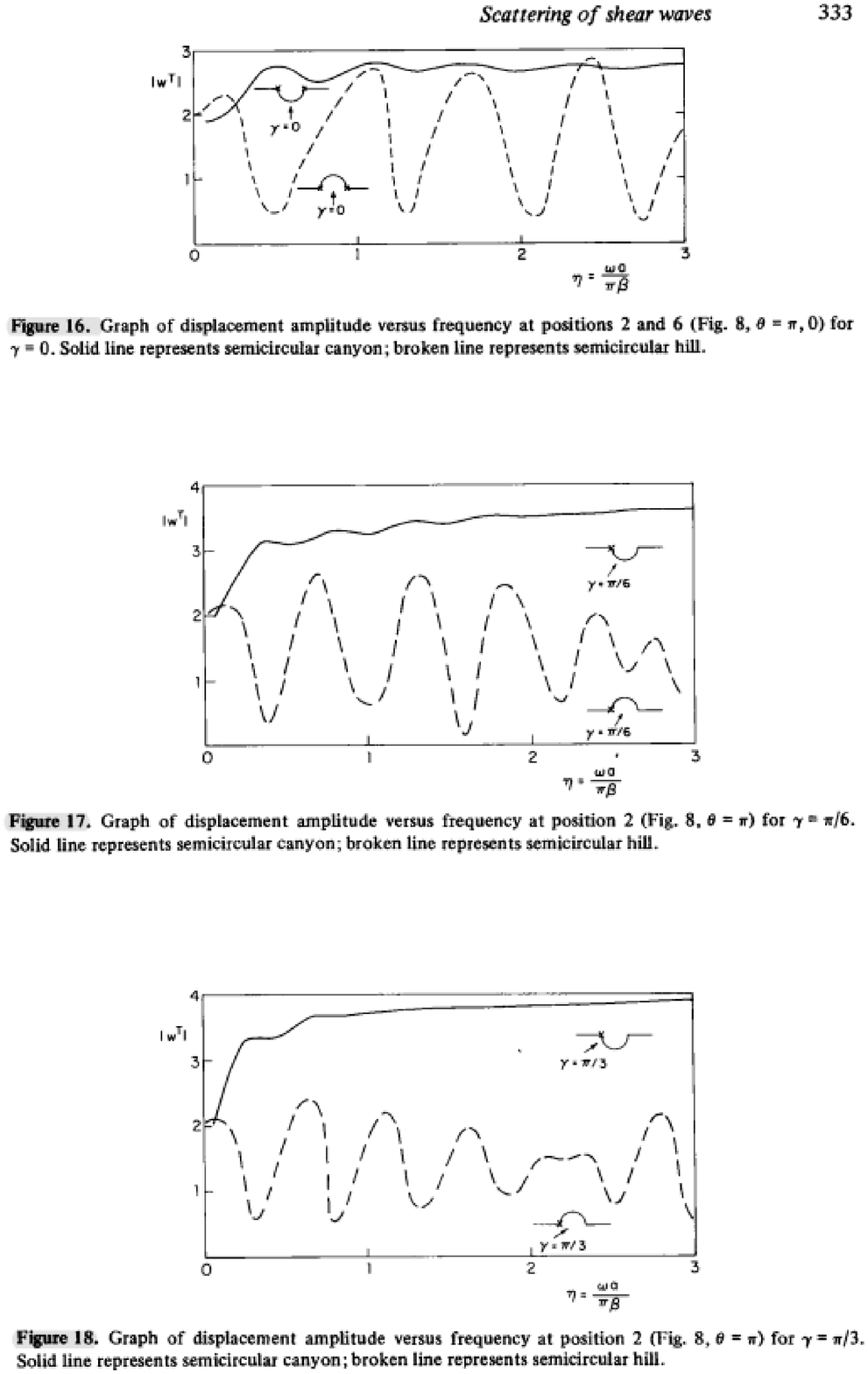}
\caption{Figs. 16-18 in \cite{si78}. Transfer functions, computed via a boundary-integral equation, for shear-horizontal plane wave incidence on a traction-free, semi-circular (radius $a$) cylindrical hill (dashed curves) and canyon (full curves). The medium underneath these surface irregularities is linear, homogeneous and isotropic, with $\beta$ the bulk shear velocity therein. The abscissa $\omega a/(\pi\beta)$ is the dimensionless frequency, $\omega=2\pi f$  the angular frequency and $f$ the frequency. The ordinate represents the total SH displacement field response at the indicated points (top center for upper left-hand panel, left-hand corner for bottom left-hand panel, and right-hand corner for bottom right-hand panel) of the boundary.}
\label{fsills}
\end{center}
\end{figure}

After searching the literature dealing with the scattering of elastic waves in general, and the scattering of seismic waves from surface irregularities in particular, I found two articles \cite{tr73,si78} (see fig. \ref{fsills}) herein) which show that the spectral  response (i.e., transfer function) of unfilled below-ground features is much smoother than that of the similar above-ground feature (e.g., the latter being the semi-circular mirror-image of the below-ground feature) and, in any case not evocative of resonant response.

This finding was later corroborated in \cite{wj82,be87,tc08,ea09,tc10,zg12,cl90a,cl90b,le90,lc89,ml96,ss87,sg80,tc08,ts10,wo82,zg12,sk10} for canyons of various other  (e.g., elliptical, triangular, parabolic, gaussian) shapes (however, the results in \cite{rs11} are less evident). Often, articles dealing with the response to elastic waves of surface or subsurface features such as (unfilled) cracks, canyons, dams, trenches, tunnels, etc.  do not even contain the transfer functions by which resonant effects can be made apparent \cite{qc20,fp14,ka88,ll09,ss81,sr79,wt74,lb15,wo82,zc06}. There also exist papers  on this subject that exhibit transfer functions with strange resonant features but which did not elicit discussion by their authors as to their origin \cite{dw12,ll19}.
%

On the other hand, as early as 1988,  Nowak \cite{no88} sent a signal to the elastic wave community as to the possible existence of 'artificial' resonances (in the sense that  the latter are the result of the numerical method employed for predicting the response rather than being of physical origin). A few years later,  Nowak and Hall \cite{nh93} gave more evidence of this phenomenon, but their study has all but been forgotten (even in  review articles and books such as \cite{lm11,md16,md15}) in spite of the fact that it was published in one of the leading journals of the elastic wave community. Thus, it would appear that, at present, the consensus  is that all elastic wave resonances are either of well-known physical origin (e.g., Love, Rayleigh (1D variety) resonances, Bard-Bouchon (2D variety) resonances \cite{bb85},...) or non-existent. However, the results in \cite{no88,nh93} leave some room for doubt.

\begin{figure}[ht]
\begin{center}
\includegraphics[width=0.5\textwidth]{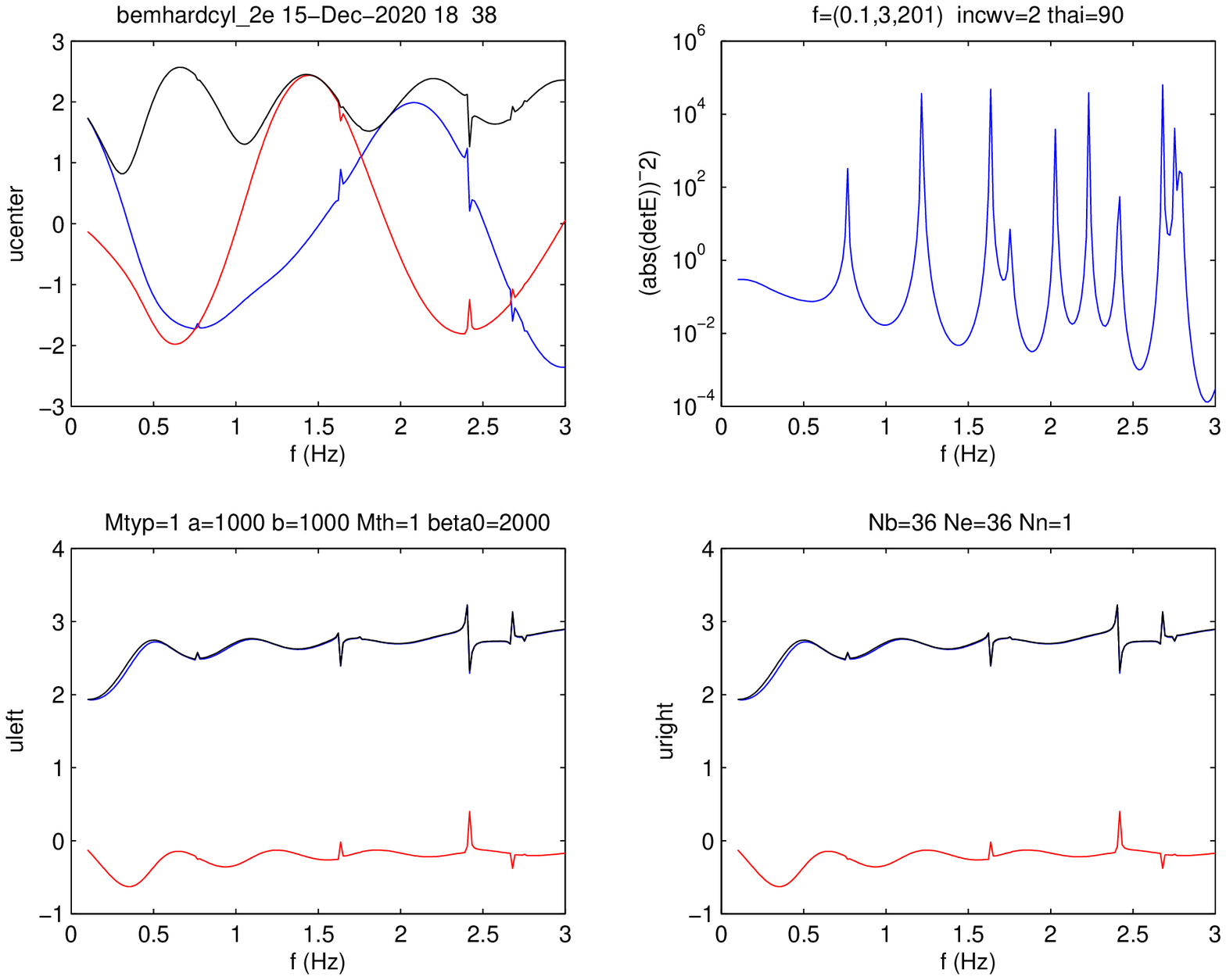}
\caption{Same  scattering  problem as in \cite{si78} and fig. 1. Transfer functions, computed via a boundary-integral equation, for shear-horizontal plane wave normal incidence on a semi-circular (radius $a$) traction-free cylindrical canyon. The ordinates represent: blue curves  for the real part, red curves  for the imaginary part and  black curves  for the modulus of the total displacement at the middle top point (upper left-hand panel), left-hand corner point (lower left panel) and right-hand corner point (lower right-hand panel). The medium underneath these surface irregularities is linear, homogeneous and isotropic, with $\beta$ the bulk shear velocity therein. The abscissa  is as in fig. \ref{fsills} herein since my choice of $a$ and $\beta$ is such that $f=\omega a/(\pi\beta)$.  The ordinate represents the total SH displacement field response at the indicated points of the boundary. The upper right-hand panel represents the modulus of the determinant of the matrix equation involved in the computation. The information content of this panel is discussed further on.}
\label{mysills}
\end{center}
\end{figure}

My interest in this issue was further stimulated by a computation I recently made to verify the results of Sills \cite{si78} depicted herein in fig. \ref{fsills}. I chose the same scattering configuration as in the upper panel of fig. \ref{fsills} herein and essentially the same BIE (boundary integral equation) numerical algorithm as Sills to obtain the solutions of fig. \ref{mysills}.

It is readily-observed that the overall response at the two corners is the same as that predicted by Sills, but this response is also marked by what appears as resonant features (spikes) similar to those e.g., found by Nowak and Hall (their fig. 5 in \cite{nh93}) as well as by Bendali and Fares (their figs. 2 and 3 in \cite{bf07}.

For this, and the previously-mentioned reasons, I thought it to be necessary to undertake the present study, whose purpose  is to find out, by theoretical and numerical means whether unfilled below-ground surface features submitted to elastic waves are able to produce resonant response, and if so, is this response real or simply the result of some theoretical/numerical misplay that must, and can, be eliminated.

This study will hinge predominantly on the example of the scattering of elastic waves by a cylindrical canyon of semi-circular shape, and by extension, that of scattering of elastic waves by a circular cylinder. To demonstrate the universal nature of my demonstration, I will treat not only the stress-free (homogenous Neumann) boundary condition but also the rigid (homogeneous Dirichlet) boundary condition. Both of these problems have obvious counterparts in the fields of fluid acoustics \cite{sc68,mh04,tg01,bf07,cc01a,cc01b}, and electromagnetism \cite{bt70,bt73,mh77,mh78,ma99,sm83,wi02,xh05,ss84,lm11} where they have been, and continue to be,  intensely studied (in fact, much more so than in the elastic wave community), increasingly with the support of applied mathematicians \cite{bw65,bm71,cl06,am87,ad07,ch20,za00,zz15}. This involvement of mathematicians explains why the subject of what turns out to be that of spurious resonances is becoming more and more abstract and therefore not necessarily familiar to engineers and geophysicists. This is the reason why my study is largely restricted to a single canonical problem that can be solved, by well-known techniques (most of which are explained, and placed in their historical context, in the classical work of Mow and Pao \cite{mp71}), in closed form, and thus able to reveal its subtle features in a relatively-simple manner.
\section{Description of the the canonical elastic wave scattering problem}\label{desc}
\begin{figure}[ht]
\begin{center}
\includegraphics[width=0.5\textwidth]{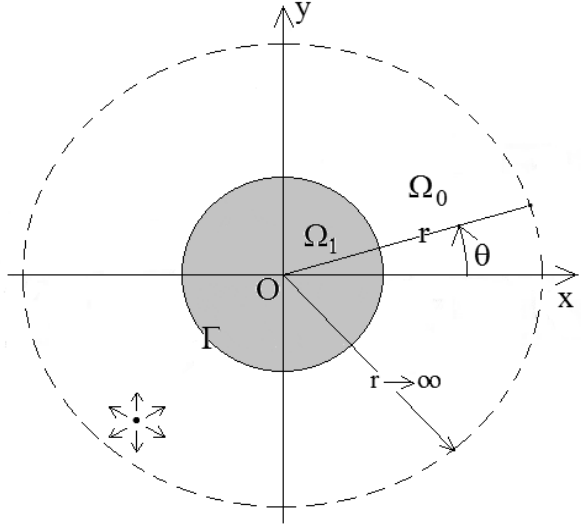}
\caption{Sagittal (i.e., cross-section) plane view of 2D scattering configuration of an impervious cylinder subjected to the wave radiated by  line sources (here only one is depicted). The dashed curve is the virtual boundary at infinity of the region exterior to the cylindrical object. }
\label{basin}
\end{center}
\end{figure}
The canonical elastic wave problem is that of the scattering, by a  cylindrical object, of the shear-horizontal (SH) wave radiated by one or two line sources (parallel to the cylinder and located outside the object; some examples will also be given of plane wave solicitation). The boundary of the object is circular (in the cross-section plane) and the locus of either a homogeneous Neumann (stress-free body in the context of elastic waves) or homogeneous Dirichlet (rigid body in the context of elastic waves) condition.  This problem is two dimensional, with means that the elastic wavefield does not depend on the $z$ coordinate of a cartesian system $Oxyz$ with origin $O$ at the center of the circular boundary (i.e., the $z$-axis is the axis of the cylinder and the line source is parallel to $z$ axis). The medium in the region exterior to the cylinder is linear, homogeneous and isotropic and the bulk shear wavespeed therein is $\beta$. The wavefield is nil within the cylinder due to the nature of the assumed boundary conditions. Since the exterior region is of infinite extent, a sort of 'boundary' condition must be specified at points infinitely-distant from the cylinder boundary. This is the so-called radiation condition which states that the scattered field behaves like an outgoing wave at these points. The relation of one of these problems (the one related to the Neumann boundary condition) to the problem of elastic wave scattering by a canyon will be explained further on.
\section{The scattering problem  in the frequency domain for the Dirichlet boundary condition}
\subsection{Governing equations}\label{goveq}
These equations are \cite{wi19}:
\begin{equation}\label{1-010}
u(\mathbf{x})=u^{i}(\mathbf{x})+u^{s}(\mathbf{x})
~,
\end{equation}
\begin{equation}\label{1-020}
\big(\nabla\cdot\nabla+k^{2}\big)u(\mathbf{x})=-s(\mathbf{x})~;~\forall\mathbf{x}\in\Omega_{0}~,
\end{equation}
\begin{equation}\label{1-030}
u^{s}(\mathbf{x})\sim {\text{outgoing wave}};~\|\mathbf{x}\|\rightarrow\infty~,
\end{equation}
\begin{equation}\label{1-040}
u(\mathbf{x})=0~;~\mathbf{x}\in\Gamma~,
\end{equation}
wherein:\\
a) $\mathbf{x}$ is a vector in the $x-y$ (cross-section) plane  directed from the origin $0$ to an arbitrary point $(x,y)$ in cartesian coordinates or $r,\theta$ in polar coordinates,\\
b)  as concerns the displacement frequency domain fields:  $u^{i}(\mathbf{x})$ is shorthand for $u_{z}^{i}(\mathbf{x};\omega)$,  $u^{s}(\mathbf{x})$ is shorthand for $u_{z}^{s}(\mathbf{x};\omega)$,  $u(\mathbf{x})$ is shorthand for $u_{z}(\mathbf{x};\omega)$, with $\omega=2\pi f$ the angular frequency and $f$ the frequency,\\
c) $u^{i}(\mathbf{x})$ is the wave (called 'incident wave'), radiated by the source of density $s(\mathbf{x})$, that exists in the configuration in which the body is absent,\\
d) $u^{s}(\mathbf{x})$ is the scattered field,\\
e) $u(\mathbf{x})$ is the total field in the region $\Omega_{0}$ exterior to the body, the interior of the latter being denoted by $\Omega_{1}$,\\
f) $\Gamma$ is the closed curve delineating the boundary between $\Omega_{0}$ and $\Omega_{1}$, and, at present, I take this curve to be a circle of radius $a$ (note that $\Omega_{0}$, $\Omega_{1}$, and $\Gamma$ are geometric entities in the $x-y$ plane),\\
g) the frequency domain field is related to the time $(t)$ domain field by the relation $u(\mathbf{x};t)=2\Re\int_{0}^{\infty}u(\mathbf{x};\omega)\exp(-i\omega t)d\omega$,\\
h) $k=\omega/\beta$ is the (positive real) wavenumber.\\\\
Note that I am dealing with a forward-scattering problem, i.e., $s(\mathbf{x})$ and therefore $u^{i}$, $\beta$, $a$, $\omega$ are assumed to be known and the problem is to determine $u^{s}$ and/or $u$.
\subsection{The free-space Green's function}\label{fsgf}
 The material in this section is  of general nature (i.e., not dependent on the presence of a scattering body) and can be found in more detail in \cite{mf53,wi19} . The free-space Green's function $G(\mathbf{x};\mathbf{x}')$ satisfies (\ref{1-020})-(\ref{1-030}), in which $s(\mathbf{x})=\delta(\mathbf{x}-\mathbf{x}')$, $\delta(~)$ is the Dirac delta distribution and $\mathbf{x}'=(x',y')=(r',\theta')$. It then turns out that
\begin{equation}\label{1-050}
G(\mathbf{x};\mathbf{x})=\frac{i}{4}H_{0}^{(1)}(k\|\mathbf{x}-\mathbf{x}'\|)~,
\end{equation}
wherein $H_{l}^{(1)}(~)$ is the $l$-th-order Hankel function of the first kind and $\mathbf{x}'=(x',y')=(r',\theta')$. The polar representation of $G$ is:
\begin{equation}\label{1-060}
G(\mathbf{x};\mathbf{x})=\frac{i}{4}\sum_{l=-\infty}^{\infty}\left[H(r-r')H_{l}^{(1)}(kr)J_{l}(kr')+
H(r'-r)H_{l}^{(1)}(kr')J_{l}(kr)\right]\exp[il(\theta-\theta')]~,
\end{equation}
in which $J_{l}(~)$ is the $l$-th-order Bessel function, $H_{l}^{(1)}(~)$ is the $l$-th-order Hankel function of the first kind, and $H(~)$ the 1-dimensional Heaviside distribution.
\subsection{The relation of $u^{i}$ to $s$}\label{uitus}
The material in this section is likewise of general nature (i.e., not dependent on the presence of a scattering body). As shown in \cite{wi19}
\begin{equation}\label{1-070}
u^{i}(\mathbf{x})=\int_{\mathbb{R}^{2}}G(\mathbf{x};\mathbf{x})s(\mathbf{x}')d\varpi(\mathbf{x}')~,
\end{equation}
wherein $d\varpi$ is the differential surface element in the $x-y$ plane. Note that for a line source located at $\mathbf{x}^{s}=(r^{s},\theta^{s})$, $s(\mathbf{x})=\delta(\mathbf{x}-\mathbf{x}^{s})$, so that
\begin{equation}\label{1-075}
u^{i}(\mathbf{x})=\int_{\mathbb{R}^{2}}G(\mathbf{x};\mathbf{x})\delta(\mathbf{x}'-\mathbf{x}^{s})d\varpi(\mathbf{x}')=G(\mathbf{x};\mathbf{x}^{s})=
\frac{i}{4}H_{0}^{(1)}(k\|\mathbf{x}-\mathbf{x}^{s}\|)~.
\end{equation}
More generally, always assuming that the source is in $\Omega_{0}\subset \mathbb{R}^{2}$,  but of finite support $\Omega_{s}$, then the latter  is entirely within $\Omega_{0}$, so that
\begin{equation}\label{1-080}
u^{i}(\mathbf{x})=\int_{\Omega_{0}}G(\mathbf{x};\mathbf{x})s(\mathbf{x}')d\varpi(\mathbf{x}')=
\int_{\Omega_{s}}G(\mathbf{x};\mathbf{x})s(\mathbf{x}')d\varpi(\mathbf{x}')~.
\end{equation}
If the (finite) support of the source is between $r=r^{-}$ and $r=r^{+}>r^{-}$ in terms of $r$ and between $\theta=\theta^{-}$ and $\theta=\theta^{+}>\theta{-}$ in terms of $\theta$, with the understanding that $r^{-}>a$, then
\begin{multline}\label{1-090}
u^{i}(\mathbf{x})=\frac{i}{4}
\sum_{l=-\infty}^{\infty}\int_{\theta^{-}}^{\theta^{+}}d\theta'\exp[il(\theta-\theta')]\times\\
\int_{r^{-}}^{r^{+}}dr'r'\left[H(r-r')H_{l}^{(1)}(kr)J_{l}(kr')+
H(r'-r)J_{l}(kr)H_{l}^{(1)}(kr')\right]~
s(r',\theta')~~;~~\forall~\mathbf{x}\in
\mathbb{R}^{3}~,
\end{multline}
so that, from the definition of the Heaviside distribution,
\begin{multline}\label{1-100}
u^{i}(r>r^{+},\theta)=
\sum_{l=-\infty}^{\infty}H_{l}^{(1)}(kr)\exp[il\theta]\frac{i}{4}\int_{\theta^{-}}^{\theta^{+}}d\theta'\exp[-il\theta']
\int_{r^{-}}^{r^{+}}dr'r'J_{l}(kr')~
s(r',\theta'):=\\
\sum_{l=-\infty}^{\infty}B_{l}H_{l}^{(1)}(kr)\exp[il\theta]~~;~~\forall~\theta\in[0,2\pi[~,
\end{multline}
with
\begin{equation}\label{1-110}
B_{l}=
\frac{i}{4}\int_{\theta^{-}}^{\theta^{+}}d\theta'\exp[-il\theta']
\int_{r^{-}}^{r^{+}}dr'r'J_{l}(kr')s(r',\theta')
~.
\end{equation}
and
\begin{multline}\label{1-120}
u^{i}(r<r^{-},\theta)=
\sum_{l=-\infty}^{\infty}J_{l}(kr)\exp[il\theta]\frac{i}{4}\int_{\theta^{-}}^{\theta^{+}}d\theta'\exp[-il\theta']
\int_{r^{-}}^{r^{+}}dr'r'H_{l}^{(1)}(kr')~
s(r',\theta'):=\\
\sum_{l=-\infty}^{\infty}A_{l}J_{l}(kr)\exp[il\theta]~~;~~\forall~\theta\in[0,2\pi[~,
\end{multline}
with
\begin{equation}\label{1-130}
A_{l}=
\frac{i}{4}\int_{\theta^{-}}^{\theta^{+}}d\theta'\exp[-in\theta']
\int_{r^{-}}^{r^{+}}dr'r'H_{l}^{(1)}(kr')s(r',\theta')~
~.
\end{equation}
Consequently, the representation of the incident wave, due to applied sources in $\Omega_{s}$, that I have to take into account in the boundary condition (\ref{1-040}), is
\begin{equation}\label{1-140}
u^{i}(r<r^{-},\theta)=\sum_{l=-\infty}^{\infty}A_{l}J_{l}(kr)\exp(il\theta)~;~\forall~\theta\in[0,2\pi[~.
\end{equation}
I assumed that the source is linear and located at $(r^{s},\theta^{s})$  {\it exterior} to the cylinder, so that $r^{-}=r^{s}>a$, whence the  associated field on $\Gamma$ is
\begin{equation}\label{1-150}
u^{i}(a,\theta)=\sum_{l=-\infty}^{\infty}A_{l}J_{l}(ka)\exp(il\theta)~;~\forall~\theta\in[0,2\pi[~,
\end{equation}
wherein
\begin{equation}\label{1-160}
A_{l}=\frac{i}{4}H_{l}(kr^{s})\exp(-il\theta^{s})~.
\end{equation}
More generally, but always in the case of a linear source,
\begin{equation}\label{1-163}
u^{i}(r,\theta)=\sum_{l=-\infty}^{\infty}\left[H(r-r^{s})B_{m}H_{l}^{(1)}(kr)+H(r^{s}-r)A_{l}J_{l}(kr)\right]e^{il\theta}~,
\end{equation}
with
\begin{equation}\label{1-165}
B_{l}=\frac{i}{4}J_{l}(kr^{s})\exp(-il\theta^{s})~.
\end{equation}
\subsection{The separation of variables (SOV) solution (i.e., DSOV) for the Dirichlet-boundary body}
The well-known SOV technique consists (for 2D problems such as mine) in assuming that the solution (actually just a representation thereof)  can be expressed as the product of two functions, each of which depends on only one of the two chosen coordinates, whereupon the partial differential (wave) equation (\ref{1-020}) separates into two independent ordinary differential equations the solution of which can be expressed in terms of elementary functions.

I choose the $r,\theta$ coordinates so that the $\theta$ differential equation turns out to have solutions $\exp(in\theta)~;~n\in \mathbb{Z}$ whereas the $r$ differential equation has solutions $J_{n}(kr)~;~n\in\mathbb{Z}$ on the one hand, and $H_{n}^{(1)}(kr)~;~n\in\mathbb{Z}$ on the other hand. The solutions in terms of the Bessel functions can be ruled out in the region $\Omega_{0}$ exterior to the scattering object because of the radiation condition (\ref{1-030}) so that the SOV representation of the scattered field in $\Omega_{0}$ becomes
\begin{equation}\label{1-170}
u^{s}(r,\theta)=\sum_{n\in\mathbb{Z}}C_{n}H_{n}^{(1)}(kr)\exp(in\theta)~~;~~\forall~\theta\in[0,2\pi[~.
\end{equation}
The actual SOV solution to the scattering problem requires the invocation of the boundary condition (\ref{1-040}) and (\ref{1-010})
\begin{equation}\label{1-180}
u^{i}(a,\theta)+u^{s}(a,\theta)=\sum_{n\in\mathbb{Z}}\left[A_{n}J_{n}(ka)+C_{n}H_{n}^{(1)}(ka)\right]\exp(in\theta)=0
~~;~~\forall~\theta\in[0,2\pi[~.
\end{equation}
The solution for $\{C_{n}\}$ is quite obvious (recall that $\{A_{n}\}$ is known via (\ref{1-160}) and from the fact that $r^{s},\theta^{s}$ are known), but I wish to bring to the fore a feature that will be useful further on.  Thus, I choose to project (\ref{1-180}) as follows:
\begin{equation}\label{1-190}
\int_{0}^{2\pi}\sum_{n\in\mathbb{Z}}\left[A_{n}J_{n}(ka)+C_{n}H_{n}^{(1)}(ka)\right]\exp(in\theta)\exp(-im\theta)d\theta=0~;\forall m\in\mathbb{Z}
~,
\end{equation}
which, after interchanging the integral and the sum, and making use of the identity (in which $\delta_{mn}$ is the Kronecker delta symbol)
\begin{equation}\label{1-200}
\int_{0}^{2\pi}\exp[i(n-m)\theta]d\theta=2\pi\delta_{mn}~;\forall m\in\mathbb{Z}
~,
\end{equation}
yields
\begin{equation}\label{1-210}
\sum_{n\in\mathbb{Z}}\left[-H_{n}^{(1)}(ka)\delta_{mn}\right]C_{n}=A_{n}J_{n}(ka)~;\forall m\in\mathbb{Z}
~,
\end{equation}
which is an infinite-order matrix equation in which the matrix $[~]$ is diagonal and non-singular for real frequencies $f$ due to the fact that the Hankel function is complex and its real and imaginary parts vanish for different values of $ka$ \cite{as68}.
It follows, by simple matrix inversion, that
\begin{equation}\label{1-220}
C_{n}=-A_{n}\frac{J_{n}(ka)}{H_{n}^{(1)}(ka)}~;\forall n\in\mathbb{Z}
~.
\end{equation}
Thus, on account of (\ref{1-010}) and (\ref{1-163})
\begin{multline}\label{1-230}
u(\mathbf{x})=\sum_{n\in\mathbb{Z}}\left\{H(r-r^{s})B_{n}H_{n}^{(1)}(kr)+
A_{n}\left[H(r^{s}-r)J_{n}(kr)-\frac{J_{n}(ka)}{H_{n}^{(1)}(ka)}H_{n}^{(1)}(kr)\right]\right\}\times\\
\exp(in\theta)~;~\forall\mathbf{x}\in\Omega_{0}
~,
\end{multline}
wherein the $A_{n}$ and $B_{n}$ are given in (\ref{1-160}) and (\ref{1-165}) respectively.

Eq. (\ref{1-230}) can be considered as the exact solution to the scattering problem. This solution for $u$  shows no sign of resonances.

As I show further on, it is of some interest to determine a function related to the traction (in the context of elastic wave problems) on the scattering boundary. This function is
\begin{equation}\label{1-240}
v(\mathbf{x})=\frac{1}{k}\boldsymbol{\nu}\cdot\nabla u(\mathbf{x})\big |_{\Gamma}
~,
\end{equation}
wherein $\boldsymbol{\nu}$ is the inner-directed unit vector normal to $\Gamma$. At present, this function is
\begin{equation}\label{1-245}
v(a,\theta)=-\frac{1}{k}\frac{\partial u(r,\theta)}{\partial r}\Big |_{r=a}
~,
\end{equation}
so that making use of (\ref{1-230}) gives
\begin{equation}\label{1-250}
v(a,\theta))=-\sum_{n\in\mathbb{Z}}A_{n}\left[\dot{J}_{n}(ka)-
\frac{J_{n}(ka)}{H_{n}^{(1)}(ka)}\dot{H}_{n}^{(1)}(ka)\right]\exp(in\theta)~;\theta\in [0,2\pi[
~,
\end{equation}
where $\dot{Z}_{n}(z)=\frac{dZ_{n}}{dz}$. I now make use of the identity (9.1.16) in \cite{as68}
\begin{equation}\label{1-260}
\dot{H}_{n}^{(1)}(z)J_{n}(z)-H_{n}^{(1)}(z)\dot{J}_{n}(z)=\frac{2i}{\pi z}
~,
\end{equation}
to finally obtain
\begin{equation}\label{1-270}
v(a,\theta))=\sum_{n\in\mathbb{Z}}\left[A_{n}\left(\frac{2i}{\pi ka}\right)\left(\frac{1}{H_{n}^{(1)}(ka)}\right)\right]\exp(in\theta)~;~\theta\in [0,2\pi[
~,
\end{equation}
This solution for $v$ shows no sign of resonances either.
\subsection{Some consequences of Green's second identity}
The material in this section does  not depend on the specific conditions on the boundary  of the scattering body.

As previously, consider an incident wave $u^{i}$ (now not necessarily that radiated by a line source, but outgoing from the location of the source) impinging on a cylindrical closed body whose boundary $\Gamma$ in the cross-section plane (now not necessarily circular) separates the inner region $\Omega_{1}$ of finite extent from the outer region $\Omega_{0}$ of inifinite extent, both of these regions being subsets of $\mathbb{R}^{2}$. Let $\boldsymbol{\nu}(\mathbf{x}')$ designate the unit vector normal to $\Gamma$ at point $\mathbf{x}'\in\Gamma$, directed towards the inside of $\Omega_{1}$ and therefore towards the outside of $\Omega_{0}$. I now address the  problem defined by (\ref{1-010})-(\ref{1-040}), without specifying, for the moment, the boundary condition on $\Gamma$.

As shown in \cite{mp71,wi19}, on account(\ref{1-010})-(\ref{1-030}), Green's second identity leads  to the expression
\begin{equation}\label{1-300}
\mathcal{H}_{\Omega_{0}}(\mathbf{x})u(\mathbf{x})=u^{i}(\mathbf{x})+
\int_{\Gamma}\left[G(\mathbf{x};\mathbf{x}')\boldsymbol{\nu}(\mathbf{x}')\cdot\nabla (\mathbf{x}')u(\mathbf{x}')-
u(\mathbf{x}')\boldsymbol{\nu}(\mathbf{x}')\cdot\nabla (\mathbf{x}')G(\mathbf{x};\mathbf{x}')\right]d\gamma(\mathbf{x}')
~,
\end{equation}
wherein:\\
 a) $d\gamma$ is the differential element of arc length along $\Gamma$,\\
 b) $\mathcal{H}_{\Omega_{0}}(\mathbf{x})=1~;~\mathbf{x}\in\Omega_{0}$,
 $\mathcal{H}_{\Omega_{0}}(\mathbf{x})=0~;~\mathbf{x}\in\Omega_{1}$ is the 2D Heaviside distribution.\\
  A question of some importance is what value should be attributed to $\mathcal{H}_{\Omega_{0}}(\mathbf{x})$ when $\mathbf{x}\in\Gamma$. The answer is not clear-cut unless one asks the same question regarding the integral involving the normal derivative of the Green's function. Following common usage, I attribute the value $1/2$ to $\mathcal{H}_{\Omega_{0}}(\mathbf{x}\in\Gamma)$ provided the integral involving the normal derivative of $G$ is evaluated in the sense of a Cauchy principal value, the designation of which hereafter is $pv$. Thus, the three consequences of (\ref{1-300}) are:
\begin{equation}\label{1-310}
u(\mathbf{x})=u^{i}(\mathbf{x})+
\int_{\Gamma}\left[G(\mathbf{x};\mathbf{x}')\boldsymbol{\nu}(\mathbf{x}')\cdot\nabla (\mathbf{x}')u(\mathbf{x}')-
u(\mathbf{x}')\boldsymbol{\nu}(\mathbf{x}')\cdot\nabla (\mathbf{x}')G(\mathbf{x};\mathbf{x}')\right]d\gamma(\mathbf{x}')
~;~\forall\mathbf{x}\in\Omega_{0}~,
\end{equation}
\begin{equation}\label{1-320}
\frac{1}{2}u(\mathbf{x})=u^{i}(\mathbf{x})+
\int_{\Gamma}G(\mathbf{x};\mathbf{x}')\boldsymbol{\nu}(\mathbf{x}')\cdot\nabla (\mathbf{x}')u(\mathbf{x}')d\gamma(\mathbf{x}')-
pv\int_{\Gamma}u(\mathbf{x}')\boldsymbol{\nu}(\mathbf{x}')\cdot\nabla (\mathbf{x}')G(\mathbf{x};\mathbf{x}')d\gamma(\mathbf{x}')
~;~\forall\mathbf{x}\in\Gamma~,
\end{equation}
\begin{equation}\label{1-330}
0=u^{i}(\mathbf{x})+
\int_{\Gamma}\left[G(\mathbf{x};\mathbf{x}')\boldsymbol{\nu}(\mathbf{x}')\cdot\nabla (\mathbf{x}')u(\mathbf{x}')-
u(\mathbf{x}')\boldsymbol{\nu}(\mathbf{x}')\cdot\nabla (\mathbf{x}')G(\mathbf{x};\mathbf{x}')\right]d\gamma(\mathbf{x}')
~;~\forall\mathbf{x}\in\Omega_{1}~.
\end{equation}
The object of what follows is obviously to apply any one of these boundary integral (BI) expressions, or combinations thereof, to solve the various boundary-value problems mentioned in the Introduction.
\subsection{The three BI expressions for the case of a Dirichlet boundary condition}
These are:
\begin{equation}\label{1-340}
u(\mathbf{x})=u^{i}(\mathbf{x})+
\int_{\Gamma}G(\mathbf{x};\mathbf{x}')\boldsymbol{\nu}(\mathbf{x}')\cdot\nabla (\mathbf{x}')u(\mathbf{x}')d\gamma(\mathbf{x}')
~;~\forall\mathbf{x}\in\Omega_{0}~,
\end{equation}
\begin{equation}\label{1-350}
0=u^{i}(\mathbf{x})+
\int_{\Gamma}G(\mathbf{x};\mathbf{x}')\boldsymbol{\nu}(\mathbf{x}')\cdot\nabla (\mathbf{x}')u(\mathbf{x}')d\gamma(\mathbf{x}')
~;~\forall\mathbf{x}\in\Gamma~,
\end{equation}
\begin{equation}\label{1-360}
0=u^{i}(\mathbf{x})+
\int_{\Gamma}G(\mathbf{x};\mathbf{x}')\boldsymbol{\nu}(\mathbf{x}')\cdot\nabla (\mathbf{x}')u(\mathbf{x}')d\gamma(\mathbf{x}')
~;~\forall\mathbf{x}\in\Omega_{1}~.
\end{equation}
The first of these three only enables to determine the wavefield in the outer region {\it after} determining the normal derivative of $u$ on $\Gamma$ either by the second or third BI equation (BIE for short), or by a combination of these two BIE. Note that (\ref{1-350}) is a first-kind BIE and (\ref{1-360}) is what is frequently called an 'extended boundary condition' (EBC).
\subsection{Solution of the first kind BIE (i.e., DBIE1) for the case of a Dirichlet condition on the circular boundary}
The BIE is:
\begin{equation}\label{1-370}
0=u^{i}(\mathbf{x})+
\int_{\Gamma}kG(\mathbf{x};\mathbf{x}')v(\mathbf{x}')d\gamma(\mathbf{x}')
~;~\forall\mathbf{x}\in\Gamma~,
\end{equation}
wherein $v$ was defined in (\ref{1-240}). The circular nature of $\Gamma$ entails:
\begin{equation}\label{1-380}
0=u^{i}(a,\theta)+
\int_{0}^{2\pi}kG(a,\theta;a,\theta')v(a,\theta')ad\theta'
~;~\forall\theta\in[0,2\pi[~,
\end{equation}
and the task is henceforth to determine $v$.

The $2\pi$-periodic nature (in terms of $\theta$) of $u^{i}$ and $v$ incites one to expand these functions in terms of Fourier basis functions:
\begin{equation}\label{1-390}
u^{i}(a,\theta)=\sum_{n\in\mathbb{Z}}g_{n}\exp(in\theta)~~,~~v(a,\theta)=\sum_{n\in\mathbb{Z}}f_{n}\exp(in\theta)
~;~\forall\theta\in[0,2\pi[~,
\end{equation}
and to employ a Galerkin procedure, consisting of projecting the integral equation on the same Fourier basis set of functions so as to obtain, after sum and integral exchanges and  use of (\ref{1-200}):
\begin{equation}\label{1-400}
0=g_{m}+\sum_{n\in\mathbb{Z}}f_{n}\int_{0}^{2\pi}d\theta\exp(-im\theta)\int_{0}^{2\pi}d\theta'\frac{ka}{2\pi}G(a,\theta;a,\theta')\exp(in\theta')
~;~\forall m\in\mathbb{Z}~.
\end{equation}
I now make use of (\ref{1-060})
\begin{multline}\label{1-410}
G(a,\theta;a,\theta')=\frac{i}{4}\sum_{l=-\infty}^{\infty}\left[H(0_{+})H_{l}^{(1)}(ka)J_{l}(ka)+
H(0_{-})H_{l}^{(1)}(ka)J_{l}(ka)\right]\exp[il(\theta-\theta')]=\\
\frac{i}{4}\left[H(0_{+})+H(0_{-})\right]\sum_{l=-\infty}^{\infty}H_{l}^{(1)}(ka)J_{l}(ka)\exp[il(\theta-\theta')]~,
\end{multline}
or, by virtue of the definition of the 1D Heaviside distribution,
\begin{equation}\label{1-415}
G(a,\theta;a,\theta')=\frac{i}{4}\sum_{l=-\infty}^{\infty}H_{l}^{(1)}(ka)J_{l}(ka)\exp[il(\theta-\theta')]~,
\end{equation}
I find
\begin{equation}\label{1-420}
0=g_{m}+\sum_{n\in\mathbb{Z}}f_{n}\sum_{l\in\mathbb{Z}}\frac{ika}{8\pi}H_{l}^{(1)}(ka)J_{l}(ka)\int_{0}^{2\pi}d\theta\exp[i(l-m)\theta]\int_{0}^{2\pi}d\theta'\exp[i(n-l)\theta']
~;~\forall m\in\mathbb{Z}~.
\end{equation}
or, on account of (\ref{1-200})
\begin{equation}\label{1-430}
0=g_{m}+\sum_{n\in\mathbb{Z}}f_{n}\sum_{l\in\mathbb{Z}}\frac{ika\pi}{2}H_{l}^{(1)}(ka)J_{l}(ka)\delta_{ml}\delta_{ln}
~;~\forall m\in\mathbb{Z}~.
\end{equation}
Employment of the sifting properties of the Kronecker delta, finally leads to
\begin{equation}\label{1-440}
0=g_{m}+\sum_{n\in\mathbb{Z}}f_{n}\frac{ika\pi}{2}H_{n}^{(1)}(ka)J_{n}(ka)\delta_{mn}
~;~\forall m\in\mathbb{Z}~.
\end{equation}
which can be re-written as the  matrix equation
\begin{equation}\label{1-450}
\sum_{n\in\mathbb{Z}}E_{mn}f_{n}=g_{m}~;~\forall m\in\mathbb{Z}~,
\end{equation}
wherein
\begin{equation}\label{1-460}
E_{mn}=\frac{-ika\pi}{2}H_{n}^{(1)}(ka)J_{n}(ka)\delta_{mn}~;~\forall m,n\in\mathbb{Z}~.
\end{equation}
Once again, I have to deal with a diagonal infinite-order matrix, thus enabling, in theory, the obtention of a closed-form solution for ${f_{n}}$. But I forsee a major problem due to the fact that now this matrix vanishes for certain real frequencies, this being due to fact that  the Bessel functions are equal to zero at an infinite discrete set of their real arguments \cite{as68}. Be this as it may, at real frequencies not in the neigborhood of the indicated frequencies, it is legitimate to invert $\mathbf{E}=\{E_{mn}\}$ whence
\begin{equation}\label{1-470}
\mathbf{f}=\mathbf{E}^{-1}\mathbf{g}~~\Rightarrow~~
f_{m}=\left[\frac{-ika\pi}{2}H_{m}^{(1)}(ka)J_{m}(ka)\right]^{-1}g_{m}~;~\forall m\in\mathbb{Z}~.
\end{equation}
If I recall that for my line source
\begin{equation}\label{1-480}
u^{i}(a,\theta)=\sum_{m\in\mathbf{Z}}A_{m}J_{m}(ka)\exp(im\theta)=\sum_{m\in\mathbf{Z}}g_{m}\exp(im\theta)~,
\end{equation}
then
\begin{equation}\label{1-485}
g_{m}=A_{m}J_{m}(ka)~,
\end{equation}
whence
\begin{equation}\label{1-490}
\mathbf{f}=\mathbf{E}^{-1}\mathbf{g}~~\Rightarrow~~
f_{m}=A_{m}\left[\frac{-ika\pi}{2}H_{m}^{(1)}(ka)\right]^{-1}~;~\forall m\in\mathbb{Z}~,
\end{equation}
which, by virtue of (\ref{1-390}), agrees with the SOV exact solution (\ref{1-270} for $v(a,\theta)$. However, it is important to recall that this solution for $f_{m}$ is only applicable for  real frequencies that are not in the neighborhood for which $J_{n}(ka)=0; \forall n\in \mathbb{Z}$.
\subsection{The field outside the object obtained by using the 'solution' of the first kind BIE for the case of a Dirichlet condition on the circular boundary}
The field outside the object is obtainable via (\ref{1-340})
\begin{equation}\label{6-400}
u(\mathbf{x})=u^{i}(\mathbf{x})+
\int_{\Gamma}kG(\mathbf{x};\mathbf{x}')v(\mathbf{x}')d\gamma(\mathbf{x}')
~;~\forall\mathbf{x}\in\Omega_{0}~,
\end{equation}
Note that this is not a BIE but rather a boundary-integral representation (BIR) of the field (in the region $\Omega_{0}$). The solution for the latter field is obtained by merely introducing the previously-found $v$ into the integrand. In polar coordinates, the BIR is
\begin{equation}\label{6-410}
u(r,\theta)=u^{i}(r,\theta)+\int_{0}^{2\pi}G(r,\theta;a,\theta')v(a,\theta)ad\theta
~;~ r>a~, \forall\theta\in[0,2\pi[~,
\end{equation}
I make use of
\begin{equation}\label{6-420}
G(r>a,\theta;a,\theta)=\frac{i}{4}\sum_{l\in\mathbb{Z}}H_{l}^{(1)}(kr)J_{l}(ka)\exp[il(\theta-\theta')]
~,
\end{equation}
and previous expansions to obtain
\begin{multline}\label{6-430}
u(r,\theta)=\sum_{n\in\mathbb{Z}}\left[H(r-r^{s})B_{n}H_{n}^{(1)}(kr)+H(r^{s}-r)A_{n}J_{n}(kr)\right]\exp(in\theta)+\\
\sum_{n\in\mathbb{Z}}f_{n}\sum_{l\in\mathbb{Z}}\frac{ika}{4\pi}\sum_{l\in\mathbb{Z}}H_{l}^{(1)}(kr)J_{l}(ka)\exp[il\theta)\int_{0}^{2\pi}\exp[i(n-l)\theta']d\theta'
~;~ r>a~, \forall\theta\in[0,2\pi[~,
\end{multline}
or
\begin{multline}\label{6-440}
u(r,\theta)=\sum_{n\in\mathbb{Z}}\left\{H(r-r^{s})B_{n}H_{n}^{(1)}(kr)+
A_{n}\left[H(r^{s}-r)J_{n}(kr)+f_{n}\frac{ika\pi}{2}H_{l}^{(1)}(kr)J_{l}(ka)\right]\right\}\times\\
\exp[in\theta)
~;~ r>a~, \forall\theta\in[0,2\pi[~,
\end{multline}
which, after the introduction of (\ref{1-470}), becomes
\begin{multline}\label{6-450}
u(r,\theta)=\sum_{n\in\mathbb{Z}}\left\{H(r-r^{s})B_{n}H_{n}^{(1)}(kr)+A_{n}\left[H(r^{s}-r)J_{n}(kr)-
\frac{J_{l}(ka)}{H_{l}^{(1)}(ka)}H_{l}^{(1)}(kr)\right]\right\}\times\\
\exp[in\theta)
~;~ r>a~, \forall\theta\in[0,2\pi[~,
\end{multline}
which agrees with the exact SOV solution (\ref{1-230}). As before, I call attention to the fact that this solution relies on a 'solution' for $v$ that can only be obtained at real frequencies  that are not in the neighborhood for which $J_{l}(ka)=0$.
\subsection{Determination of $v$ via the extended boundary condition integral equation (i.e., DEBC)for the circular object with Dirichlet boundary condition}
I recall the EBC integral equation expressed in (\ref{1-360})
\begin{equation}\label{6-460}
0=u^{i}(\mathbf{x})+
\int_{\Gamma}kG(\mathbf{x};\mathbf{x}')v(\mathbf{x}')d\gamma(\mathbf{x}')
~;~\forall\mathbf{x}\in\Omega_{1}~.
\end{equation}
I choose to sample this equation on $\Gamma_{in}\subset\Omega_{1}$, where $\Gamma_{in}$ is a circle, with center at the origin $O$, of radius $b<a$. Consequently, the polar coordinate expression of (\ref{6-460}) is
\begin{equation}\label{6-470}
0=u^{i}(b,\theta)+\int_{0}^{2\pi}G(b,\theta;a,\theta')v(a,\theta')ad\theta'~;~\forall\theta\in[0,2\pi[
~.
\end{equation}
I employ the following expressions of the Green's function and $u^{i}(b,\theta)$ (on account of the fact that $b<a<r^{s})$
\begin{equation}\label{6-480}
G(b,\theta;a,\theta')=\frac{i}{4}\sum_{l\in\mathbb{Z}}H_{l}^{(1)}(ka)J_{l}(kb)\exp[il(\theta-\theta')]~,~
u^{i}(b,\theta)=\sum_{n\in\mathbb{Z}}A_{n}J_{n}(kb)\exp(in\theta)
~,
\end{equation}
to obtain, by the usual Galerkin procedure
\begin{equation}\label{6-490}
0=h_{m}+\sum_{n\in\mathbb{Z}}f_{n}\left[\frac{ika\pi}{2}H_{n}^{(1)}(ka)J_{n}(kb)\right]\delta_{mn}
~;~\forall m\in\mathbb{Z}
~,
\end{equation}
wherein $h_{m}=A_{m}J_{m}(kb)$. As previously, I am confronted with a matrix equation, the matrix of which is of infinite order, diagonal, and singular at a denumerable, infinite set of frequencies for which $J_{n}(kb)=0~;~\forall n\in\mathbb{Z}$ so that this matrix cannot be inverted at these frequencies. At real frequencies not in the neighborhood of these singular frequencies, the solution is, as before
\begin{equation}\label{6-495}
f_{n}=A_{n}\left[\frac{ika\pi}{2}H_{n}^{(1)}(ka)\right]^{-1}~;~\forall n\in\mathbb{Z}
~,
\end{equation}
which is nothing other than the exact SOV solution. It ensues, that at these frequencies the field is as previously within $\Omega_{0}$.
\subsection{A fourth (second-kind) BIE  (i.e., DBIE2) for the case of a circular cylinder with a Dirichlet condition on its boundary}
It is generally thought \cite{wa11} that second kind BIE's are less prone than first kind BIE's to ill-conditioning problems. This is why I expose the way to treat the elastic wave response of a  circular cylinder with Dirichlet boundary condition by means of a second-kind BIE.

The point of departure is (\ref{1-300})
\begin{equation}\label{1-500}
\mathcal{H}_{\Omega_{0}}(\mathbf{x})u(\mathbf{x})=u^{i}(\mathbf{x})+
\int_{\Gamma}\left[G(\mathbf{x};\mathbf{x}')\boldsymbol{\nu}(\mathbf{x}')\cdot\nabla (\mathbf{x}')u(\mathbf{x}')-
u(\mathbf{x}')\boldsymbol{\nu}(\mathbf{x}')\cdot\nabla (\mathbf{x}')G(\mathbf{x};\mathbf{x}')\right]d\gamma(\mathbf{x}')
~;~\forall \mathbf{x}\in\mathbb{R}^{2}~,
\end{equation}
which, for a homogeneous Dirichlet boundary condition becomes
\begin{equation}\label{1-510}
\mathcal{H}_{\Omega_{0}}(\mathbf{x})u(\mathbf{x})=u^{i}(\mathbf{x})+
\int_{\Gamma}G(\mathbf{x};\mathbf{x}')\boldsymbol{\nu}(\mathbf{x}')\cdot\nabla (\mathbf{x}')u(\mathbf{x}')d\gamma(\mathbf{x}')
~;~\forall \mathbf{x}\in\mathbb{R}^{2}~,
\end{equation}
to which I have added the recollection that this expression is valid for arbitray points in the $x-y$ plane. With this in mind, the idea is to take the normal derivative of (\ref{1-510}) so as to obtain (assuming that it is valid to interchange the integral and gradient)
\begin{multline}\label{1-520}
\mathcal{H}_{\Omega_{0}}(\mathbf{x})\boldsymbol{\nu}(\mathbf{x})\cdot\nabla (\mathbf{x})u(\mathbf{x}')+u(\mathbf{x})\boldsymbol{\nu}(\mathbf{x})\cdot\nabla (\mathbf{x})\mathcal{H}_{\Omega_{0}}(\mathbf{x})=\\
\boldsymbol{\nu}(\mathbf{x})\cdot\nabla u^{i}(\mathbf{x})+
\int_{\Gamma}\boldsymbol{\nu}(\mathbf{x})\cdot\nabla(\mathbf{x})G(\mathbf{x};\mathbf{x}')\boldsymbol{\nu}(\mathbf{x}')\cdot\nabla (\mathbf{x}')u(\mathbf{x}')d\gamma(\mathbf{x}')
~.
\end{multline}
or, with the previous definition of the traction $v$ and application of the Dirichlet boundary condition (since $\boldsymbol{\nu}(\mathbf{x})\cdot\nabla (\mathbf{x})\mathcal{H}_{\Omega_{0}}(\mathbf{x})$  behaves like a Dirac delta distribution  that is nil everywhere except on $\Gamma$)
\begin{equation}\label{1-530}
\mathcal{H}_{\Omega_{0}}(\mathbf{x})v(\mathbf{x})=
v^{i}(\mathbf{x})+
\int_{\Gamma}\boldsymbol{\nu}(\mathbf{x})\cdot\nabla(\mathbf{x})G(\mathbf{x};\mathbf{x}')v(\mathbf{x}')d\gamma(\mathbf{x}')
~.
\end{equation}
It follows, after appealing to previous considerations, that
\begin{equation}\label{1-540}
\frac{1}{2}v(\mathbf{x})=
v^{i}(\mathbf{x})+
pv\int_{\Gamma}\boldsymbol{\nu}(\mathbf{x})\cdot\nabla(\mathbf{x})G(\mathbf{x};\mathbf{x}')v(\mathbf{x}')d\gamma(\mathbf{x}')
~;~\forall \mathbf{x}\in\Gamma~.
\end{equation}
which is the sought-for second-kind BIE. Note that until now no restrictions have been made on the shape of the boundary.

Henceforth, I return to the case of the circular boundary $r=a$. In polar coordinates, the BIE is
\begin{equation}\label{1-550}
\frac{1}{2}v(a,\theta)=
v^{i}(a,\theta)-
pv\int_{0}^{2\pi}\frac{\partial}{\partial r}G(a,\theta;a,\theta')v(a,\theta')ad\theta'
~;~\forall\theta\in[0,2\pi[~.
\end{equation}
Again, I appeal to a Galerkin technique for solving the BIE, now via the expansions on a Fourier basis
\begin{equation}\label{1-560}
v(a,\theta)=\sum_{n\in\mathbb{Z}}f_{n}\exp(in\theta)~~,~~v^{i}(a,\theta)=-\sum_{n\in\mathbb{Z}}g_{n}\exp(in\theta)=\sum_{n\in\mathbb{Z}}A_{n}\dot{J}_{n}(ka)\exp(in\theta)
~,
\end{equation}
so that, after projection on the same Fourier basis, I obtain
\begin{equation}\label{1-570}
\frac{1}{2}f_{m}=g_{m}
-
\sum_{n\in\mathbb{Z}}f_{n}\int_{0}^{2\pi}d\theta\exp(-im\theta)\left[pv\int_{0}^{2\pi}d\theta'\frac{a}{2\pi}\frac{\partial}{\partial r}G(a,\theta;a,\theta')\exp(in\theta')\right]
~;~\forall m\in \mathbb{Z}~.
\end{equation}
I make use of:
\begin{equation}\label{1-580}
\frac{\partial}{\partial r}G(a,\theta;a,\theta')=\frac{ik}{4}H(0)\sum_{l\in\mathbb{Z}}\left[\dot{H}_{l}^{(1)}(ka)J_{l}(ka)+
H_{l}^{(1)}(ka)\dot{J}_{l}(ka)\right]\exp[il(\theta-\theta')]~,
\end{equation}
which, within  the $pv$ integral (since $H(0)=1/2$ therein), and on account of the identity (\ref{1-260}), takes the form
\begin{equation}\label{1-590}
\frac{\partial}{\partial r}G(a,\theta;a,\theta')=\frac{ik}{8}\sum_{l\in\mathbb{Z}}\left[\frac{2i}{\pi ka}+2H_{l}^{(1)}(ka)\dot{J}_{l}(ka)\right]\exp[il(\theta-\theta')]~.
\end{equation}
The introduction of this expression into (\ref{1-570}) gives rise to
\begin{multline}\label{1-600}
\frac{1}{2}f_{m}=g_{m}+
\sum_{n\in\mathbb{Z}}f_{n}\sum_{l\in\mathbb{Z}}\left(\frac{-ika}{8}\right)\left[\frac{i}{\pi ka}+H_{l}^{(1)}(ka)\dot{J}_{l}(ka)\right]\times\\
\int_{0}^{2\pi}d\theta\exp[i(l-m)\theta]\int_{0}^{2\pi}d\theta'\exp[i(n-l)\theta')
~;~\forall m\in \mathbb{Z}~,
\end{multline}
or
\begin{equation}\label{1-610}
\frac{1}{2}f_{m}=g_{m}+\frac{1}{2}f_{m}+
\sum_{n\in\mathbb{Z}}f_{n}\left[\frac{-ika\pi}{2}H_{n}^{(1)}(ka)\dot{J}_{n}(ka)\right]\delta_{mn}
~,
\end{equation}
which reduces to the matrix equation
\begin{equation}\label{1-620}
\sum_{n\in\mathbb{Z}}f_{n}\left[\frac{ika\pi}{2}H_{n}^{(1)}(ka)\dot{J}_{n}(ka)\right]\delta_{mn}=g_{m}~;~\forall m\in\mathbb{Z}
~.
\end{equation}
Once again, the matrix is of infinite order, diagonal, but singular at certain real frequencies. The latter are those that correspond to the zeros of the derivative of the Bessel function (real), i.e., $\dot{J}_{n}(ka)=0~;~\forall n\in\mathbb{Z}$. Thus, the matrix cannot be inverted in general. However, at real frequencies not near the neighborhood of these singular frequencies, the matrix is invertible so that the solution for $f_{n}$ is:
\begin{equation}\label{1-630}
f_{n}=\left[\frac{ika\pi}{2}H_{n}^{(1)}(ka)\dot{J}_{n}(ka)\right]^{-1}g_{n}=A_{n}\left[\frac{-ika\pi}{2}H_{n}^{(1)}(ka)\right]^{-1}~;~\forall n\in\mathbb{Z}
~,
\end{equation}
which is identical to the exact SOV solution. It follows that at these frequencies the displacement field within $\Omega_{0}$ is also identical to the SOV solution for this field.
\subsection{Numerical details}
The material of the preceding sections (the same will be true for those devoted to the Neumann boundary condition and to the methods of cures) showed that everything ends up with the problem of determining the vector $\mathbf{f}$ of a matrix equation of the type $\mathbf{E}\mathbf{f}=\mathbf{g}$. The evaluation of the elements of $\mathbf{E}$ and $\mathbf{g}$ do not pose any particular problem here since they involve well-known elementary functions such as exponentials, Bessel and Hankel functions (they do pose some problems in the usual discretization methods since they require numerical quadratures of integrands that are weakly or strongly singular). The real difficulty arises due to the fact that the number of equations and unknowns corresponding to $\mathbf{E}\mathbf{f}=\mathbf{g}$ is infinite.

The way I handle this problem is to reduce the matrix equation to one of finite-order $2N+1<\infty$, find the solution $\mathbf{f}^{(N)}$ of this matrix equation, increase $N$, again find the solution of the matrix equation,....until the normed-difference between successive thus-obtained approximations of $\mathbf{f}$ is smaller than some pre-defined value (I say that when this is achieved, the procedure has 'levelized'). It turns out that the required $N$ for levelization increases with frequency, but, as concerns EBC methods, levelization is never really fully-achieved \cite{bw74,bw77} (which is a good reason to prefer BIE1 and BIE2 methods, in spite of the attractive feature of EBC methods which is that the aforementioned quadratures are those of non-singular functions).

Thus, the value of $N$ given in the graphs exhibited hereafter is the one required for levelization except when it refers to EBC computations in which case it corresponds to a sort of optimum (since the successive solutions first converge and then diverge, the optimum corresponding to the moment of change of character).

Unfortunately, what will henceforth reveal itself to be a 'spurious' resonance is often an elusive entity, i.e., does not readily show up in the response curves (the form of which is $\mathbf{f}$ plotted against frequency $f$ or dimensionless wavenumber $ka$). Since, as I have shown previously, this resonance is a consequence of the singular nature of $\mathbf{E}$ at the resonance frequency $f_{R}$, and $\mathbf{E}$ is diagonal via the chosen Galerkin scheme, it suffices to plot $1/\|det(\mathbf{E}(f))\|$ ($det$ signifies determinant) and spot the resonant frequencies by the fact that they occur at the giant maxima of $1/\|det(\mathbf{E}(f))\|$. But, of course, it is more convincing for physicists and engineers to 'see' the effect of resonances in the responses (i.e., transfer functions), and as said, the fact that these resonances don't always show up in these responses is an argument against their actual existence. The way I solve this problem is to make use of the fact that the singularity of $\mathbf{E}$ translates to instability of the solutions of the matrix equation, this meaning that small perturbations of either the elements of $\mathbf{E}$ or $\mathbf{g}$ translate to large perturbations of $\mathbf{f}$, the latter then showing up as the sought-for resonant features (i.e., the perturbation scheme acts like the revealing agent in photography). To actually do this, I chose the perturbation of $\mathbf{E}$ (since this is usually the entity whose computation generates the largest error) by the introduction of random error in all of its elements. The amount (chosen by trial and error until the appearance of the resonant features in the transfer functions) of the thus-introduced random error is measured by the number $\epsilon$ which is larger the greater the amount of introduced error and nil when no error is artificially introduced.
\clearpage
\newpage
\subsection{Numerical symptoms of the disease: the appearance of 'unusual' resonances}
The following figures, i.e., \ref{fdbie1-1}-\ref{fdbie1-8}, \ref{fdbie2-1}-\ref{fdbie2-8}, and \ref{fdebc-1}-\ref{fdebc-10}, all apply to a rigid circular cylinder of radius $a=1$ (a.u.) submitted to the wave radiated by a line source situated at  $r^{s}=12$ (a.u.), $\theta^{s}=30^{\circ}$. The responses (as a function of $ka$, $k$ the wavenumber) are computed first by, DBIE1, then by DBIE2, and finally by DEBC.
%
\subsubsection{DBIE1}
\begin{figure}[ht]
\begin{center}
\includegraphics[width=0.5\textwidth]{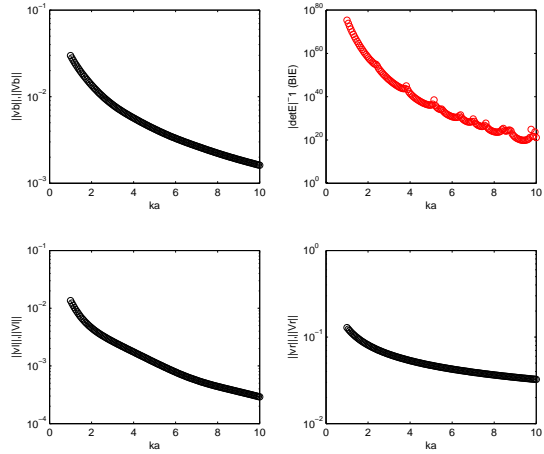}
\caption{Transfer functions of the traction at three points on the rigid boundary. The upper left-hand, lower left-hand, lower right-hand panels are for  the transfer functions at $\theta=180^{\circ}$,  $\theta=270^{\circ}$,  $\theta=360^{\circ}$, respectively. The upper right-hand panel depicts $1/\|det(\mathbf{E}(ka))\|$. Lower-case letters and circles correspond to  DBIE1 computations, upper-case letters and continuous curves to  DSOV (exact) computations.  Case $N=28$, $\epsilon=0$.}
\label{fdbie1-1}
\end{center}
\end{figure}
\begin{figure}[ptb]
\begin{center}
\includegraphics[width=0.5\textwidth]{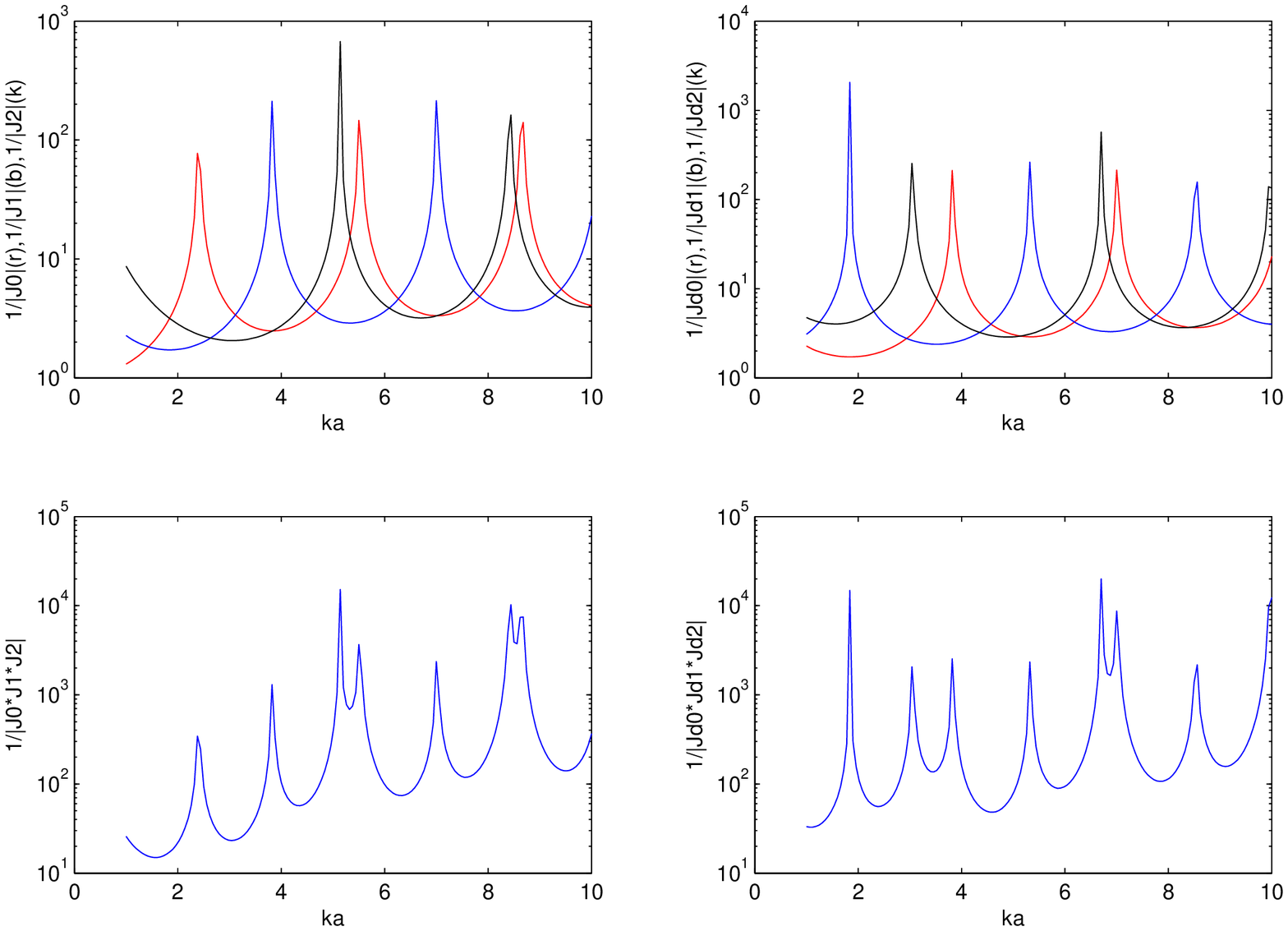}
\caption{This figure enables the connection of the observed resonance frequencies to the zeros of either $J_{n}(ka)$ (for DBIE1), $\dot{J}_{n}(ka)$ (for DBIE2) or $J_{n}(kb)$ (for DEBC). The upper left-hand panel is relative to $1/|J_{0}(ka)|$ (red), $1/|J_{1}(ka)|$ (blue),  $1/|J_{2}(ka)|$ (black) whereas the lower left-hand panel is relative to $1/|J_{0}(ka)J_{1}(ka)J_{0}(ka)|$. The upper right-hand panel is relative to $1/|\dot{J}_{0}(ka)|$ (red), $1/|\dot{J}_{1}(ka)|$ (blue),  $1/|\dot{J}_{2}(ka)|$ (black) whereas the lower right-hand  panel is relative to $1/|\dot{J}_{0}(ka)\dot{J}_{1}(ka)\dot{J}_{2}(ka)|$. As expected, the positions of the lower-frequency resonant features in fig. \ref{fdbie1-1} coincide with the zeros of $J_{n}(ka)~;~n=0,1,2$ and the first few maxima of $1/\|det(\mathbf{E}(ka))\|$   in fig.  \ref{fdbie1-1} are located at the same positions as those of $1/|J_{0}(ka)J_{1}(ka)J_{2}(ka)|$ herein.}
\label{fdbie1-2}
\end{center}
\end{figure}
\begin{figure}[ptb]
\begin{center}
\includegraphics[width=0.65\textwidth]{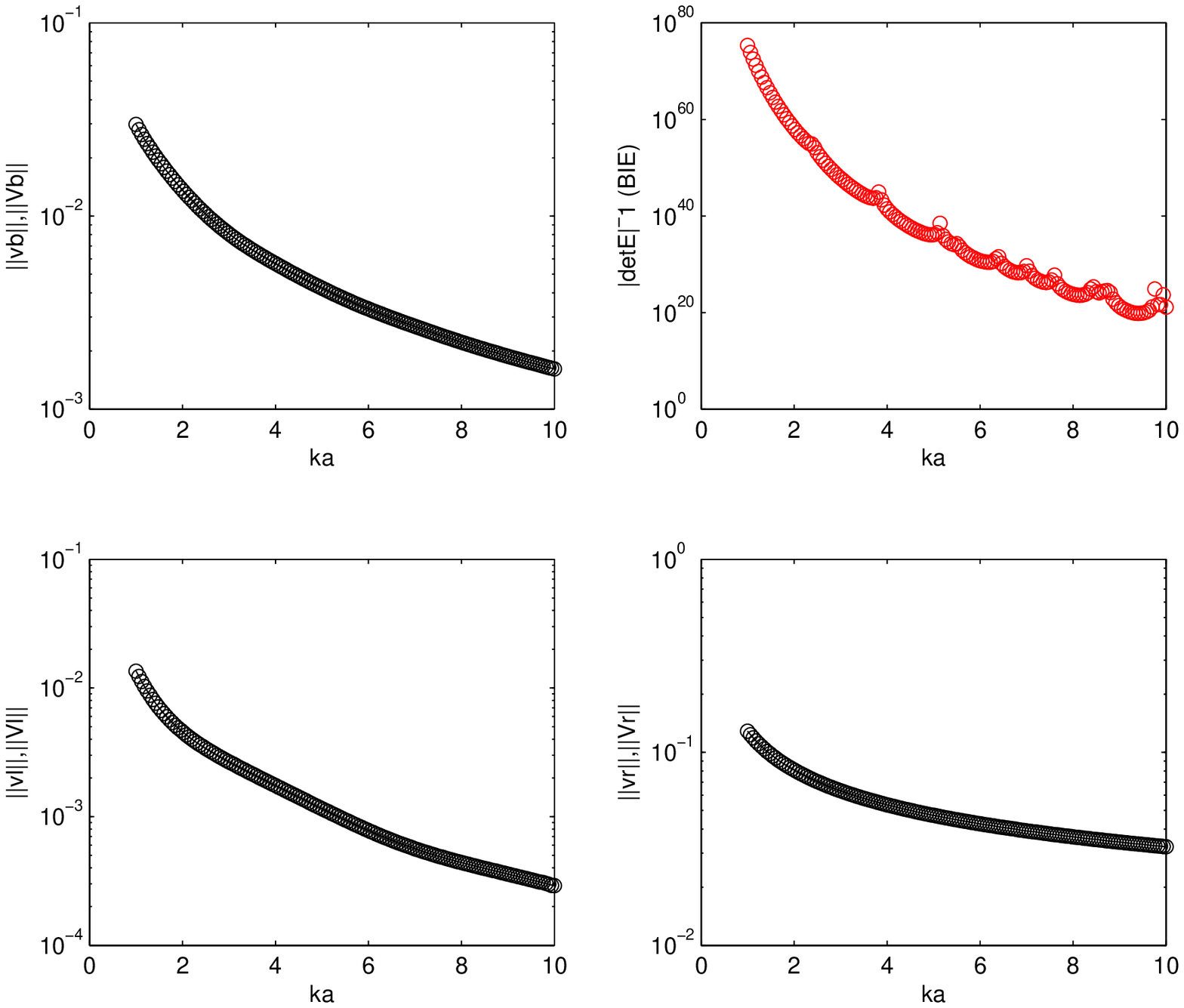}
\caption{Same as fig. \ref{fdbie1-1} except that  $N=28$, $\epsilon=10^{-6}$. I am here increasing $\epsilon$ in an effort to 'reveal' the resonances in the response curves.}
\label{fdbie1-3}
\end{center}
\end{figure}
\begin{figure}[ptb]
\begin{center}
\includegraphics[width=0.65\textwidth]{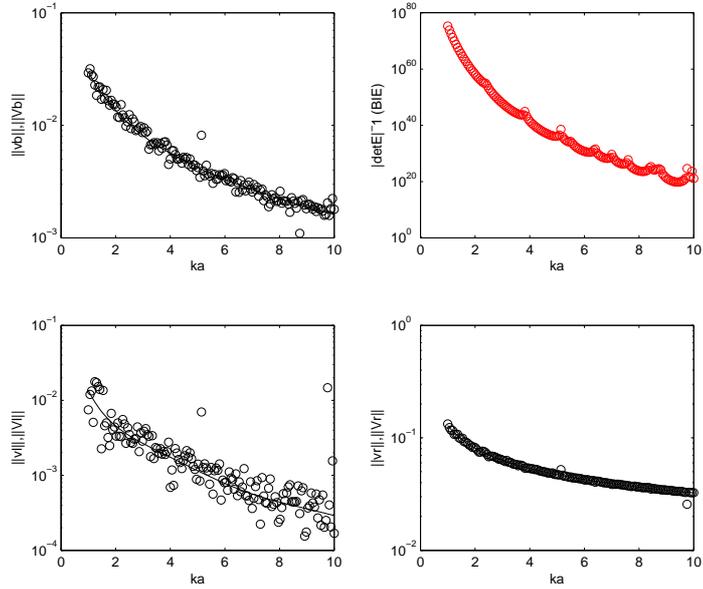}
\caption{Same as fig. \ref{fdbie1-1} except that  $N=28$, $\epsilon=10^{-3}$. I have again increased $\epsilon$ and thus finally succeeded in revealing the resonances in the response curves.}
\label{fdbie1-4}
\end{center}
\end{figure}
\begin{figure}[ptb]
\begin{center}
\includegraphics[width=0.65\textwidth]{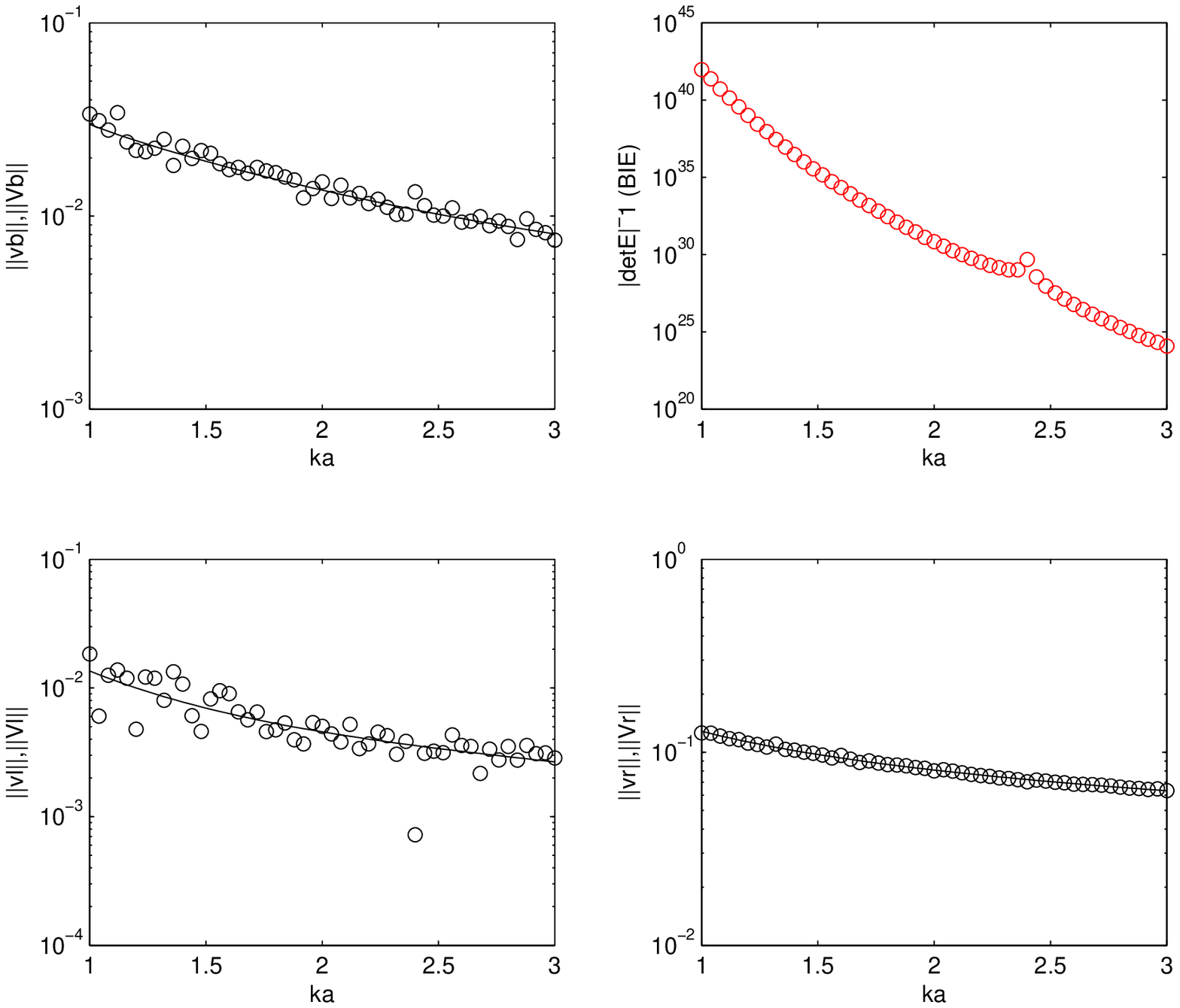}
\caption{Same as fig. \ref{fdbie1-1} except that   $N=18$, $\epsilon=10^{-3}$. This is a zoom of the preceding figure.}
\label{fdbie1-5}
\end{center}
\end{figure}
\begin{figure}[ptb]
\begin{center}
\includegraphics[width=0.65\textwidth]{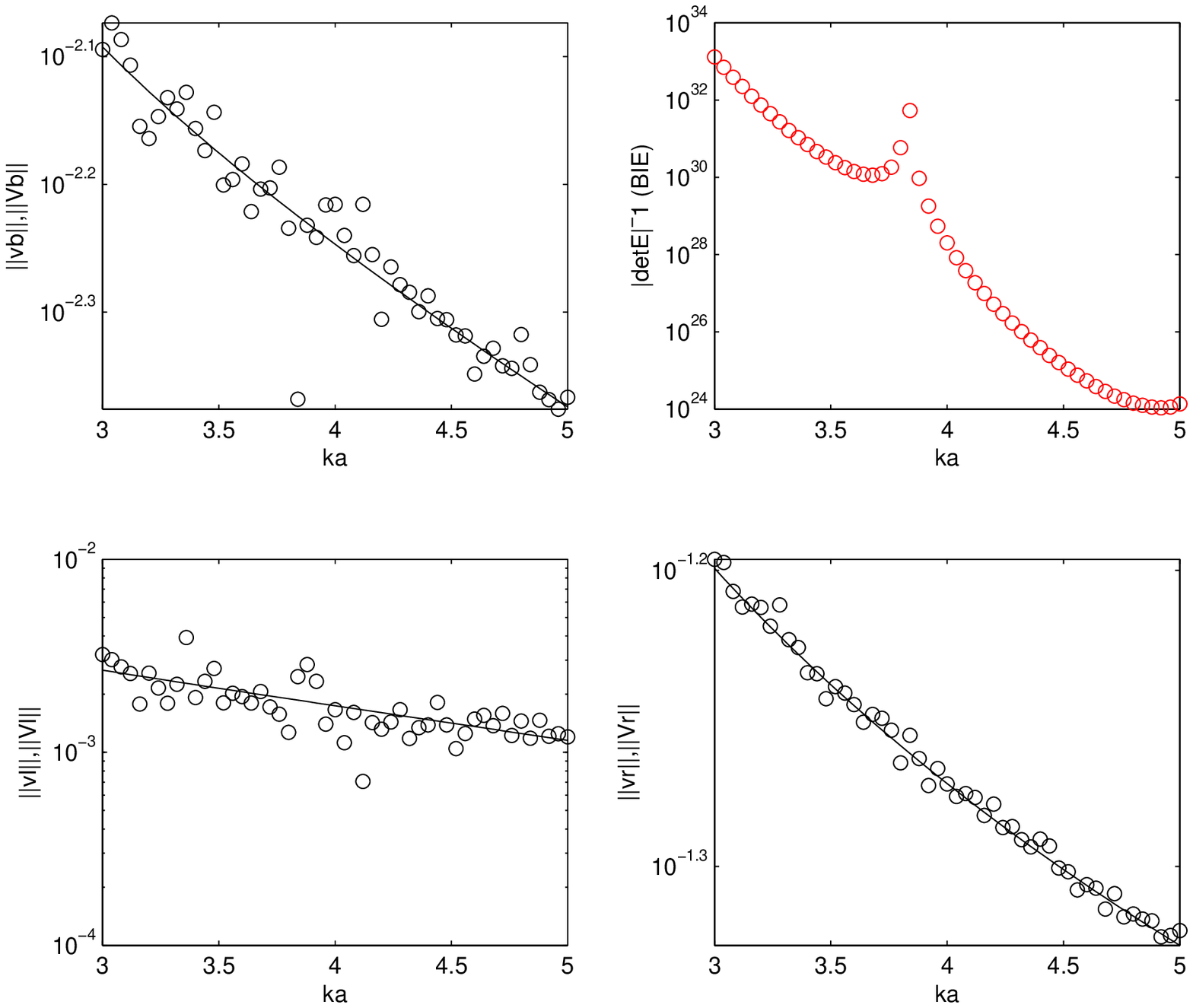}
\caption{Same as fig. \ref{fdbie1-1} except that   $N=22$, $\epsilon=10^{-3}$. This is another zoom of  fig. \ref{fdbie1-4}.}
\label{fdbie1-6}
\end{center}
\end{figure}
\begin{figure}[ptb]
\begin{center}
\includegraphics[width=0.65\textwidth]{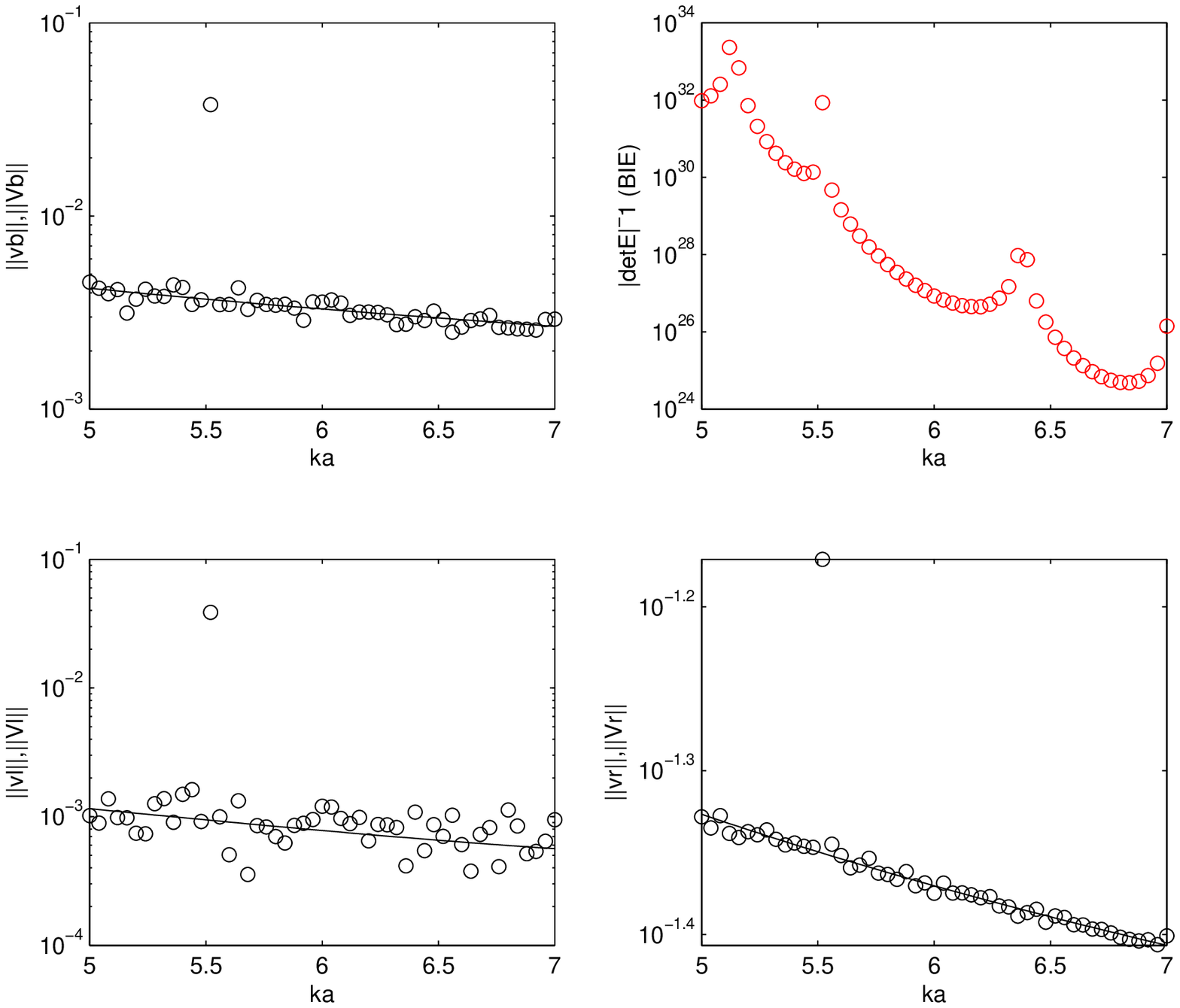}
\caption{Same as fig. \ref{fdbie1-1} except that   $N=26$, $\epsilon=10^{-3}$. This is another zoom of fig. \ref{fdbie1-4}.}
\label{fdbie1-7}
\end{center}
\end{figure}
\begin{figure}[ptb]
\begin{center}
\includegraphics[width=0.65\textwidth]{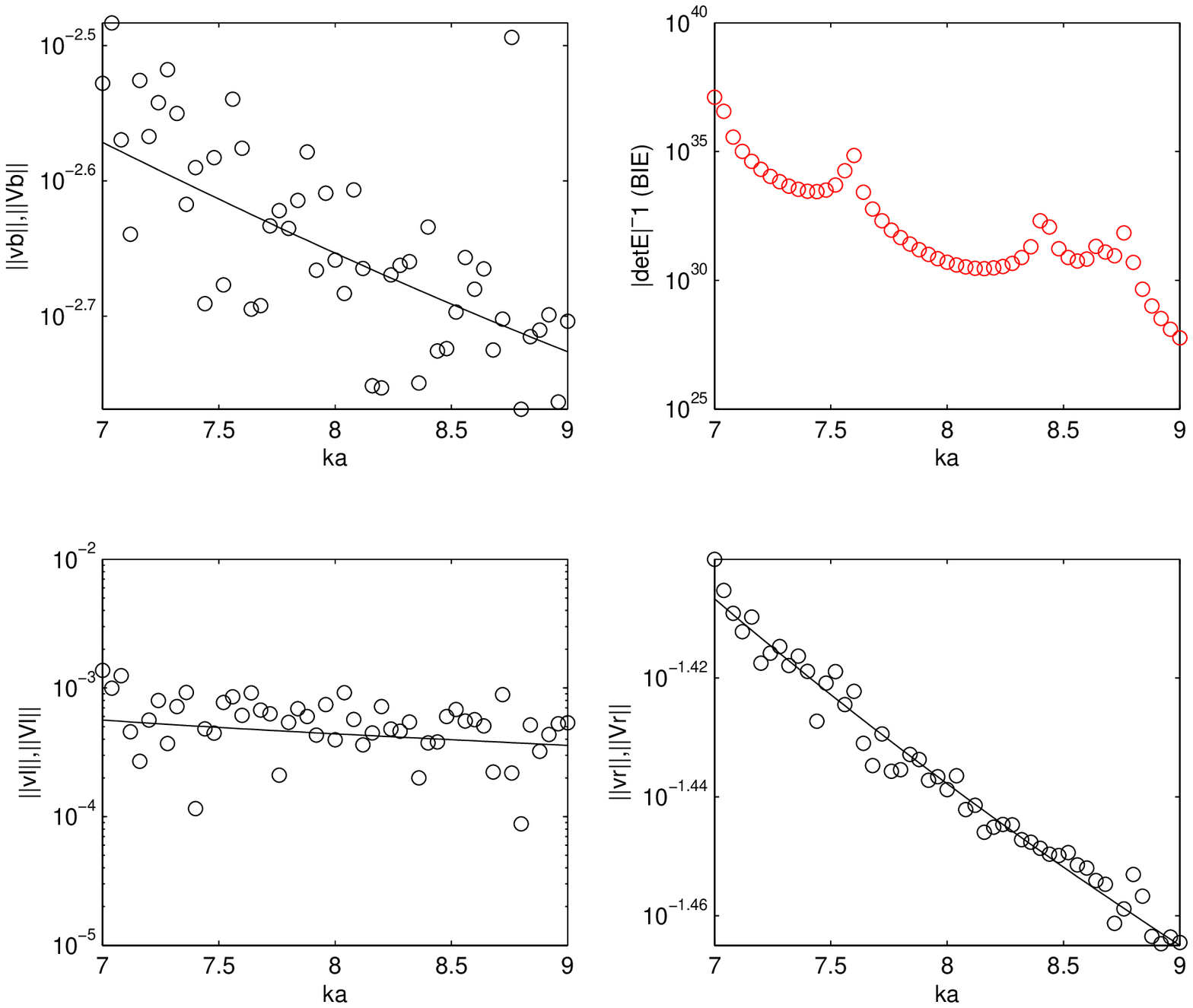}
\caption{Same as fig. \ref{fdbie1-1} except that   $N=32$, $\epsilon=10^{-3}$. This is another zoom of fig. \ref{fdbie1-4}.}
\label{fdbie1-8}
\end{center}
\end{figure}
\clearpage
\newpage
Note that all the theoretically-predicted resonances do not necessarily show up in the response curves. Moreover, a given resonance can show up in the response at one point, and not at another point, of the boundary. Finally, note the scale changes of the ordinates in going from one figure to the next.
\subsubsection{DBIE2}
\begin{figure}[ht]
\begin{center}
\includegraphics[width=0.5\textwidth]{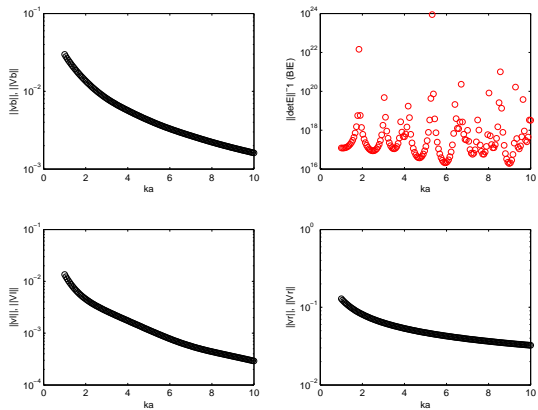}
\caption{Transfer functions of the traction at three points on the rigid boundary. The upper left-hand, lower left-hand, lower right-hand panels are for  the transfer functions at $\theta=180^{\circ}$,  $\theta=270^{\circ}$,  $\theta=360^{\circ}$, respectively. The upper right-hand panel depicts $1/\|det(\mathbf{E}(ka))\|$. Lower-case letters and circles correspond to  DBIE2 computations, upper-case letters and continuous curves to  DSOV (exact) computations.  Case $N=28$, $\epsilon=0$.}
\label{fdbie2-1}
\end{center}
\end{figure}
\begin{figure}[ht]
\begin{center}
\includegraphics[width=0.5\textwidth]{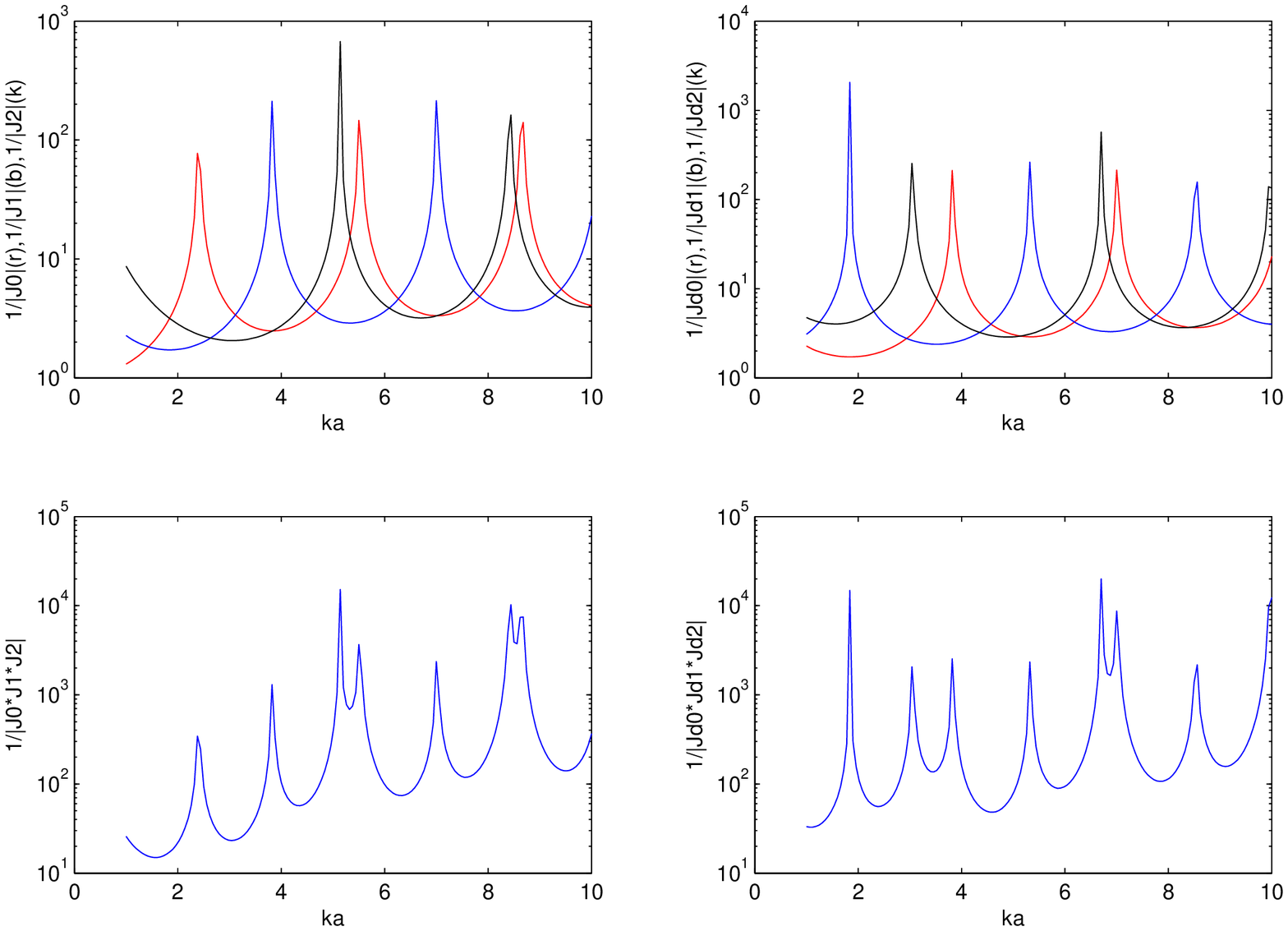}
\caption{Same as fig. \ref{fdbie1-2}. As expected, the positions of the lower-frequency resonant features in fig. \ref{fdbie2-1} coincide with the zeros of $\dot{J}_{n}(ka)~;~n=0,1,2$ and the first few maxima of $1/\|det(\mathbf{E}(ka))\|$   in fig.  \ref{fdbie2-1} are located at the same positions as those of $1/|\dot{J}_{0}(ka)\dot{J}_{1}(ka)\dot{J}_{2}(ka)|$ herein.}
\label{fdbie2-2}
\end{center}
\end{figure}
\begin{figure}[ptb]
\begin{center}
\includegraphics[width=0.7\textwidth]{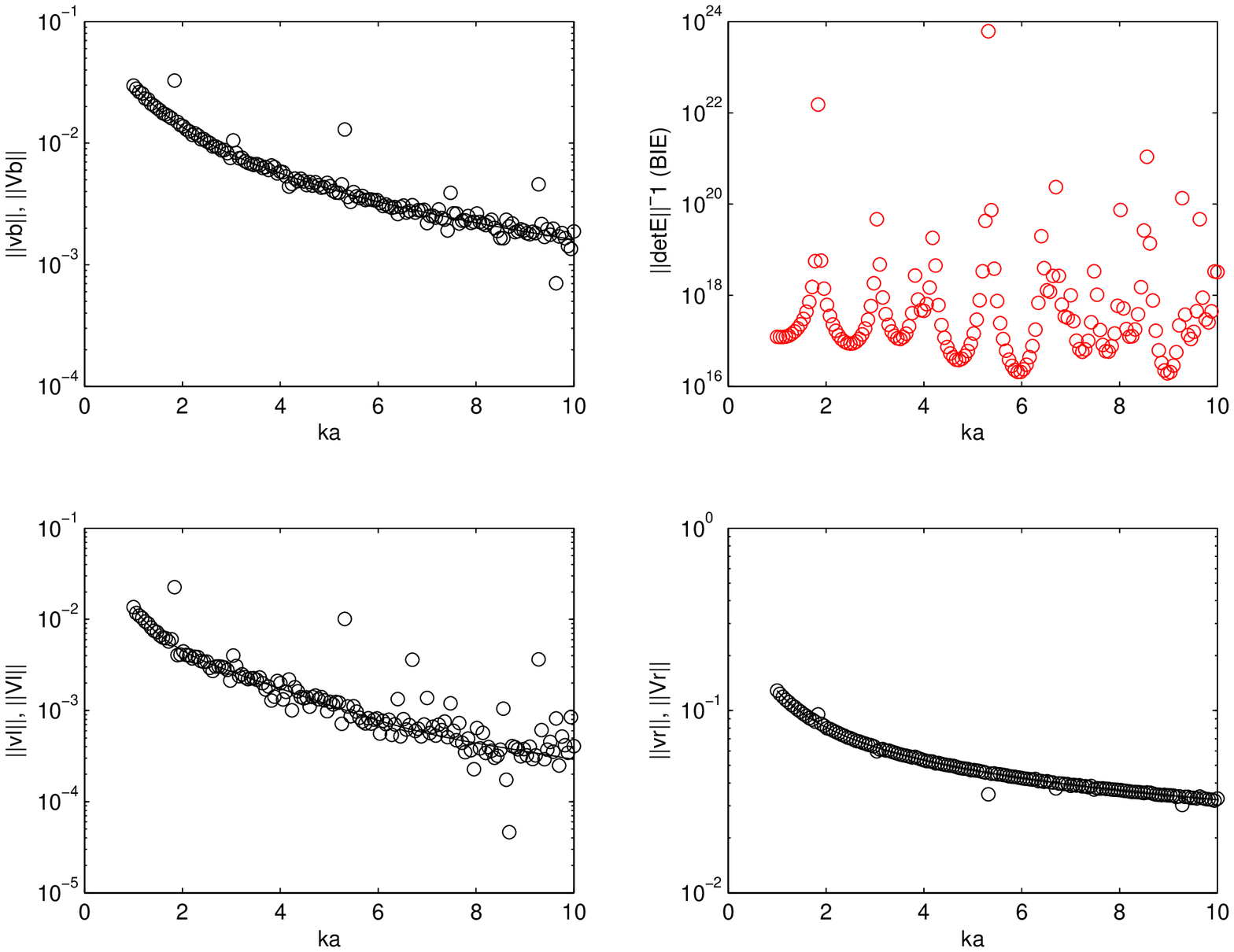}
\caption{Same as fig. \ref{fdbie2-1} except that  $N=28$, $\epsilon=10^{-3}$. I have here increased $\epsilon$ in an effort to 'reveal' the resonances in the response curves.}
\label{fdbie2-4}
\end{center}
\end{figure}
\begin{figure}[ptb]
\begin{center}
\includegraphics[width=0.7\textwidth]{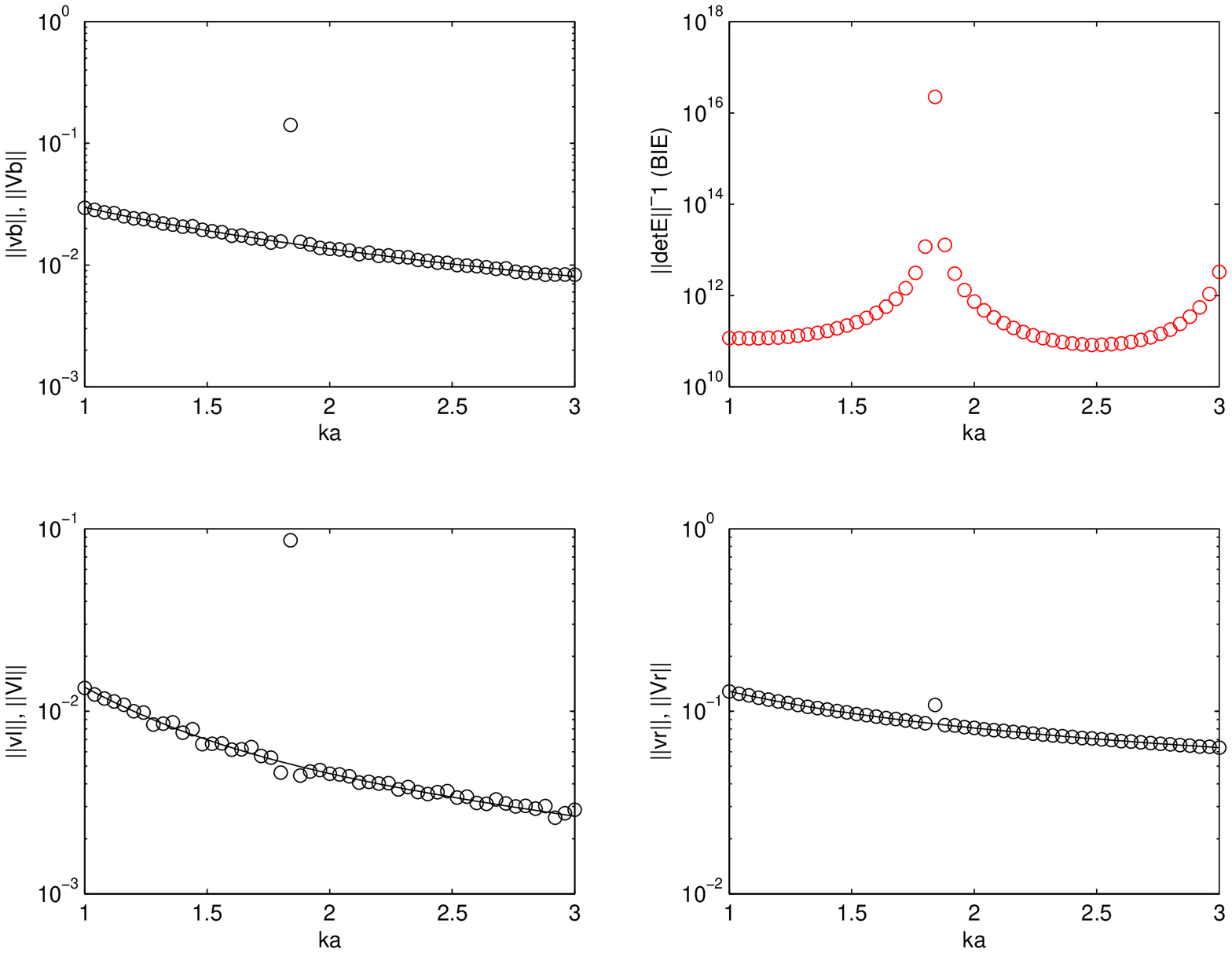}
\caption{Same as fig. \ref{fdbie2-1} except that   $N=18$, $\epsilon=10^{-3}$. This is a zoom of the preceding figure.}
\label{fdbie2-5}
\end{center}
\end{figure}
\begin{figure}[ptb]
\begin{center}
\includegraphics[width=0.7\textwidth]{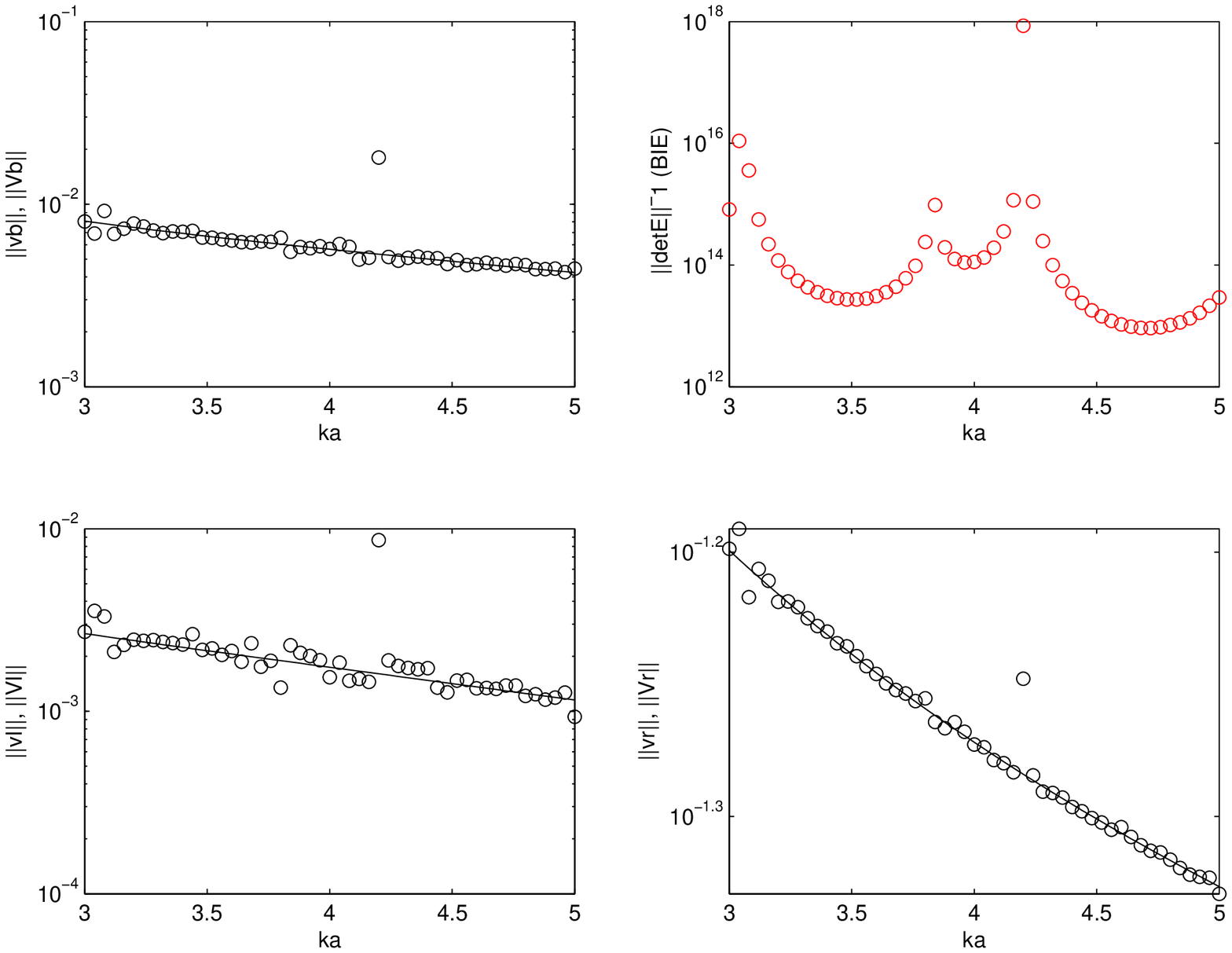}
\caption{Same as fig. \ref{fdbie2-1} except that   $N=22$, $\epsilon=10^{-3}$. This is another zoom of  fig. \ref{fdbie2-4}.}
\label{fdbie2-6}
\end{center}
\end{figure}
\begin{figure}[ptb]
\begin{center}
\includegraphics[width=0.7\textwidth]{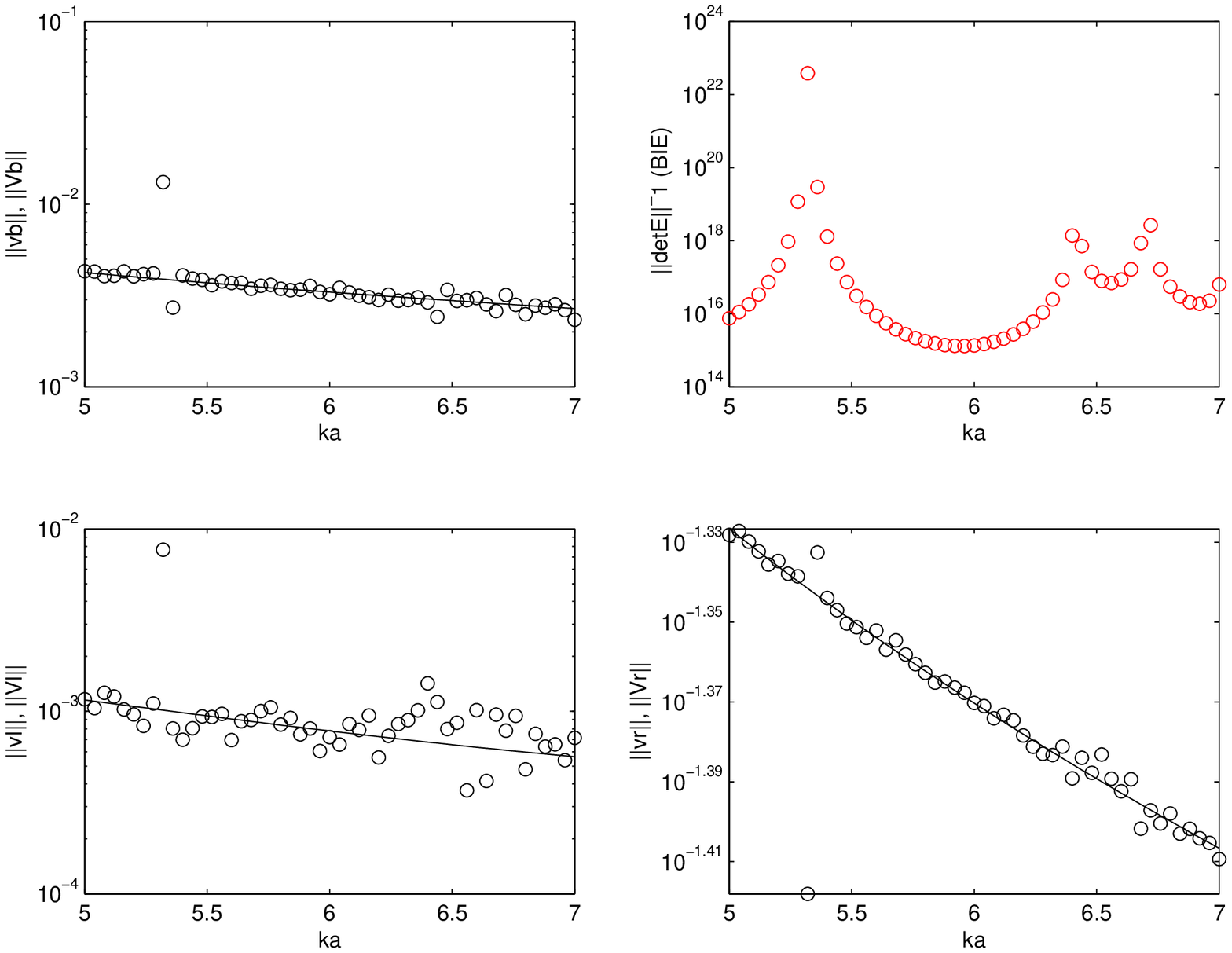}
\caption{Same as fig. \ref{fdbie2-1} except that   $N=26$, $\epsilon=10^{-3}$. This is another zoom of  fig. \ref{fdbie2-4}.}
\label{fdbie2-7}
\end{center}
\end{figure}
\begin{figure}[ptb]
\begin{center}
\includegraphics[width=0.7\textwidth]{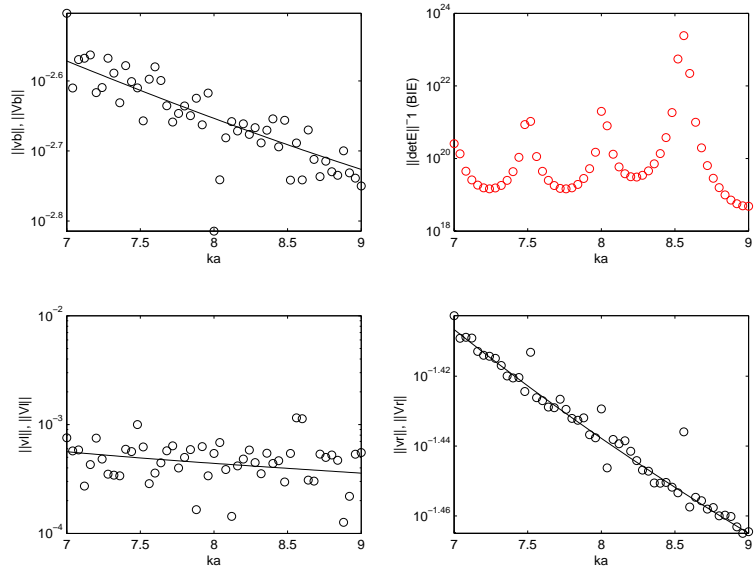}
\caption{Same as fig. \ref{fdbie2-1} except that   $N=30$, $\epsilon=10^{-3}$. This is another zoom of  fig. \ref{fdbie2-4}.}
\label{fdbie2-8}
\end{center}
\end{figure}
\clearpage
\newpage
Note that all the theoretically-predicted resonances do not necessarily show up in the response curves. Moreover, a given resonance can show up in the response at one point, and not at another point, of the boundary. Note especially that the positions of these DBIE2 resonances differ from those of the DBIE1 resonances which is contrary to the hypothesis that both sets are 'real' resonances since they both occur for the same scattering problem. Finally, note the scale changes of the ordinates in going from one figure to the next.
\subsubsection{DEBC}
\begin{figure}[ht]
\begin{center}
\includegraphics[width=0.5\textwidth]{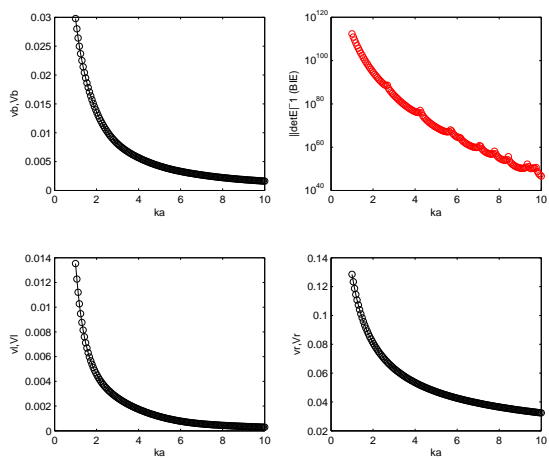}
\caption{Transfer functions of the traction at three points on the rigid boundary. The upper left-hand, lower left-hand, lower right-hand panels are for  the transfer functions at $\theta=180^{\circ}$,  $\theta=270^{\circ}$,  $\theta=360^{\circ}$, respectively. The upper right-hand panel depicts $1/\|det(\mathbf{E}(ka))\|$. Lower-case letters and circles correspond to  DEBC computations, upper-case letters and continuous curves to  DSOV (exact) computations.  Case $N=28$, $\epsilon=0$.}
\label{fdebc-1a}
\end{center}
\end{figure}
\begin{figure}[ht]
\begin{center}
\includegraphics[width=0.5\textwidth]{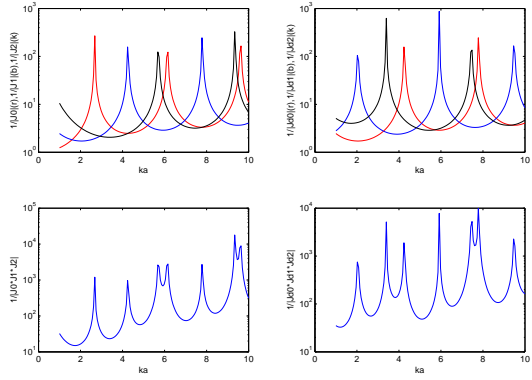}
\caption{Same as fig. \ref{fdbie1-2}. As expected, the positions of the lower-frequency resonant features in fig. \label{fdebc-1} coincide with the zeros of $J_{n}(kb)~;~n=0,1,2$ (with $b=0.9$ (a.u.)) and the first few maxima of $1/\|det(\mathbf{E}(ka))\|$   in fig. \ref{fdebc-1a} are located at the same positions as those of $1/|J_{0}(kb)J_{1}(kb)J_{2}(kb)|$ herein.}
\label{fdebc-2}
\end{center}
\end{figure}
\begin{figure}[ptb]
\begin{center}
\includegraphics[width=0.65\textwidth]{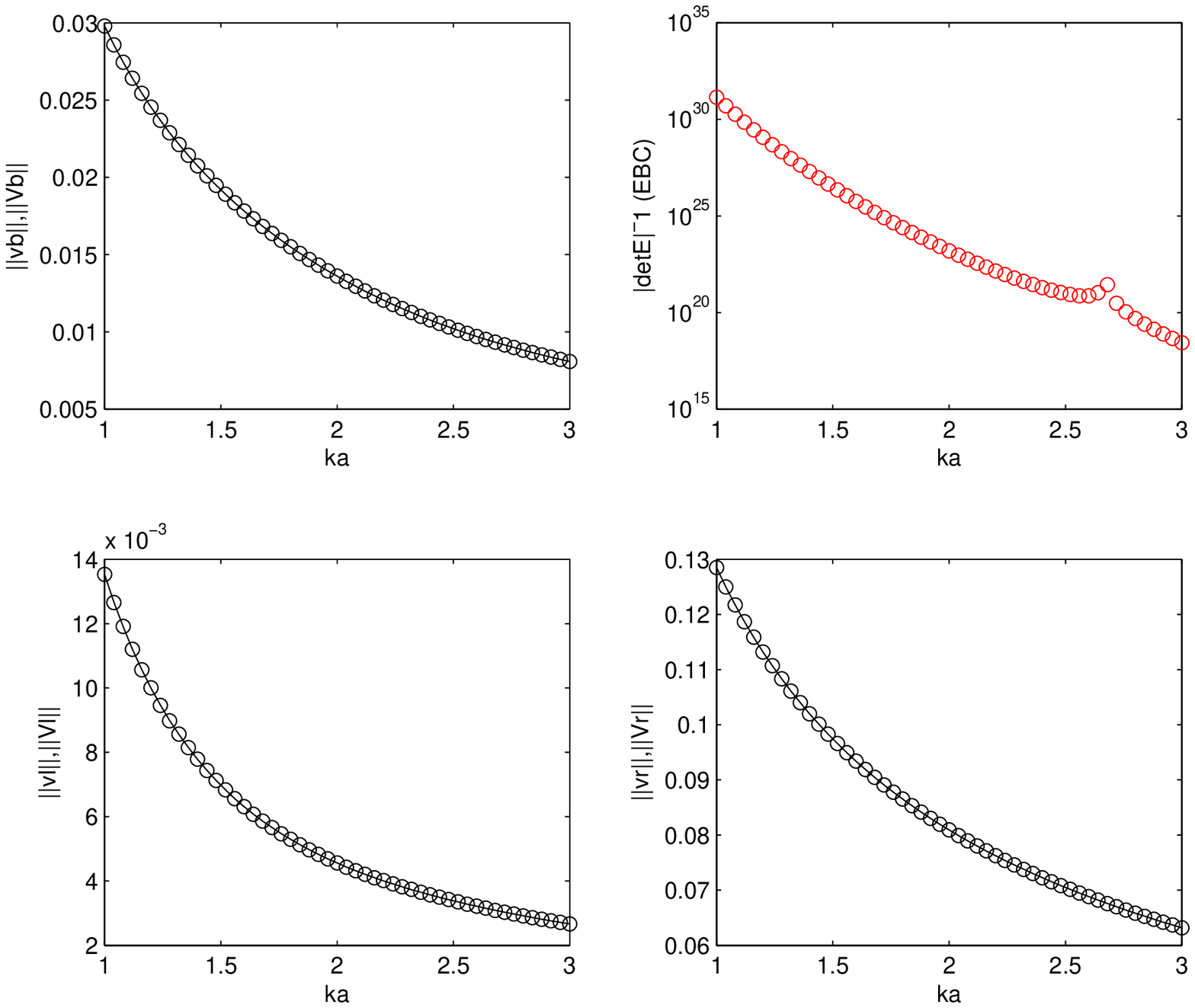}
\caption{Same as fig. \ref{fdebc-1a} except that   $N=12$, $\epsilon=10^{-7}$. Otherwise, this is a zoom of fig. \ref{fdebc-1a}.}
\label{fdebc-3}
\end{center}
\end{figure}
\begin{figure}[ptb]
\begin{center}
\includegraphics[width=0.65\textwidth]{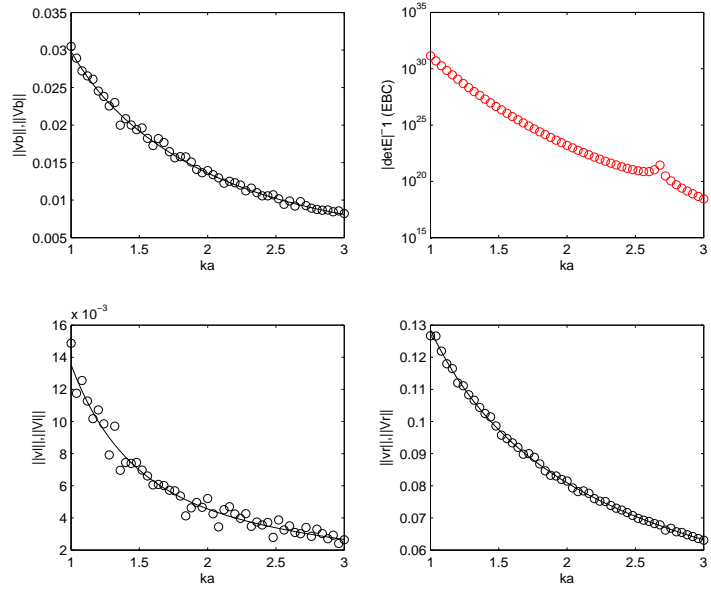}
\caption{Same as fig. \ref{fdebc-3} except that   $N=12$, $\epsilon=10^{-5}$. Here I have increased $\epsilon$ in the hope of 'revealing' the resonance.}
\label{fdebc-4}
\end{center}
\end{figure}
\begin{figure}[ptb]
\begin{center}
\includegraphics[width=0.65\textwidth]{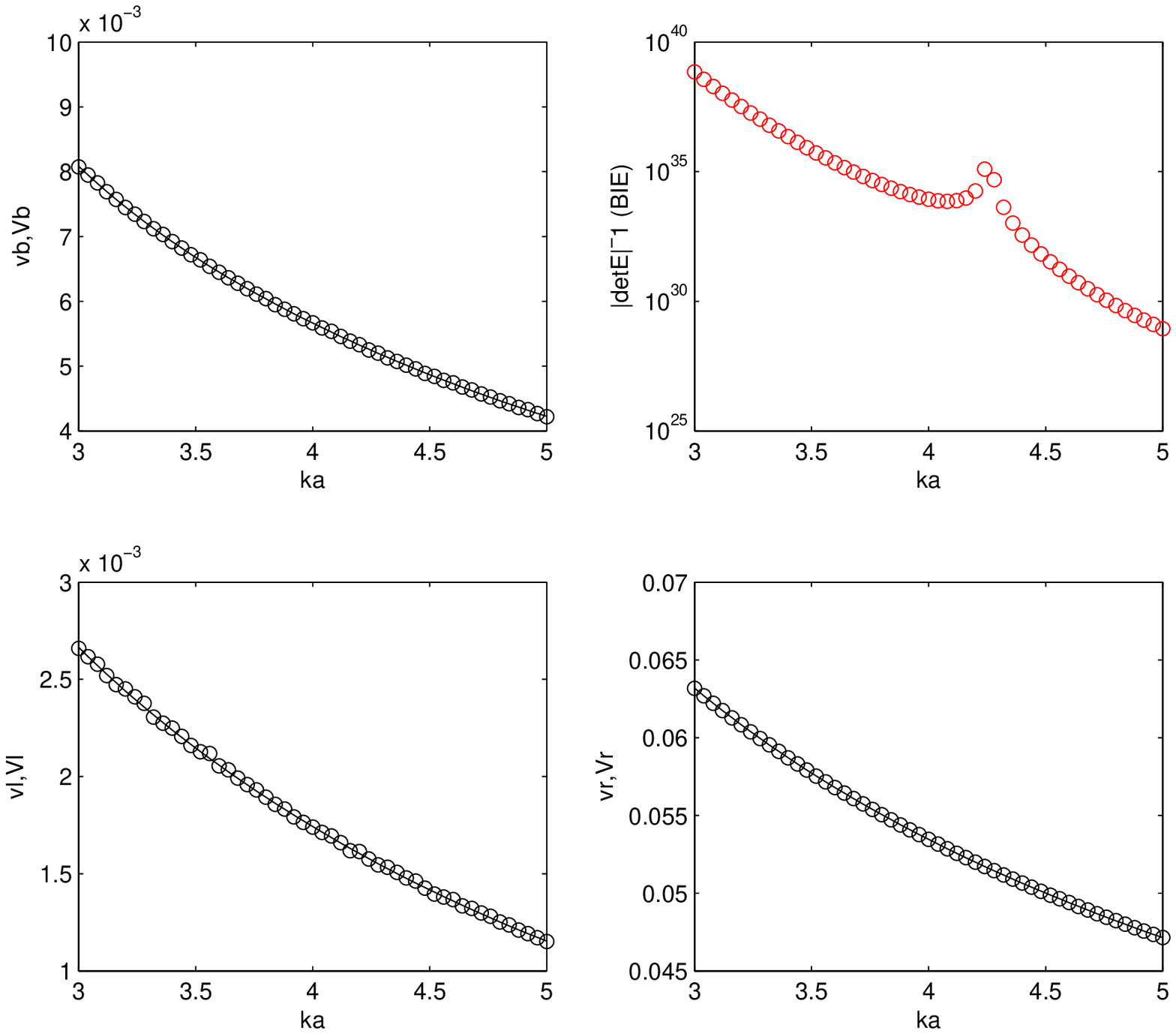}
\caption{Same as fig. \ref{fdebc-1a} except that   $N=18$, $\epsilon=10^{-7}$. Otherwise, this is another zoom of fig. \ref{fdebc-1a}.}
\label{fdebc-5}
\end{center}
\end{figure}
\begin{figure}[ptb]
\begin{center}
\includegraphics[width=0.65\textwidth]{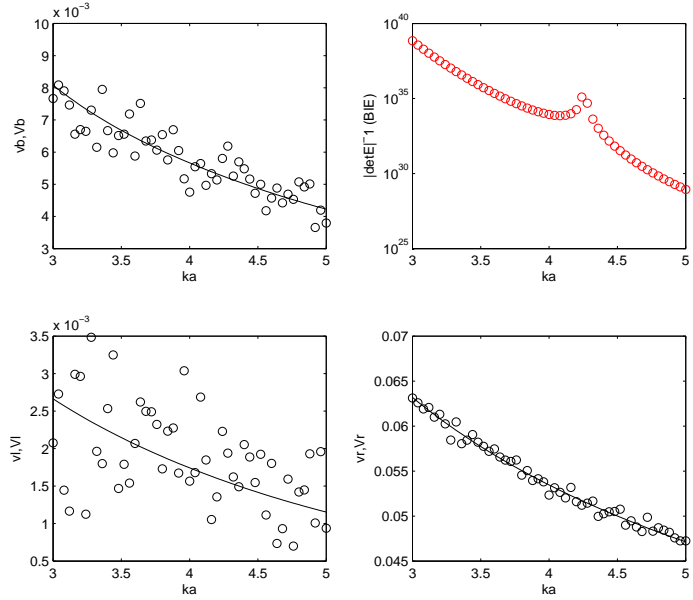}
\caption{Same as fig. \ref{fdebc-5} except that   $N=18$, $\epsilon=10^{-5}$. Here I have increased $\epsilon$ to try to 'reveal' the resonances.}
\label{fdebc-6}
\end{center}
\end{figure}
\begin{figure}[ptb]
\begin{center}
\includegraphics[width=0.65\textwidth]{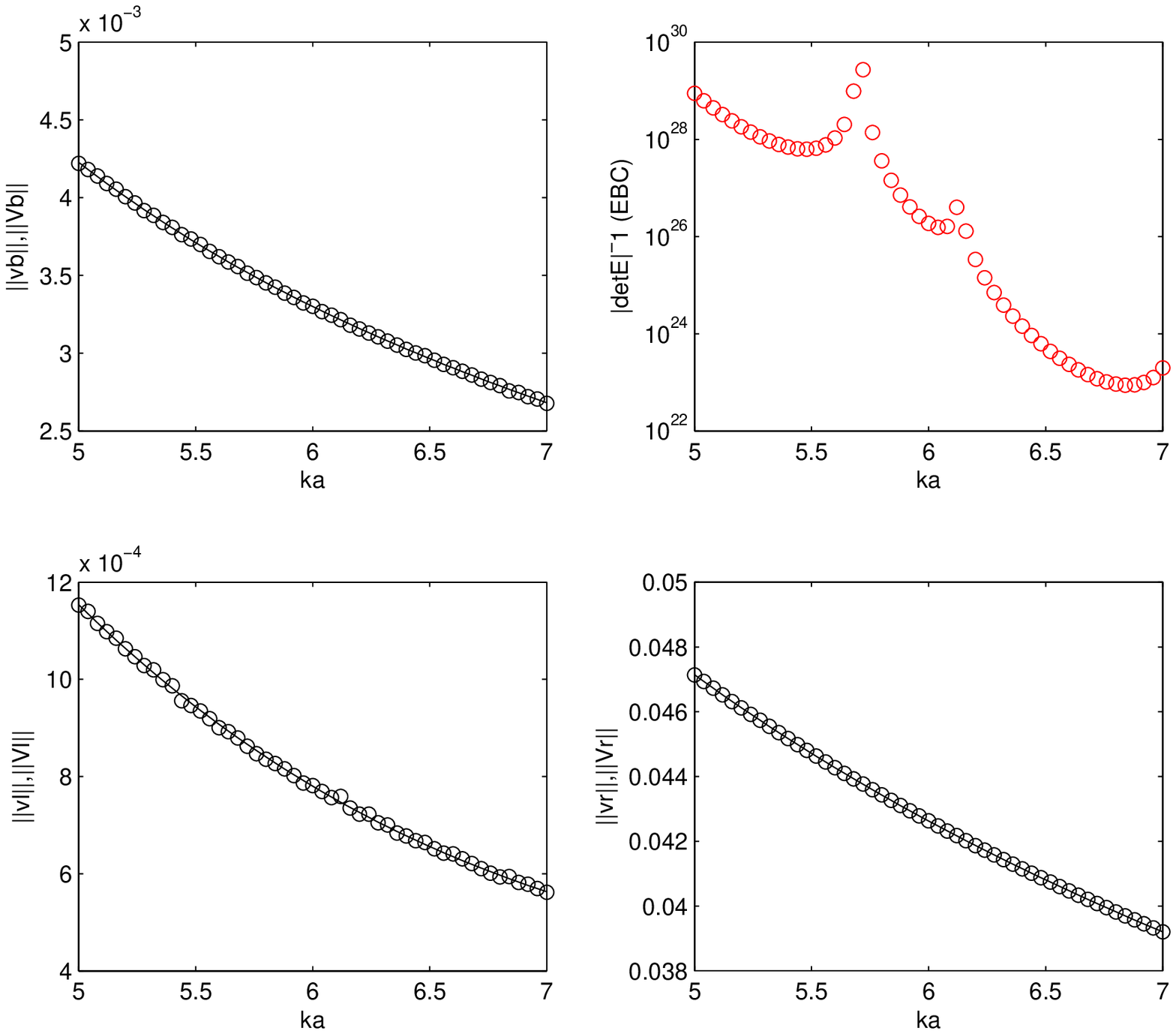}
\caption{Same as fig. \ref{fdebc-1a} except that   $N=18$, $\epsilon=10^{-7}$. Otherwise, this is another zoom of fig. \ref{fdebc-1a}.}
\label{fdebc-7}
\end{center}
\end{figure}
\begin{figure}[ptb]
\begin{center}
\includegraphics[width=0.65\textwidth]{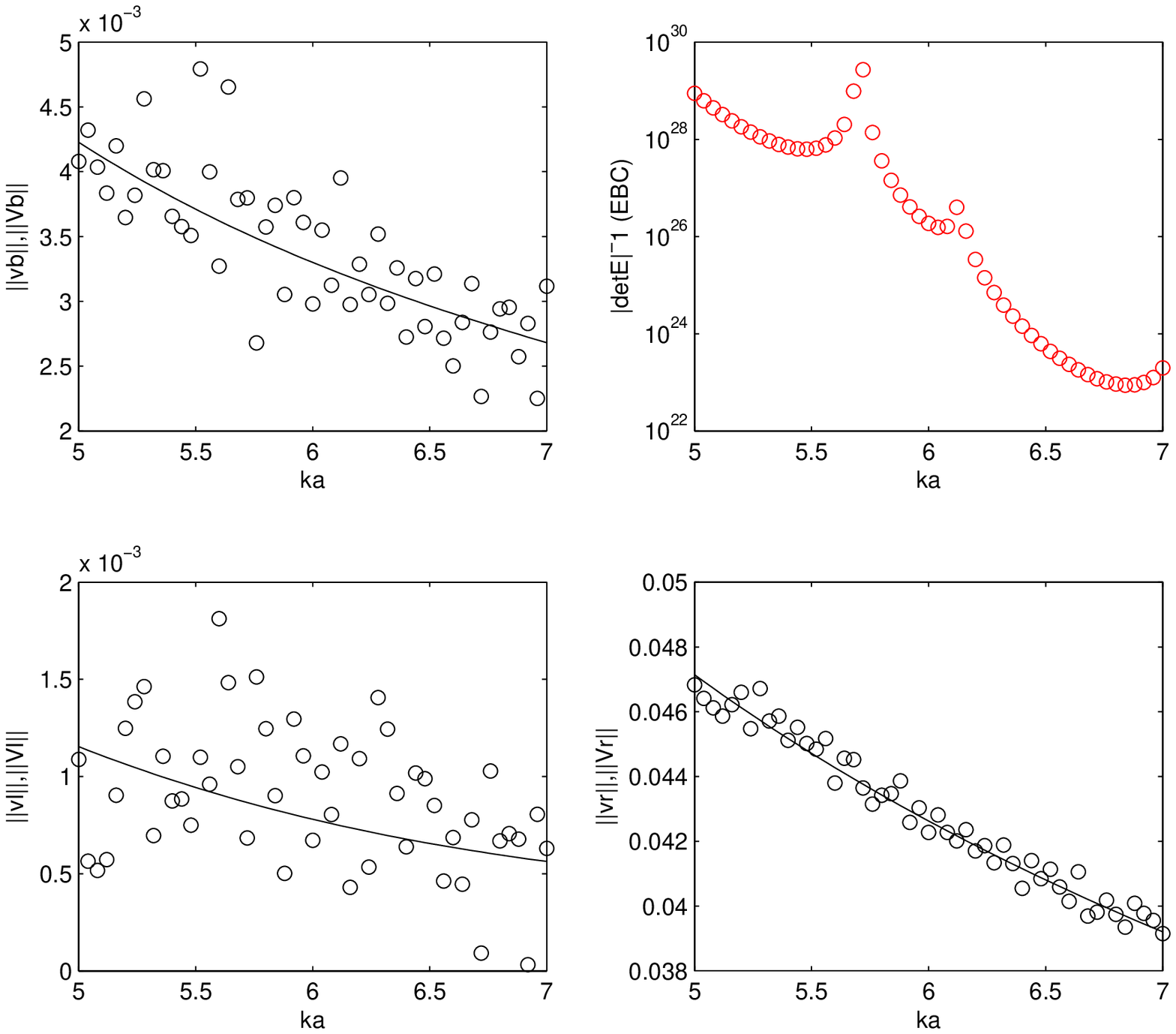}
\caption{}
\label{fdebc-8}
\end{center}
\end{figure}
\begin{figure}[ptb]
\begin{center}
\includegraphics[width=0.65\textwidth]{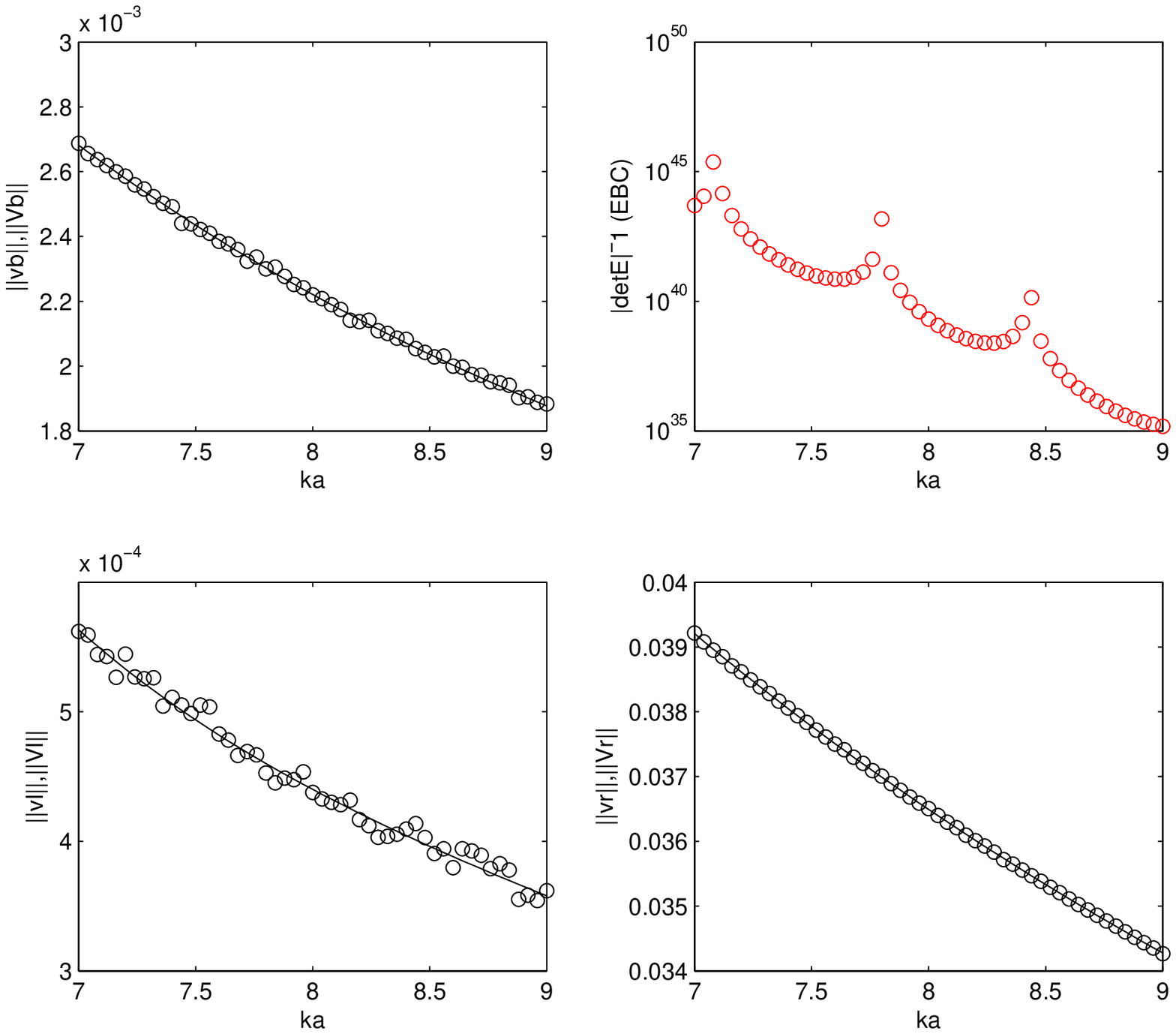}
\caption{Same as fig. \ref{fdebc-1a} except that   $N=24$, $\epsilon=10^{-7}$. Otherwise, this is another zoom of fig. \ref{fdebc-1a}.}
\label{fdebc-9}
\end{center}
\end{figure}
\begin{figure}[ptb]
\begin{center}
\includegraphics[width=0.65\textwidth]{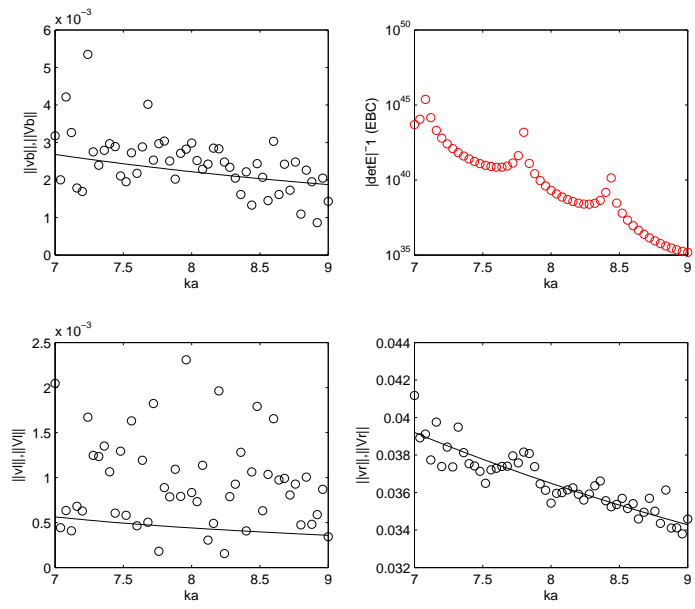}
\caption{Same as fig. \ref{fdebc-9} except that   $N=24$, $\epsilon=10^{-5}$. This is another effort to 'reveal' the resonances.}
\label{fdebc-10}
\end{center}
\end{figure}
\clearpage
\newpage
Note that most of the theoretically-predicted resonances do not  show up in the response curves.  Note especially that the positions of these DEBC resonances (which exist at least insofar as the $\mathbf{E}$ matrix exhibits peaks that betray their existence) differs from those of both the DBIE1 and DBIE2 resonances, which fact is again contrary to the hypothesis that all three sets are 'real' resonances since they all occur for the same scattering problem. Finally, note the scale changes of the ordinates in going from one figure to the next.
\subsection{The reason why some spurious resonance frequencies of the DBIE1 and DBIE2 are different, and others are identical}
\begin{figure}[ht]
\begin{center}
\includegraphics[width=0.75\textwidth]{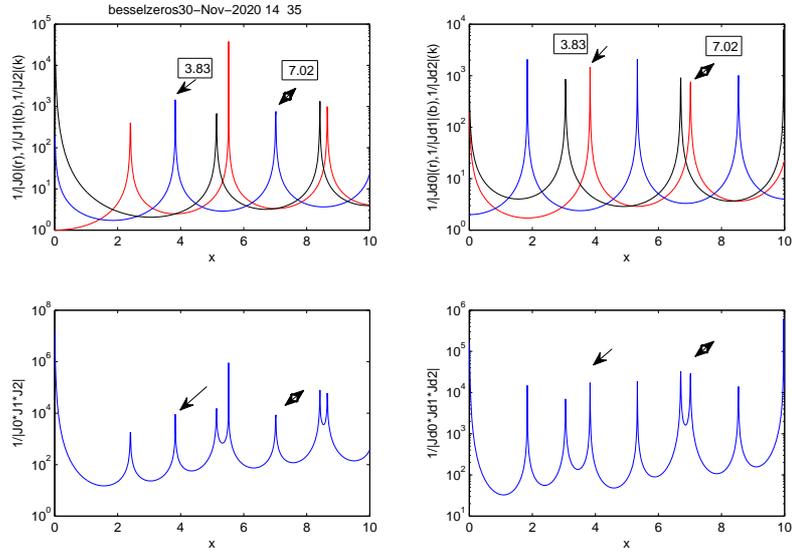}
\caption{This figure, in which the abscissa $x$ represents $ka$, shows that the resonance frequencies (i.e., frequencies for which $J_{n}=0$) of $J_{n}$ are generally different from those (i.e., frequencies for which $\dot{J}_{n}=0$) of $\dot{J}_{n}$. The only exceptions are $n=1$ for $J_{n}$ and $n=0$ for $\dot{J}_{n}$ since $\dot{J}_{0}(z)=-J(1)(z)$. Consequently the determinants for methods (e.g., DBIE1) involving products of $J_{n}~;~n=0,1,...$ vanish for frequencies that are generally different from the methods (e.g., DBIE2) involving $\dot{J}_{n}~;~n=0,1,....$, the only exceptions being the apparent resonances at $x=ka=3.83, 7.02,...$.}
\label{fdif12}
\end{center}
\end{figure}
\clearpage
\newpage
\subsection{The cure of the spurious resonance disease for the case of a circular cylinder whose boundary is the locus of a Dirichlet condition}
I first give a short review of the methods that have been proposed to cure the spurious resonance disease for {\it both Dirichlet and Neumann boundaries}. Then, I shall propose two methods of cure which can be named 'combined boundary integral equation' (CBIE) schemes for the Dirichlet boundary and further on for the Neumann boundary. Other combinations are possible and easily-recognizable from the material I am about to present.
\subsubsection{Review of the methods of cure prior to Nowak and Hall}
These methods appeared long before the ones suggested by Nowak and Hall \cite{no88,nh93}, but in the acoustical and electromagnetic wave contexts. As explained in the excellent review articles \cite{am87,bf07,za00} (see also \cite{sm83,ss84,ma99,tg01,cc01a,cc01b,wi02,mh04,cl06,ad07,wa11,ch20}), all these methods are based on employing linear combinations of the extended boundary condition, first-kind integral equation and second-kind integral equations so as to result, after discretization, in a matrix equation whose matrix is not singular at any frequency. In particular, this was the procedure adopted by Brakhage and Werner \cite{bw65}, Schenck \cite{sc68}, Bolomey and Tabbara \cite{bt70,bt73}, Burton and Miller \cite{bm71}, Mautz and Harrington \cite{mh77,mh78}, just to name a few.
\subsubsection{The Nowak scheme}
In \cite{no88},  Nowak writes, concerning the occurrence of what he terms 'artificial resonances' in his predicted seismic response of a semi-circular canyon: " The matrix equation becomes singular....To overcome this difficulty, responses are interpolated within the zones of the artificial resonances, using the undisturbed
responses outside the zones. This interpolation requires that the discretization be fine
enough to narrow the resonances enough so that the true responses can be traced." In \cite{nh93}, Nowak and Hall write, with respect to their fig. 5: "The  source  of the  artificial resonances  can  be traced  to  a  degeneracy  in  the  boundary  element  matrix  equation  at  these frequencies  as  described  in  Nowak,  1988  where  the  remedy was  to  use  a  fine enough  discretization  to  localize  the  resonances  and  allow  accurate  interpolation  of the  true  response."

\begin{figure}[ht]
\begin{center}
\includegraphics[width=0.65\textwidth]{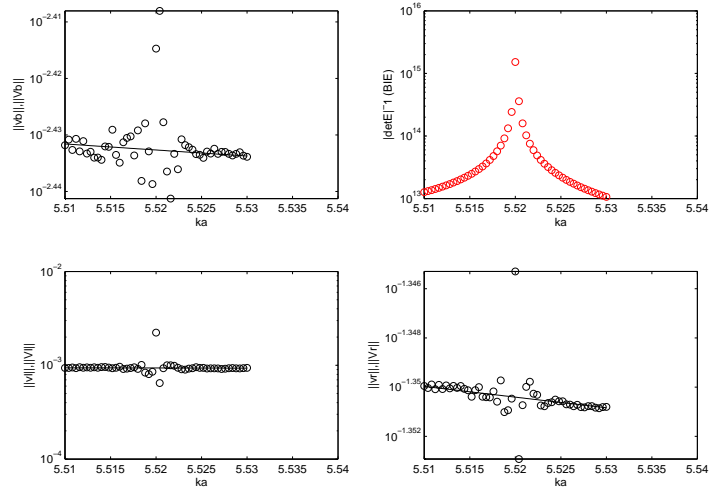}
\caption{Transfer functions of the traction at three points on the rigid boundary. The upper left-hand, lower left-hand, lower right-hand panels are for  the transfer functions at $\theta=180^{\circ}$,  $\theta=270^{\circ}$,  $\theta=360^{\circ}$, respectively. The upper right-hand panel depicts $1/\|det(\mathbf{E}(ka))\|$. Lower-case letters and circles correspond to  DBIE1 computations, upper-case letters and continuous curves to  DSOV (exact) computations.  Case $N=16$, $\epsilon=10^{-5}$}.
\label{fnow1}
\end{center}
\end{figure}

\begin{figure}[ptb]
\begin{center}
\includegraphics[width=0.65\textwidth]{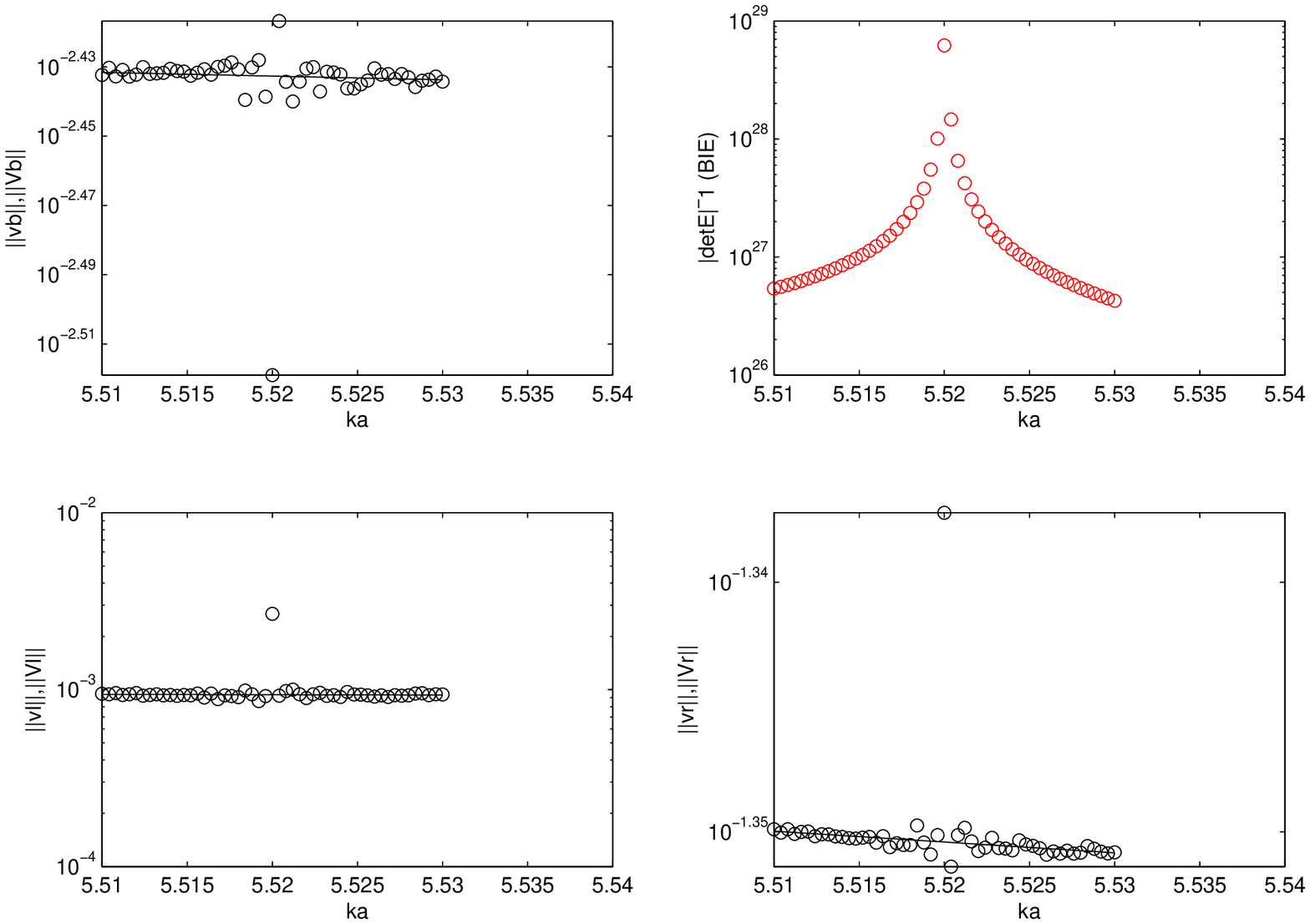}
\caption{Same as fig.\ref{fnow1} except that   $N=24$, $\epsilon=10^{-5}$.}
\label{fnow2}
\end{center}
\end{figure}

\begin{figure}[ptb]
\begin{center}
\includegraphics[width=0.65\textwidth]{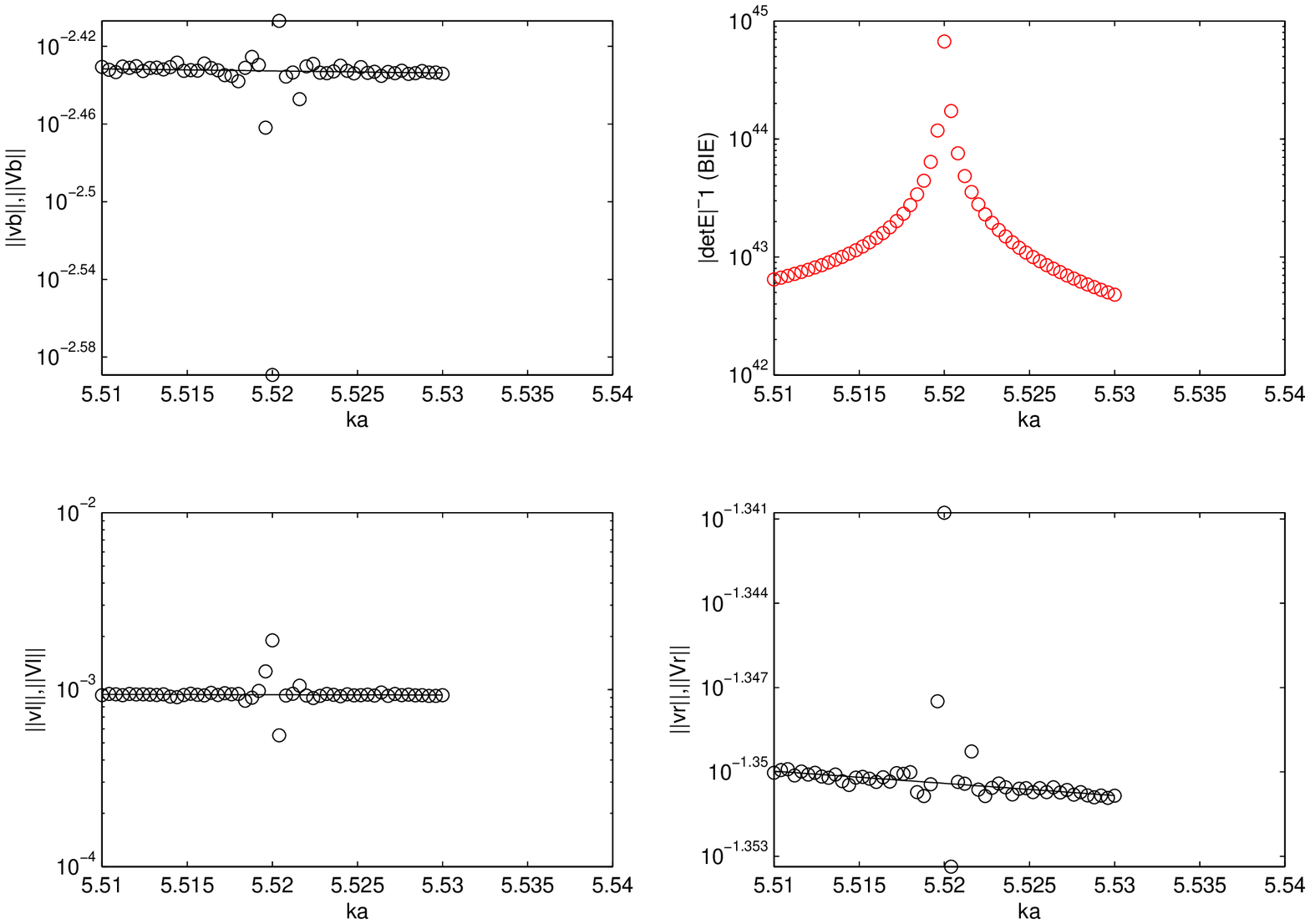}
\caption{Same as fig.\ref{fnow1} except that   $N=32$, $\epsilon=10^{-5}$.}
\label{fnow3}
\end{center}
\end{figure}

\begin{figure}[ptb]
\begin{center}
\includegraphics[width=0.65\textwidth]{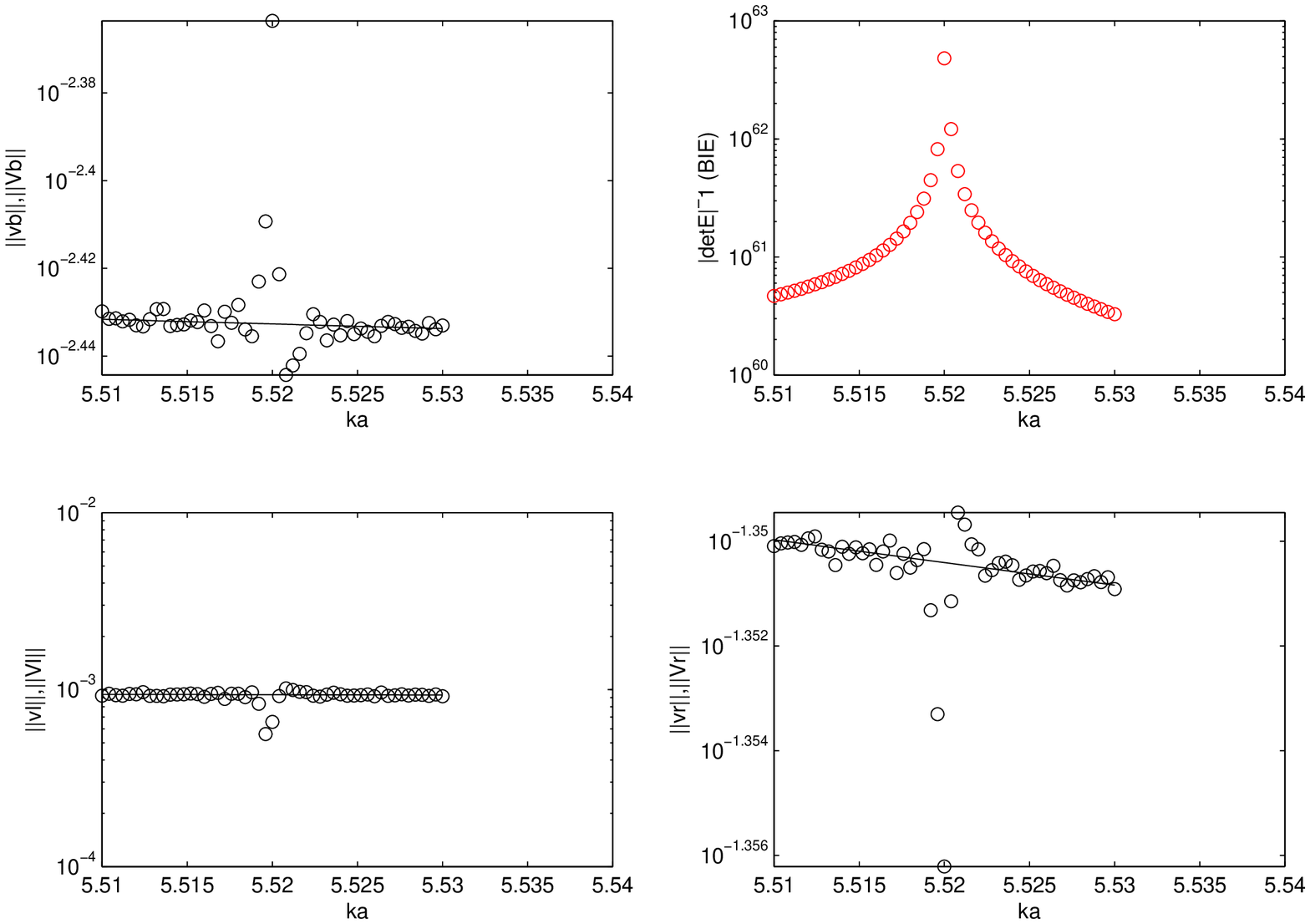}
\caption{Same as fig.\ref{fnow1} except that   $N=40$, $\epsilon=10^{-5}$.}
\label{fnow4}
\end{center}
\end{figure}

\begin{figure}[ptb]
\begin{center}
\includegraphics[width=0.65\textwidth]{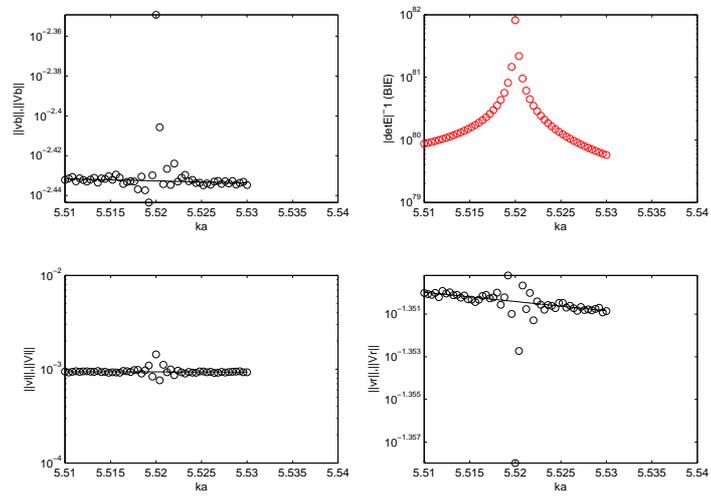}
\caption{Same as fig.\ref{fnow1} except that   $N=48$, $\epsilon=10^{-5}$.}
\label{fnow5}
\end{center}
\end{figure}

I tried to apply this scheme to obtain the results in figs. \ref{fnow1}-\ref{fnow5} herein. Since Nowak and Hall employ a discretization of the unknown boundary  functions method to solve their integral equations, whereas I appeal to a Fourier expansion of of these boundary functions, the equivalent of the Nowak scheme is, instead of refining the discretization, to increase the number ($2N+1$) of terms in the Fourier expansion.
\clearpage
\newpage
As figs. \ref{fnow1}-\ref{fnow5} (which all apply to the case of a circular rigid boundary cylinder of radius $a=1$ (a.u.) submitted to the wave radiated by a single line source located at $r^{s}=12$ (a.u.), $\theta^{s}=30^{\circ}$) show, doing this unfortunately does not result in a better localization of the resonance so as to permit the interpolation whereby the resonance is  eliminated. This is probably the reason why Nowak and Hall suggested use of another method, similar to our NCBIE (see hereafter) scheme, to suppress the resonances which  they took for granted to be 'artificial'. I say 'took for granted' because they did not actually prove that these resonances are numerical artifacts, all the more so than they trace the occurrence of the resonances to the excitation of a physically-real internal cavity  resonance (that of the cavity of which the circular arc of their canyon is a part). On the contrary, as I shall stress in the Conclusion of this contribution, my analysis in the preceding sections shows that the internal cavity resonances that are excited depend, for their appearance, on the choice of integral equation, whereas a resonance that appears in a physical response function can only be 'real' if it does not depend on the means (i.e., the type of integral equation) by which it is predicted.

\subsubsection{My first CBIE scheme (i.e., DCBIE1) appealing to DBIE1 and  DEBC}
Assuming that $\Gamma$ is the circle $r=a$ and $\Gamma_{in}$  the circle $r=b<a$, the point of departure is the two BIE's:
\begin{equation}\label{1-650}
0=u^{i}(\mathbf{x})+\int_{\Gamma}kG(\mathbf{x};\mathbf{x}')v(\mathbf{x})d\gamma(\mathbf{x'}~;~\forall \mathbf{x}\in\Gamma
~,
\end{equation}
\begin{equation}\label{1-660}
0=u^{i}(\mathbf{x})+\int_{\Gamma}kG(\mathbf{x};\mathbf{x}')v(\mathbf{x})d\gamma(\mathbf{x'}~;~\forall \mathbf{x}\in\Gamma_{in}
~,
\end{equation}
which, in polar coordinates, take the form:
\begin{equation}\label{1-670}
0=u^{i}(a,\theta)+\int_{0}^{2\pi}kaG(a,\theta;a,\theta')v(a,\theta')d\theta'~;~\forall\theta\in[0,2\pi[
~,
\end{equation}
\begin{equation}\label{1-680}
0=u^{i}(b,\theta)+\int_{0}^{2\pi}kaG(b,\theta;a,\theta')v(a,\theta')d\theta'~;~\forall\theta\in[0,2\pi[
~,
\end{equation}
Since both equations apply to the same $\theta$ intervals, I can form a linear combination of the two so as to obtain the single BIE
\begin{equation}\label{1-690}
u(a,\theta)=u^{i}(a,\theta)+\eta u^{i}(b,\theta)+\int_{0}^{2\pi}ka[G(a,\theta;a,\theta')+\eta G(b,\theta;a,\theta')]v(a,\theta')d\theta'~;~\forall\theta\in[0,2\pi[
~,
\end{equation}
wherein $\eta$ is an unspecified scalar constant for the moment.

I make the expansions
\begin{equation}\label{1-700}
v(a,\theta)=\sum_{n\in\mathbb{Z}}f_{n}\exp(in\theta)~~,~~u^{i}(a,\theta)=\sum_{n\in\mathbb{Z}}g_{n}\exp(in\theta)~~,~~
u^{i}(b,\theta)=\sum_{n\in\mathbb{Z}}h_{n}\exp(in\theta)~~;~\forall\theta\in[0,2\pi[
~,
\end{equation}
and again invoke the Galerkin procedure to obtain
\begin{equation}\label{1-710}
f_{m}=g_{m}+\eta h_{m}+\sum_{n\in\mathbb{Z}}f_{n}\int_{0}^{2\pi}d\theta\int_{0}^{2\pi}d\theta'\frac{ka}{2\pi}[G(a,\theta;a,\theta')+\eta G(b,\theta;a,\theta')]\exp[i(n-m)\theta]~;~\forall m\in\mathbb{Z}
~.
\end{equation}
By recalling previous results I find
\begin{equation}\label{1-720}
G(a,\theta;a,\theta')+\eta G(b,\theta;a,\theta')=\frac{i}{4}\sum_{l\in\mathbb{Z}}H_{l}^{(1)}(ka)[J_{l}(ka)+\eta J_{l}(kb)]\exp[i
l(\theta-\theta')]
~,
\end{equation}
so that the following matrix equation ensues
\begin{equation}\label{1-730}
\sum_{l\in\mathbb{Z}}E_{mn}f_{n}=g_{m}+\eta h_{m}=A_{m}[J_{m}(ka)+\eta J_{m}(kb)]
~,
\end{equation}
wherein
\begin{equation}\label{1-740}
E_{mn}=\frac{-ika\pi}{2}H_{n}^{(1)}(ka)[J_{n}(ka)+\eta J_{n}(kb)]\delta_{mn}
~.
\end{equation}
Again,  $\mathbf{E}=\{E_{mn}\}$ is an infinite-order, diagonal matrix, but now it is {\it not singular at any real frequency provided $\eta$ is chosen to be an imaginary scalar constant} because the Bessel functions are real at real frequencies. Consequently, with this choice of $\eta$, the inverse of $\mathbf{E}$ exists at all real frequencies so that
\begin{equation}\label{1-750}
f_{n}=A_{n}\left[\frac{-ika\pi}{2}H_{n}^{(1)}(ka)\right]^{-1}
~,
\end{equation}
which is nothing other than the exact SOV solution. Thus, this first CBIE scheme constitutes a cure for the disease that plagues traditional BIE methods (at least for scattering problems with a Dirichlet condition on a circular boundary).
\subsubsection{The second CBIE scheme (i.e., DCBIE2) appealing to DBIE1 and DBIE2}
Assuming that $\Gamma$ is the circle $r=a$ the point of departure is the two BIE's:
\begin{equation}\label{1-760}
0=u^{i}(\mathbf{x})+\int_{\Gamma}kG(\mathbf{x};\mathbf{x}')v(\mathbf{x})d\gamma(\mathbf{x'}~;~\forall \mathbf{x}\in\Gamma
~,
\end{equation}
\begin{equation}\label{1-770}
\frac{1}{2}v(\mathbf{x})=v^{i}(\mathbf{x})+
pv\int_{\Gamma}\boldsymbol{\nu}\cdot\nabla G(\mathbf{x};\mathbf{x}')v(\mathbf{x})d\gamma(\mathbf{x'}~;~\forall \mathbf{x}\in\Gamma
~,
\end{equation}
which, in polar coordinates, take the form:
\begin{equation}\label{1-780}
0=u^{i}(a,\theta)+\int_{0}^{2\pi}kaG(a,\theta;a,\theta')v(a,\theta')d\theta'~;~\forall\theta\in[0,2\pi[
~,
\end{equation}
\begin{equation}\label{1-790}
\frac{1}{2}v(a,\theta)=v^{i}(a,\theta)-pv\int_{0}^{2\pi}a\frac{\partial}{\partial  r}G(a,\theta;a,\theta')v(a,\theta')d\theta'~;~\forall\theta\in[0,2\pi[
~.
\end{equation}
Since both equations apply to the same $\theta$ intervals, I again form a linear combination of the two so as to obtain the single BIE
\begin{equation}\label{1-800}
\frac{1}{2}v(a,\theta)=u^{i}(a,\theta)+\eta v^{i}(b,\theta)+\int_{0}^{2\pi}a[kG(a,\theta;a,\theta')-\eta\frac{\partial}{\partial  r}G(a,\theta;a,\theta')]v(a,\theta')d\theta'~;~\forall\theta\in[0,2\pi[
~,
\end{equation}
wherein $\eta$ is an unspecified scalar constant for the moment and I keep in mind that the integral involving the derivative of $G$ is a principal value integral.

I make the expansions:
\begin{equation}\label{1-810}
v(a,\theta)=\sum_{n\in\mathbb{Z}}f_{n}\exp(in\theta)~~,~~u^{i}(a,\theta)=\sum_{n\in\mathbb{Z}}g_{n}\exp(in\theta)~~,~~
v^{i}(a,\theta)=\sum_{n\in\mathbb{Z}}h_{n}\exp(in\theta)~~;~\forall\theta\in[0,2\pi[
~,
\end{equation}
and again invoke the Galerkin procedure to obtain
\begin{multline}\label{1-820}
\frac{1}{2}f_{m}=g_{m}+\eta h_{m}+\\
\sum_{n\in\mathbb{Z}}f_{n}\int_{0}^{2\pi}d\theta\int_{0}^{2\pi}d\theta'\frac{a}{2\pi}[kG(a,\theta;a,\theta')-\eta \frac{\partial}{\partial  r}G(a,\theta;a,\theta')]\exp[i(n-m)\theta]~;~\forall m\in\mathbb{Z}
~.
\end{multline}
By recalling previous results I find
\begin{equation}\label{1-830}
kG(a,\theta;a,\theta')-\eta \frac{\partial}{\partial  r}G(b,\theta;a,\theta')=\frac{ik}{4}\sum_{l\in\mathbb{Z}}\left\{H_{l}^{(1)}(ka)[J_{l}(ka)-\eta \dot{J}_{l}(ka)]-\eta \frac{i}{ka\pi}\right\}\exp[i
l(\theta-\theta')]
~,
\end{equation}
so that the following matrix equation ensues
\begin{equation}\label{1-840}
\sum_{l\in\mathbb{Z}}E_{mn}f_{n}=g_{m}+\eta h_{m}=A_{m}[J_{m}(ka)-\eta \dot{J}_{m}(ka)]
~,
\end{equation}
wherein
\begin{equation}\label{1-850}
E_{mn}=\frac{-ika\pi}{2}H_{n}^{(1)}(ka)[J_{n}(ka)-\eta \dot{J}_{n}(ka)]\delta_{mn}
~.
\end{equation}
Again,  $\mathbf{E}=\{E_{mn}\}$ is an infinite-order, diagonal matrix, but now it is {\it not singular at any real frequency provided $\eta$ is chosen to be an imaginary scalar constant} because the Bessel functions and derivatives of the latter are real at real frequencies. Consequently, with this choice of $\eta$, the inverse of $\mathbf{E}$ exists at all real frequencies so that
\begin{equation}\label{1-860}
f_{n}=A_{n}\left[\frac{-ika\pi}{2}H_{n}^{(1)}(ka)\right]^{-1}
~,
\end{equation}
which is nothing other than the exact SOV solution. Thus, this second CBIE scheme constitutes another cure for the disease that plagues traditional BIE methods (at least for scattering problems with a Dirichlet condition on a circular boundary).
\clearpage
\newpage
\subsection{Numerical results for the Dirichlet boundary circular cylinder which illustrate the cure (via DCBIE1 and DCBIE2) of the spurious resonance disease}
The following figures, i.e., \ref{fdcbie1-1}-\ref{fdcbie1-12} and \ref{fdcbie2-1}-\ref{fdcbie2-9}, all apply to a rigid circular cylinder of radius $a=1$ (a.u.) submitted to the wave radiated by a line source situated at  $r^{s}=12$ (a.u.), $\theta^{s}=30^{\circ}$. The responses (as a function of $ka$, $k$ the wavenumber) are computed first by the couples (DBIE1, DCBIE1), then by the couples (DBIE1,DCBIE2). In DCBIE1 I choose $b=0.9$ (a.u.) and $\eta=0+1i$ and a randomization of the elements of the matrix $\mathbf{E}$ just like that of this matrix in DBIE1.
\subsubsection{DBIE1 cured by DCBIE1}
\begin{figure}[ht]
\begin{center}
\includegraphics[width=0.55\textwidth]{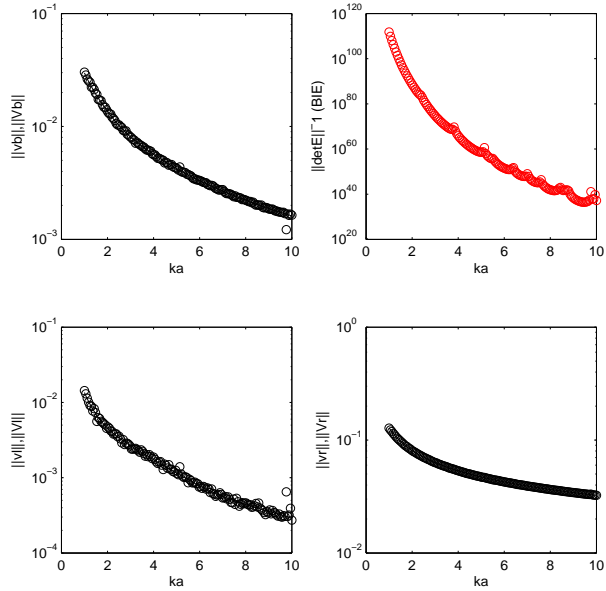}
\caption{Transfer functions of traction at three points on the rigid boundary. The upper left-hand, lower left-hand, lower right-hand panels are for  the transfer functions at $\theta=180^{\circ}$,  $\theta=270^{\circ}$,  $\theta=360^{\circ}$, respectively. The upper right-hand panel depicts $1/\|det(\mathbf{E}(ka))\|$. Lower-case letters and circles correspond to  DBIE1 computations, upper-case letters and continuous curves to  DSOV (exact) computations.  Case $N=38$, $\epsilon=10^{-4}$.}
\label{fdcbie1-1}
\end{center}
\end{figure}
\begin{figure}[ptb]
\begin{center}
\includegraphics[width=0.55\textwidth]{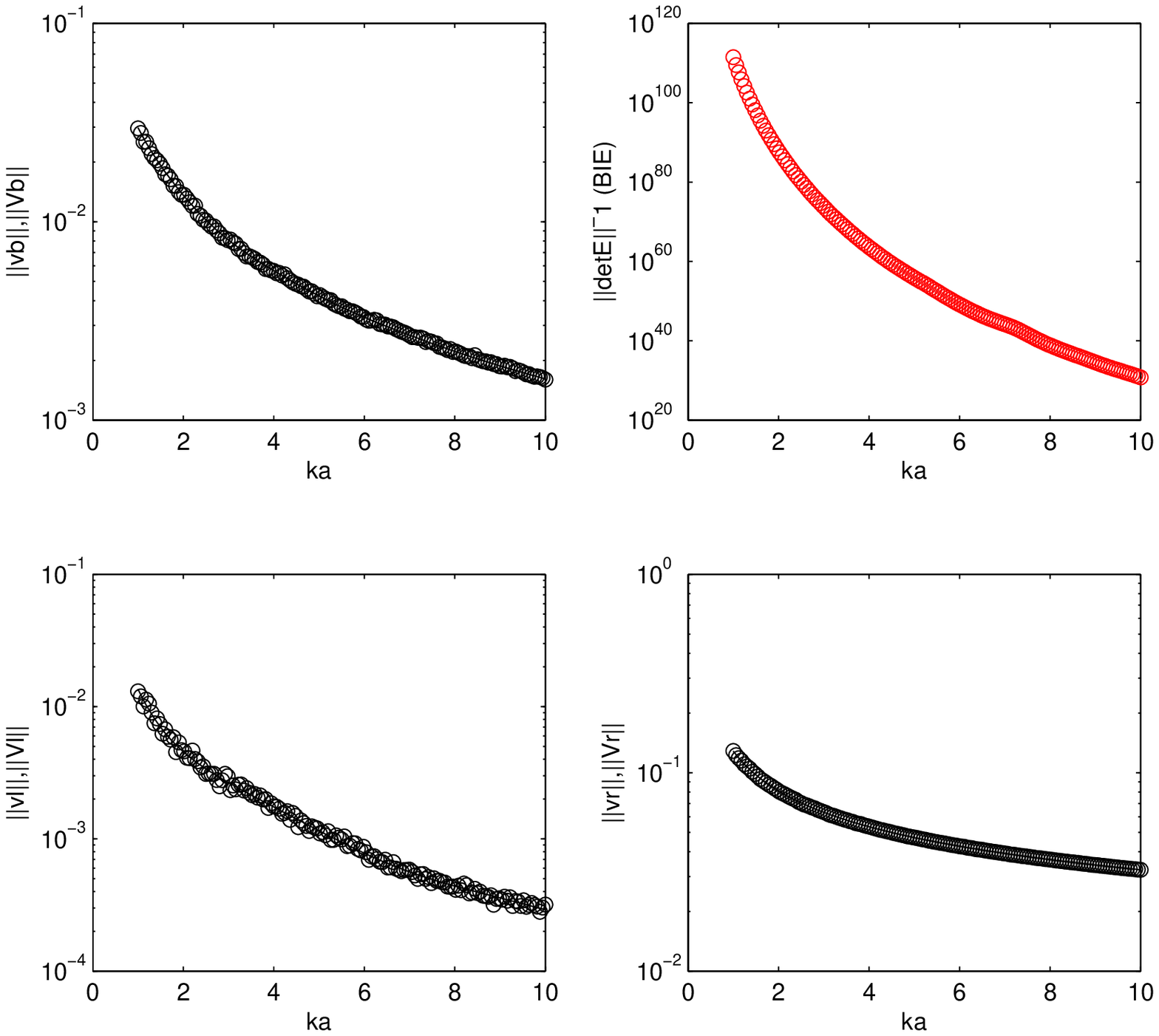}
\caption{Transfer functions of traction at three points on the rigid boundary. The upper left-hand, lower left-hand, lower right-hand panels are for  the transfer functions at $\theta=180^{\circ}$,  $\theta=270^{\circ}$,  $\theta=360^{\circ}$, respectively. The upper right-hand panel depicts $1/\|det(\mathbf{E}(ka))\|$. Lower-case letters and circles correspond to  DCBIE1 computations, upper-case letters and continuous curves to  DSOV (exact) computations. Case $N=38$, $\epsilon=10^{-4}$.}
\label{fdcbie1-2}
\end{center}
\end{figure}
\begin{figure}[ptb]
\begin{center}
\includegraphics[width=0.55\textwidth]{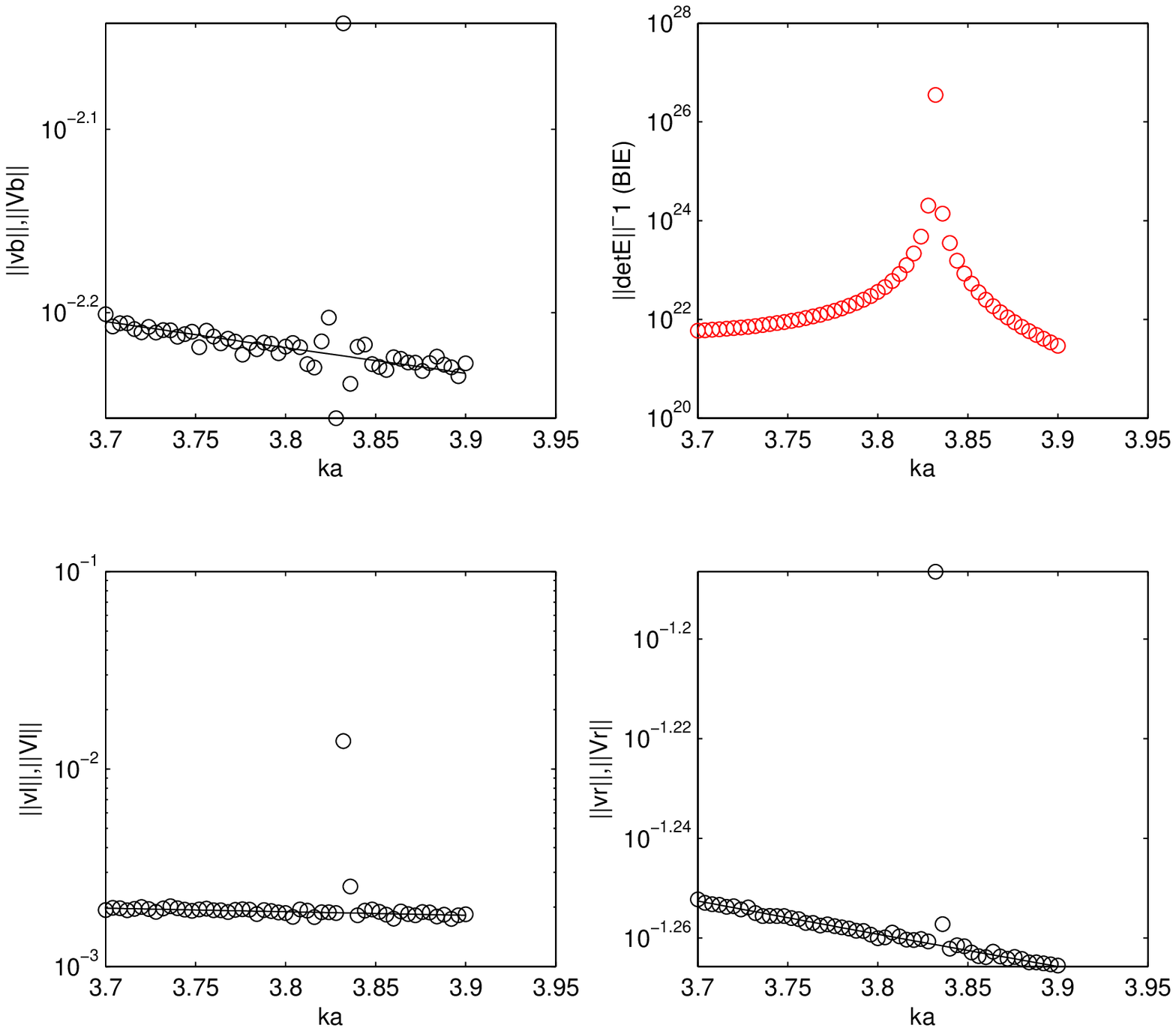}
\caption{Same as fig. \ref{fdcbie1-1} (of which the present figure is a zoom) except that  $N=18$ and  $\epsilon=10^{-4}$.}
\label{fdcbie1-3}
\end{center}
\end{figure}
\begin{figure}[ptb]
\begin{center}
\includegraphics[width=0.55\textwidth]{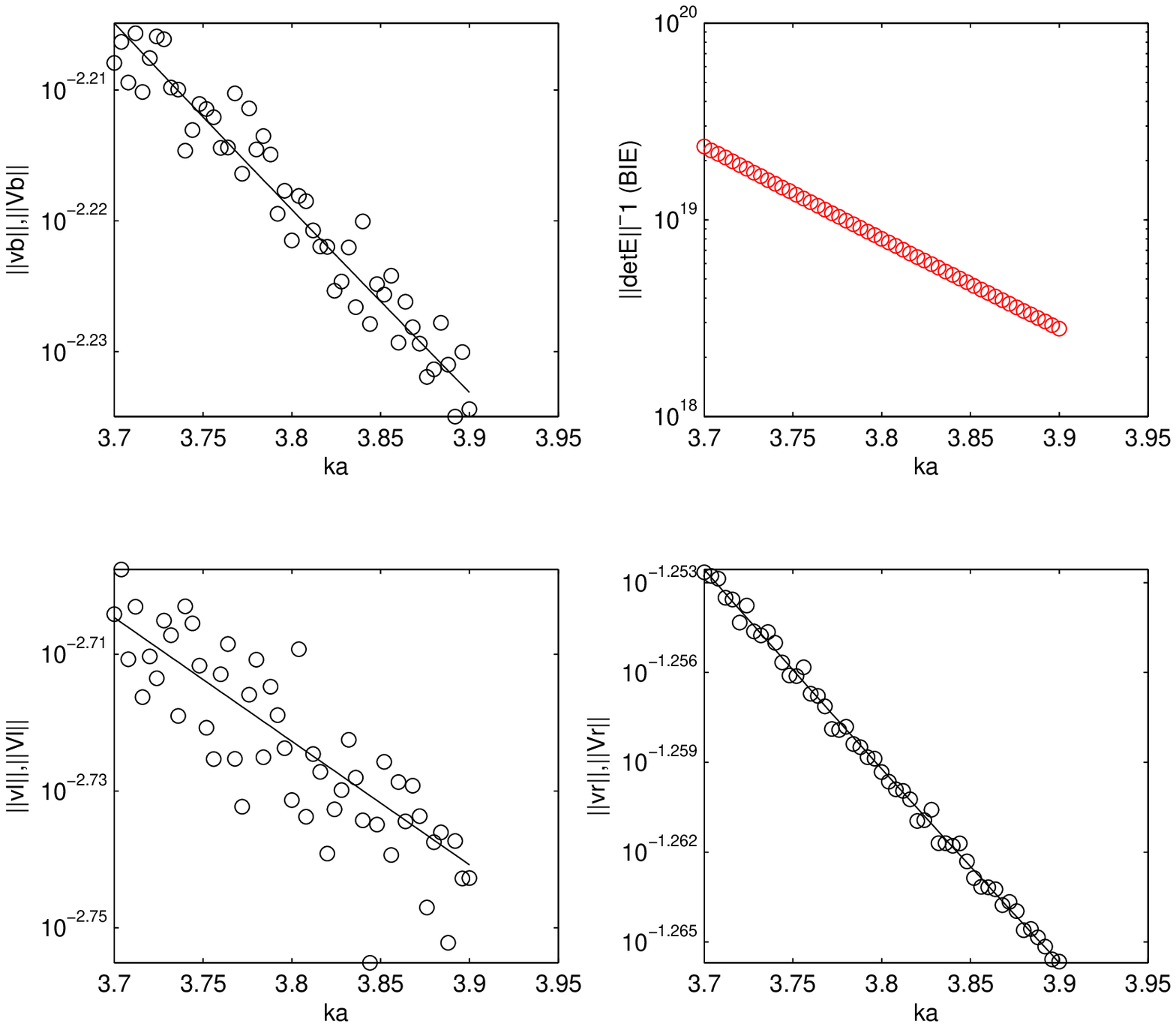}
\caption{Same as fig. \ref{fdcbie1-2} (of which the present figure is a zoom) except that  $N=18$ and  $\epsilon=10^{-4}$}
\label{fdcbie1-4}
\end{center}
\end{figure}
\begin{figure}[ptb]
\begin{center}
\includegraphics[width=0.55\textwidth]{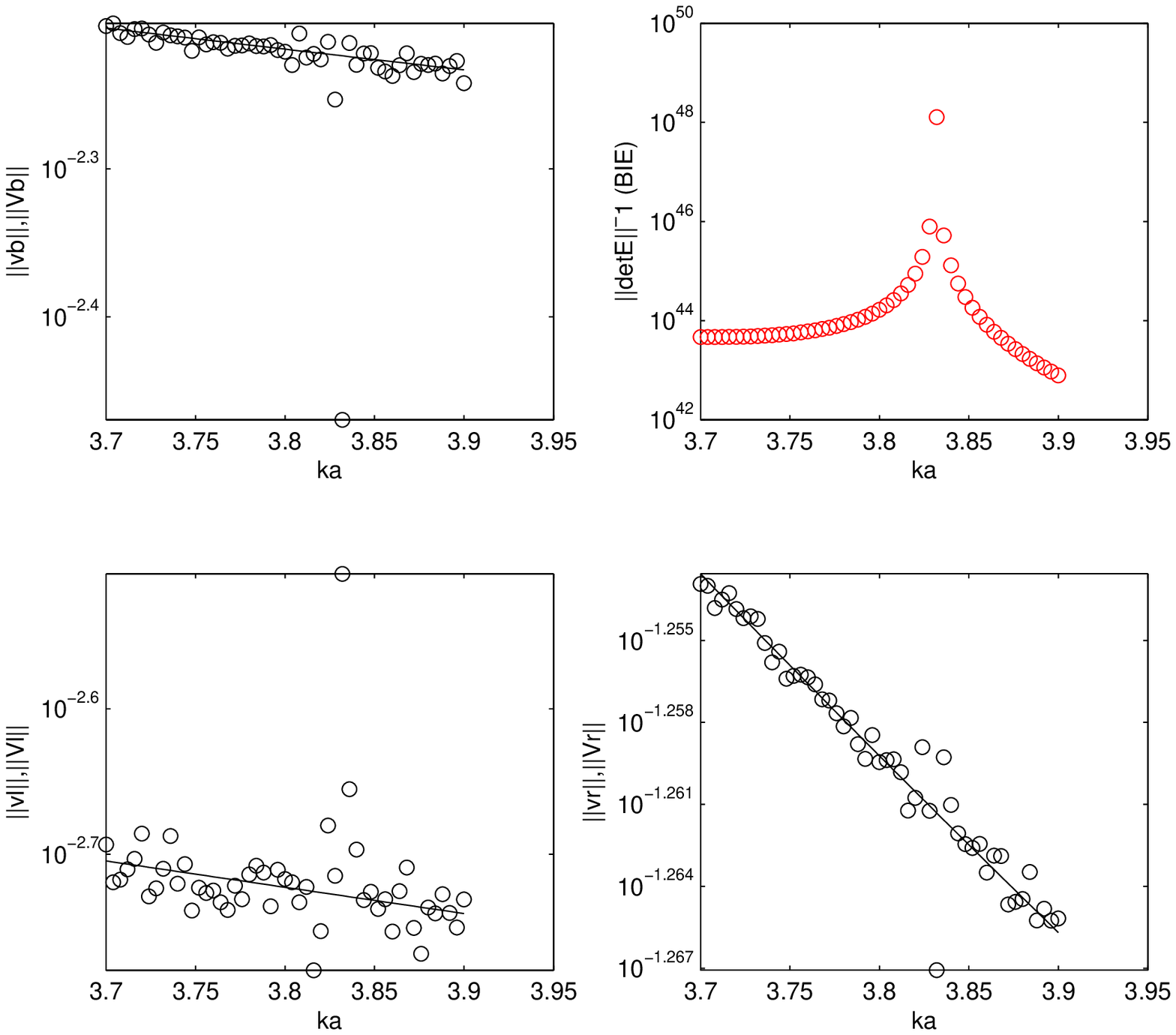}
\caption{Same as fig. \ref{fdcbie1-1} (of which the present figure is a zoom) except that  $N=28$ and  $\epsilon=10^{-4}$.}
\label{fdcbie1-5}
\end{center}
\end{figure}
\begin{figure}[ptb]
\begin{center}
\includegraphics[width=0.55\textwidth]{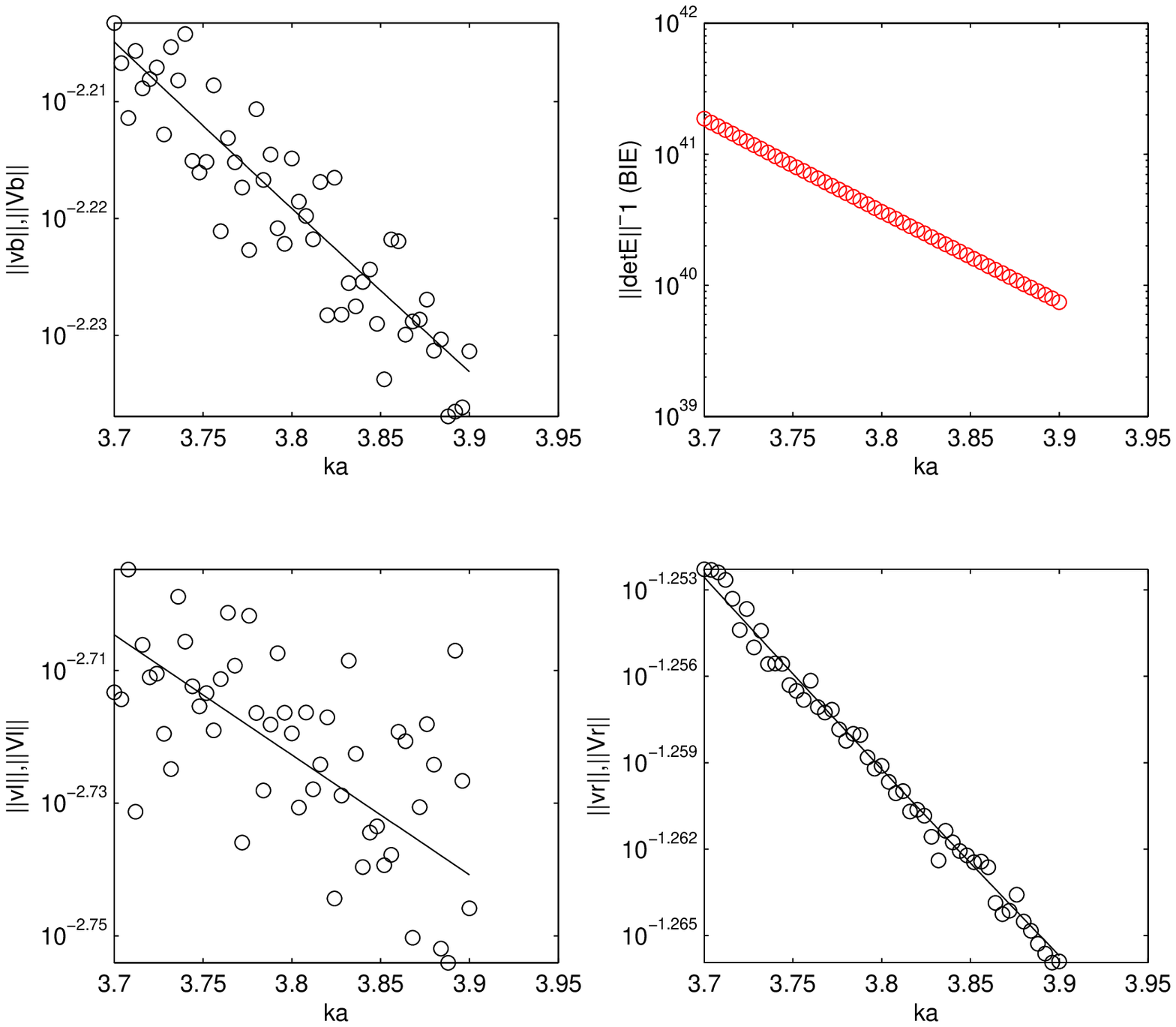}
\caption{Same as fig. \ref{fdcbie1-2} (of which the present figure is a zoom) except that  $N=28$ and  $\epsilon=10^{-4}$.}
\label{fdcbie1-6}
\end{center}
\end{figure}
\begin{figure}[ptb]
\begin{center}
\includegraphics[width=0.55\textwidth]{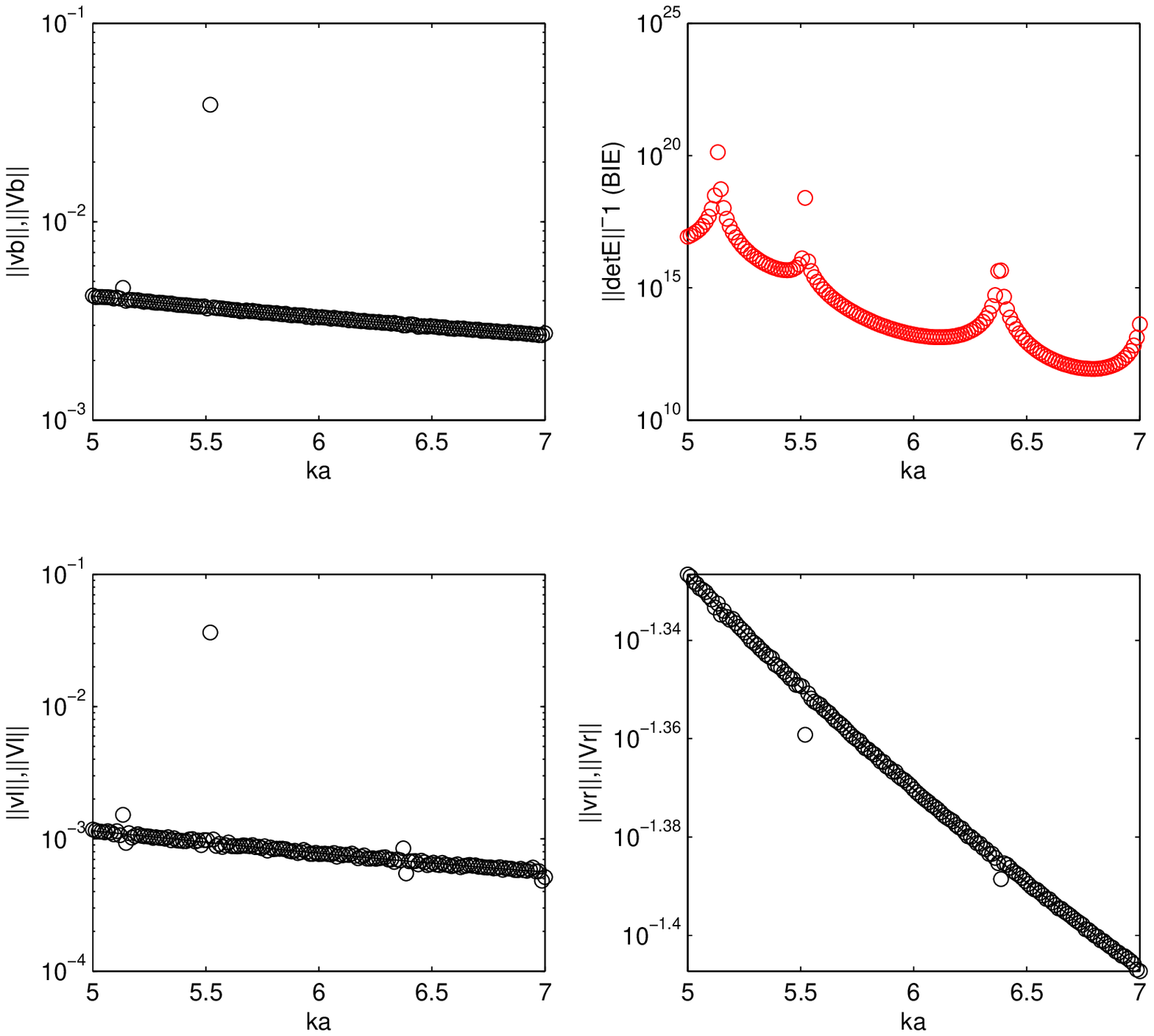}
\caption{Same as fig. \ref{fdcbie1-1} (of which the present figure is a zoom) except that  $N=18$ and  $\epsilon=10^{-4}$.}
\label{fdcbie1-7}
\end{center}
\end{figure}
\begin{figure}[ptb]
\begin{center}
\includegraphics[width=0.55\textwidth]{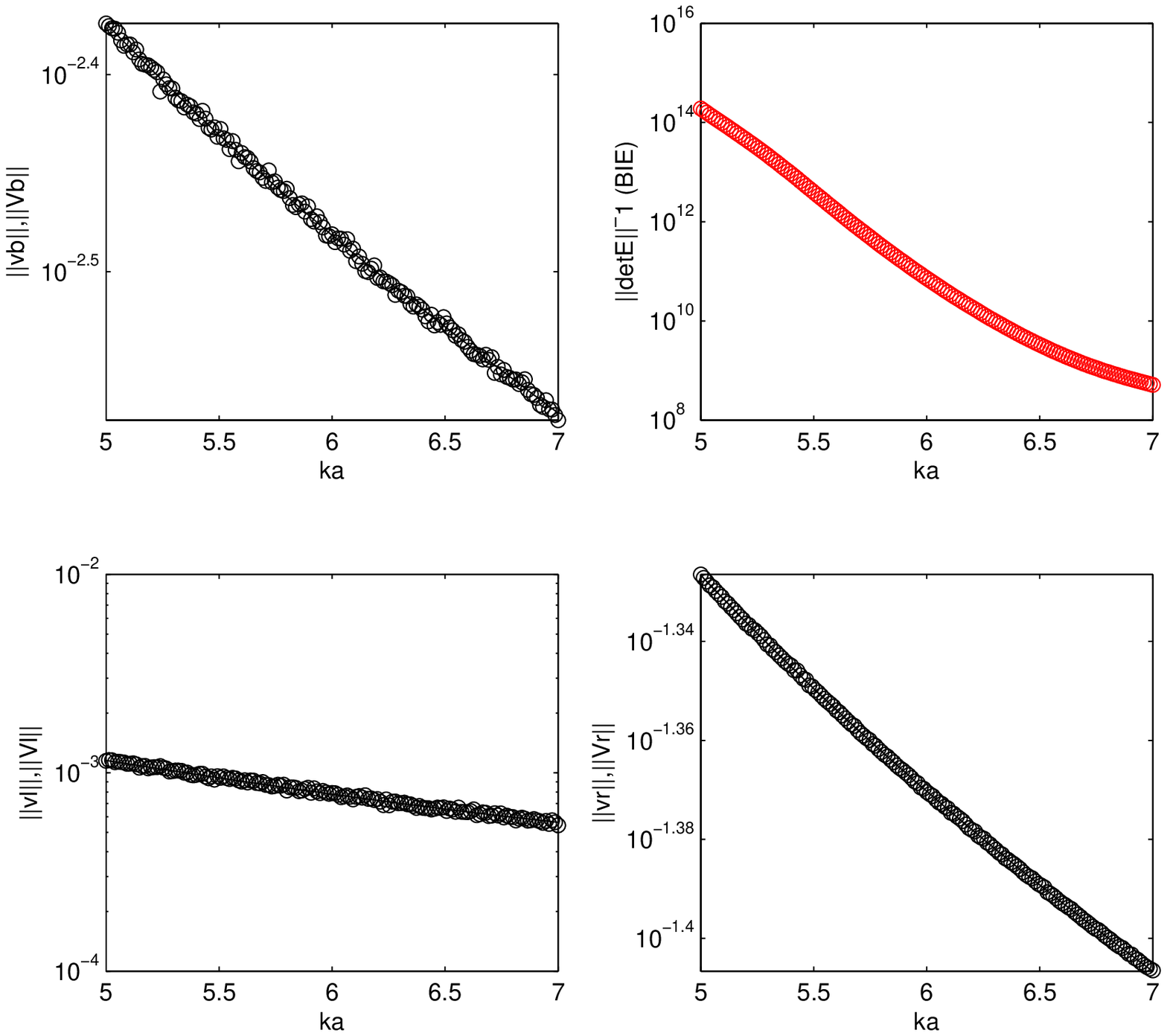}
\caption{Same as fig. \ref{fdcbie1-2} (of which the present figure is a zoom) except that  $N=18$ and  $\epsilon=10^{-4}$.}
\label{fdcbie1-8}
\end{center}
\end{figure}
\begin{figure}[ptb]
\begin{center}
\includegraphics[width=0.55\textwidth]{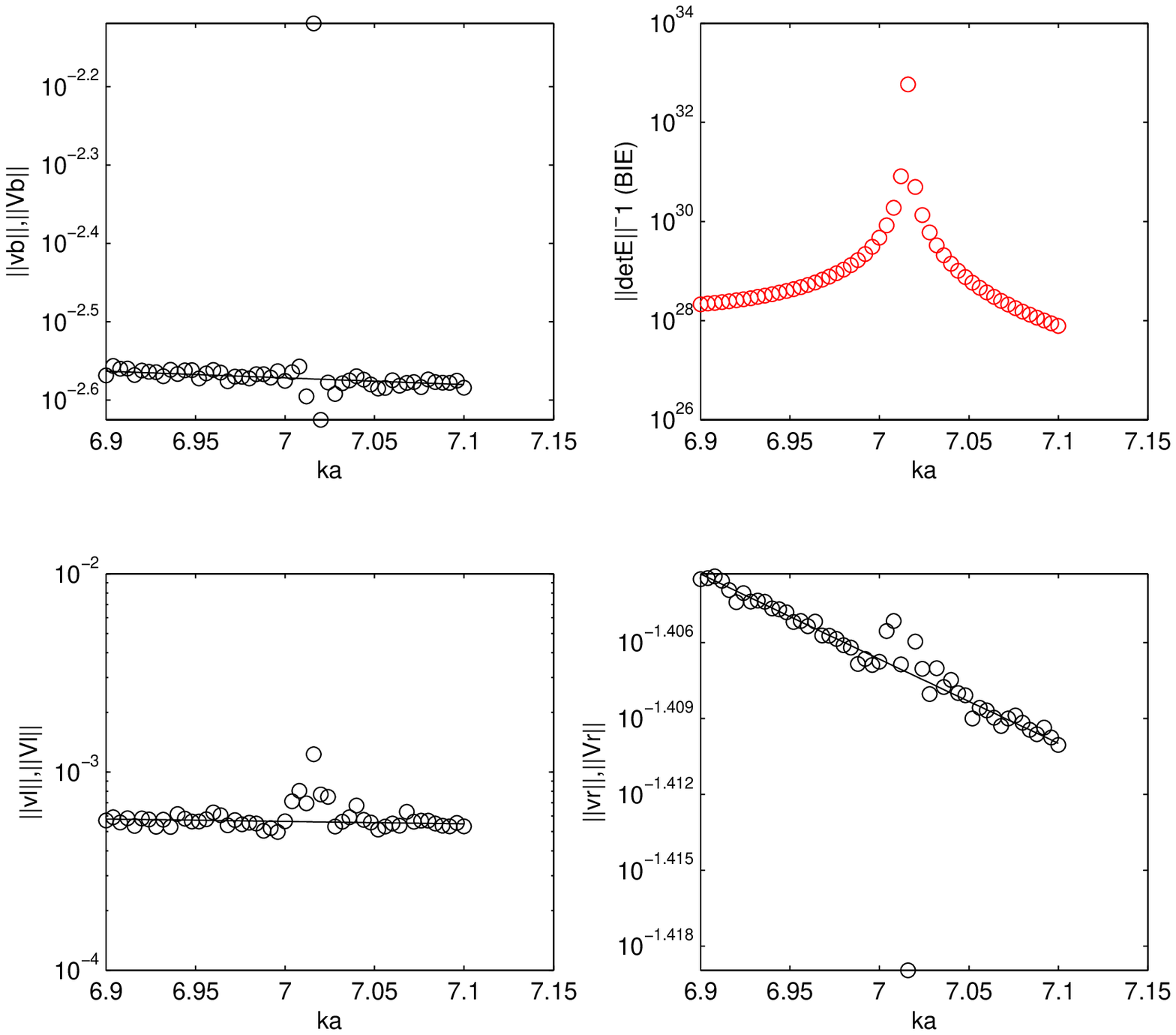}
\caption{Same as fig. \ref{fdcbie1-1} (of which the present figure is a zoom) except that  $N=28$ and  $\epsilon=10^{-4}$.}
\label{fdcbie1-9}
\end{center}
\end{figure}
\begin{figure}[ptb]
\begin{center}
\includegraphics[width=0.55\textwidth]{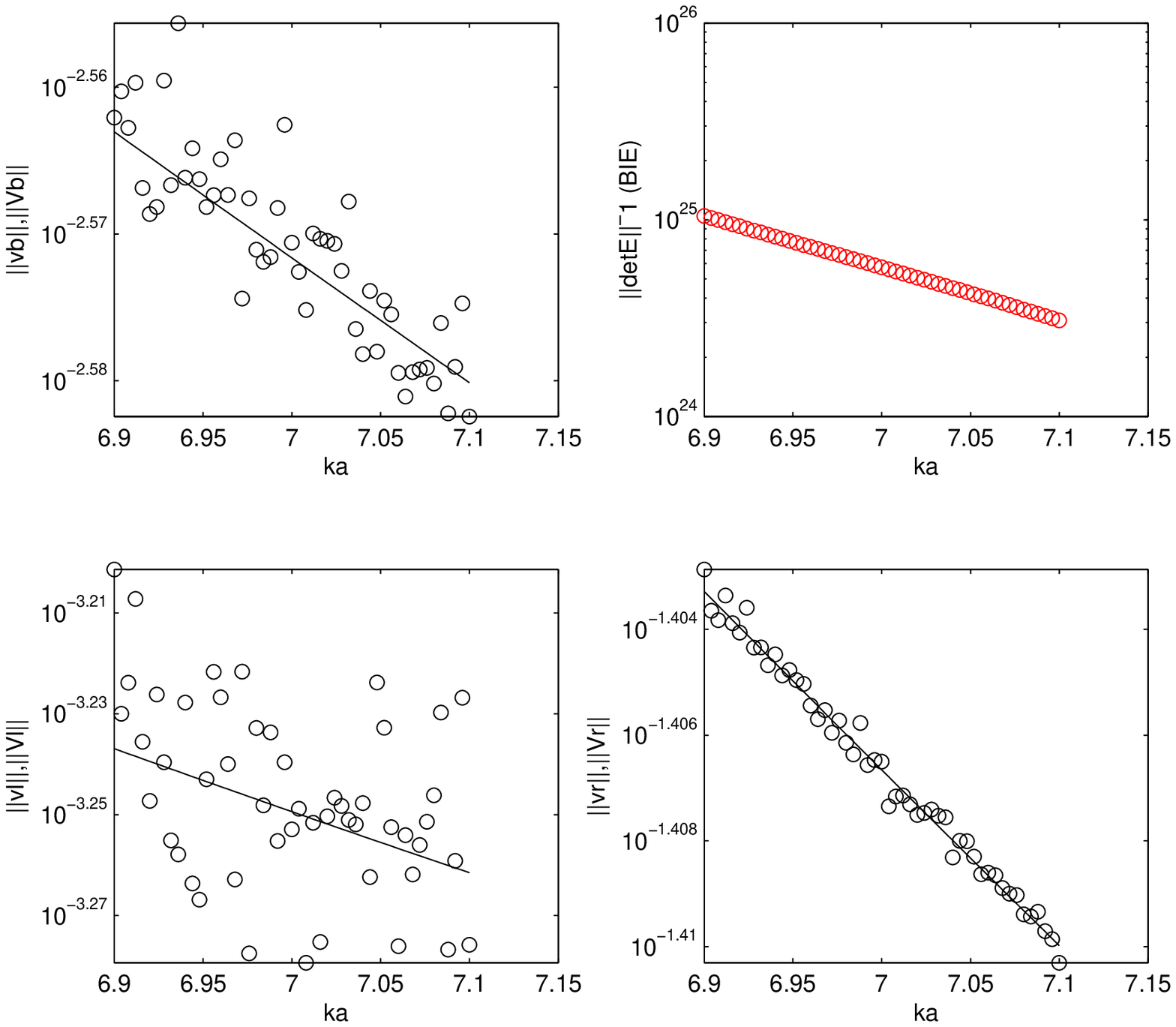}
\caption{Same as fig. \ref{fdcbie1-2} (of which the present figure is a zoom) except that  $N=28$ and  $\epsilon=10^{-4}$.}
\label{fdcbie1-10}
\end{center}
\end{figure}
\begin{figure}[ptb]
\begin{center}
\includegraphics[width=0.55\textwidth]{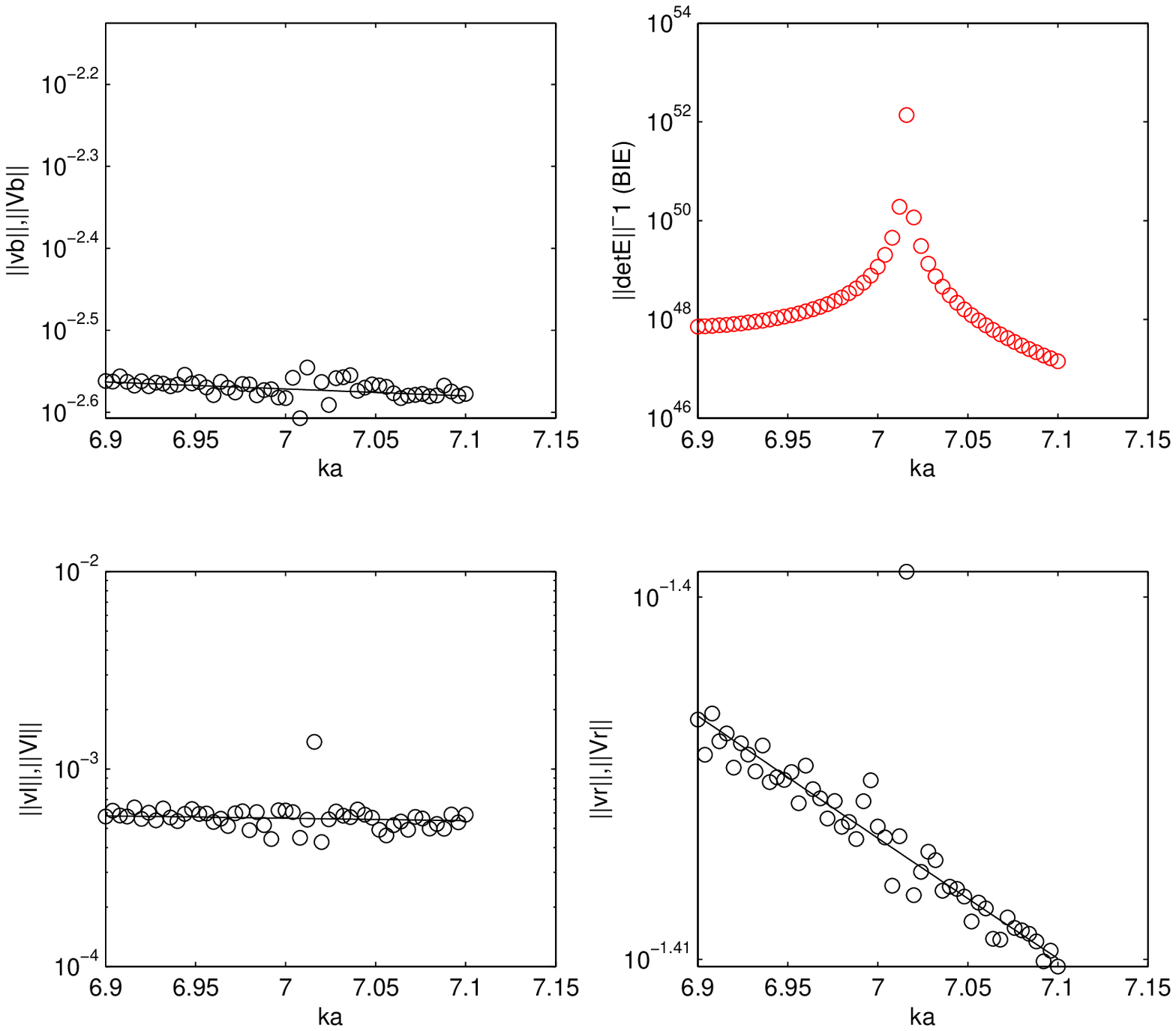}
\caption{Same as fig. \ref{fdcbie1-1} (of which the present figure is a zoom) except that  $N=38$ and  $\epsilon=10^{-4}$.}
\label{fdcbie1-11}
\end{center}
\end{figure}
\begin{figure}[ptb]
\begin{center}
\includegraphics[width=0.55\textwidth]{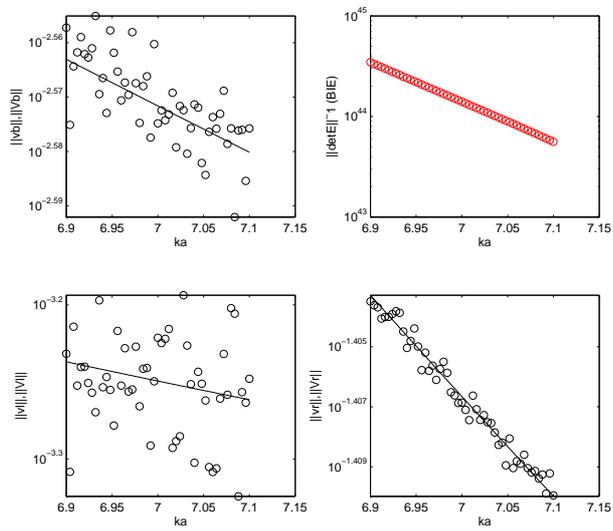}
\caption{Same as fig. \ref{fdcbie1-2} (of which the present figure is a zoom) except that  $N=38$ and  $\epsilon=10^{-4}$.}
\label{fdcbie1-12}
\end{center}
\end{figure}
\clearpage
\newpage
If account is taken of the scale changes in this set of figures, the latter shows convincingly that the method of cure DCBIE1 has enabled to eliminate all the resonances appearing in DBIE1. These results show that DCBIE1 also enables to eliminate all the resonances appearing in DBIE2 and DEBC (which, it will be recalled, occur at frequencies that  are generally-different from those at which occur the resonances appearing in DBIE1).
\subsubsection{DBIE1 cured by DCBIE2}
\begin{figure}[ht]
\begin{center}
\includegraphics[width=0.55\textwidth]{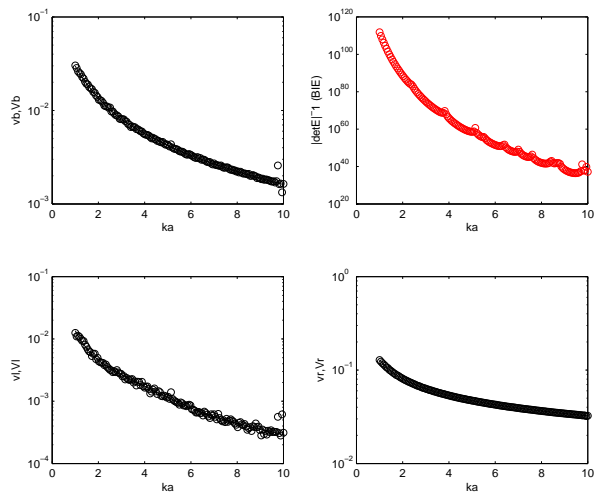}
\caption{Transfer functions of traction at three points on the rigid boundary. The upper left-hand, lower left-hand, lower right-hand panels are for  the transfer functions at $\theta=180^{\circ}$,  $\theta=270^{\circ}$,  $\theta=360^{\circ}$, respectively. The upper right-hand panel depicts $1/\|det(\mathbf{E}(ka))\|$. Lower-case letters and circles correspond to  DBIE1 computations, upper-case letters and continuous curves to  DSOV (exact) computations.  Case $N=38$, $\epsilon=10^{-4}$.}
\label{fdcbie2-1}
\end{center}
\end{figure}
\begin{figure}[ptb]
\begin{center}
\includegraphics[width=0.55\textwidth]{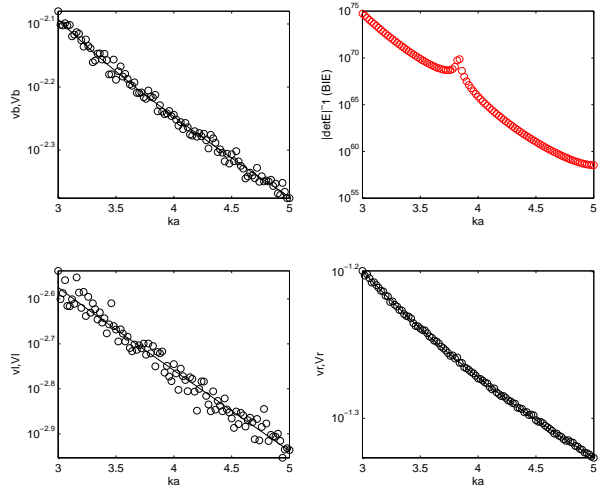}
\caption{This is a zoom of, but otherwise identical to, fig. \ref{fdcbie2-1}. Case $N=38$, $\epsilon=10^{-4}$.}
\label{fdcbie2-2}
\end{center}
\end{figure}
\begin{figure}[ptb]
\begin{center}
\includegraphics[width=0.55\textwidth]{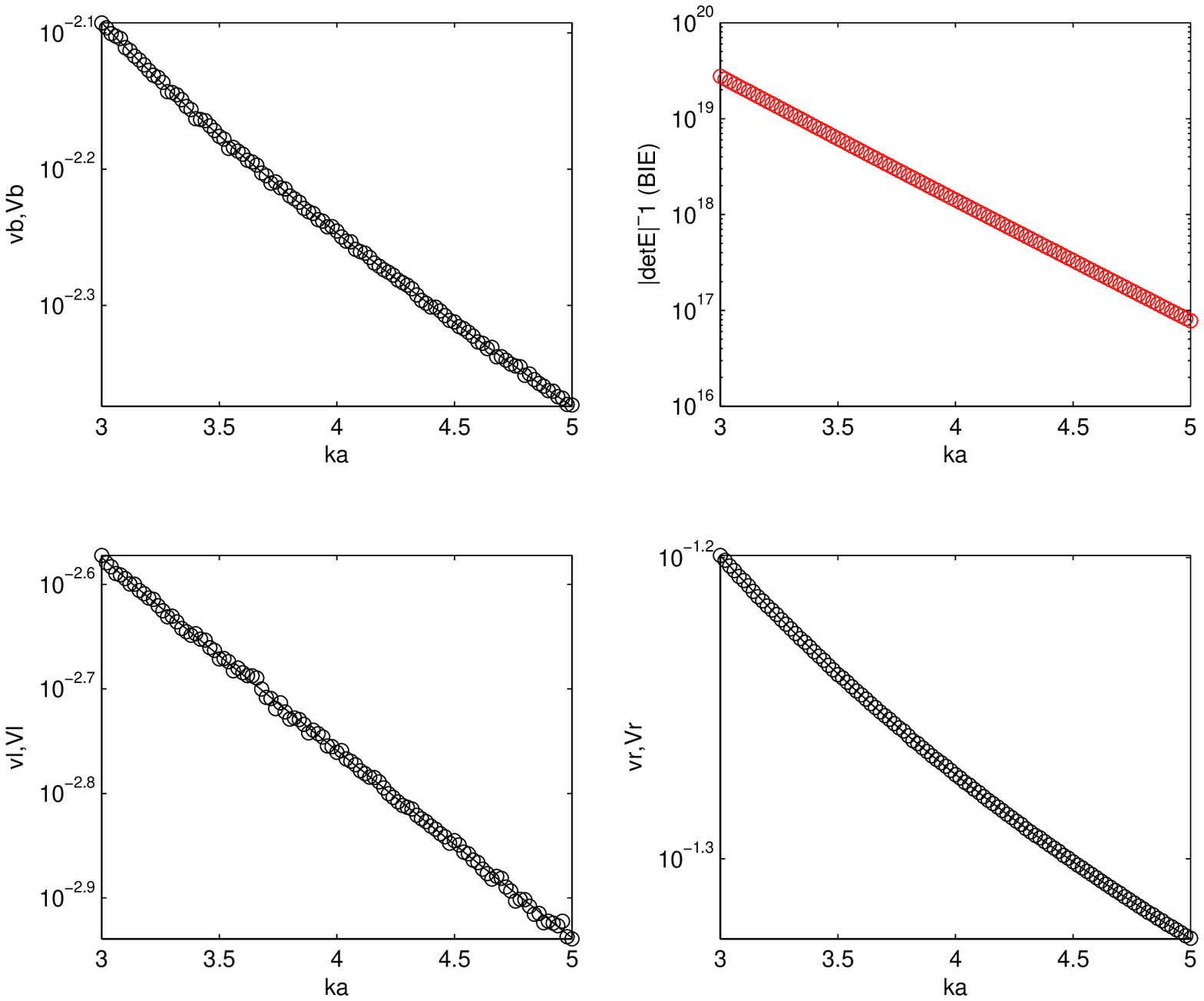}
\caption{Transfer functions of traction at three points on the rigid boundary. The upper left-hand, lower left-hand, lower right-hand panels are for  the transfer functions at $\theta=180^{\circ}$,  $\theta=270^{\circ}$,  $\theta=360^{\circ}$, respectively. The upper right-hand panel depicts $1/\|det(\mathbf{E}(ka))\|$. Lower-case letters and circles correspond to  DCBIE2 computations, upper-case letters and continuous curves to  DSOV (exact) computations.  Case $N=38$, $\epsilon=10^{-4}$.}
\label{fdcbie2-3}
\end{center}
\end{figure}
\begin{figure}[ptb]
\begin{center}
\includegraphics[width=0.55\textwidth]{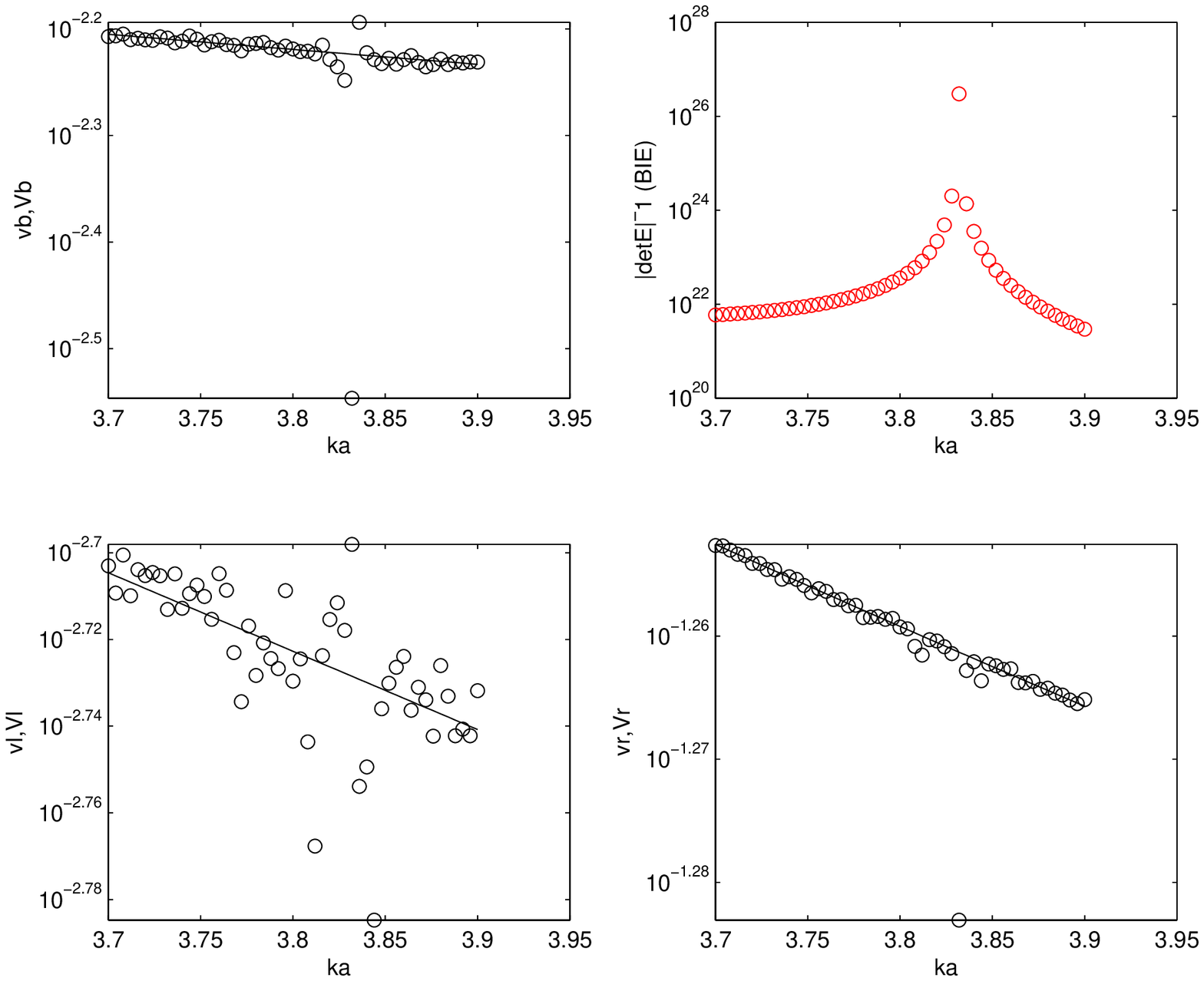}
\caption{Same as fig. \ref{fdcbie2-1} (of which the present figure is a zoom) except that  $N=18$ and  $\epsilon=10^{-4}$.}
\label{fdcbie2-4}
\end{center}
\end{figure}
\begin{figure}[ptb]
\begin{center}
\includegraphics[width=0.55\textwidth]{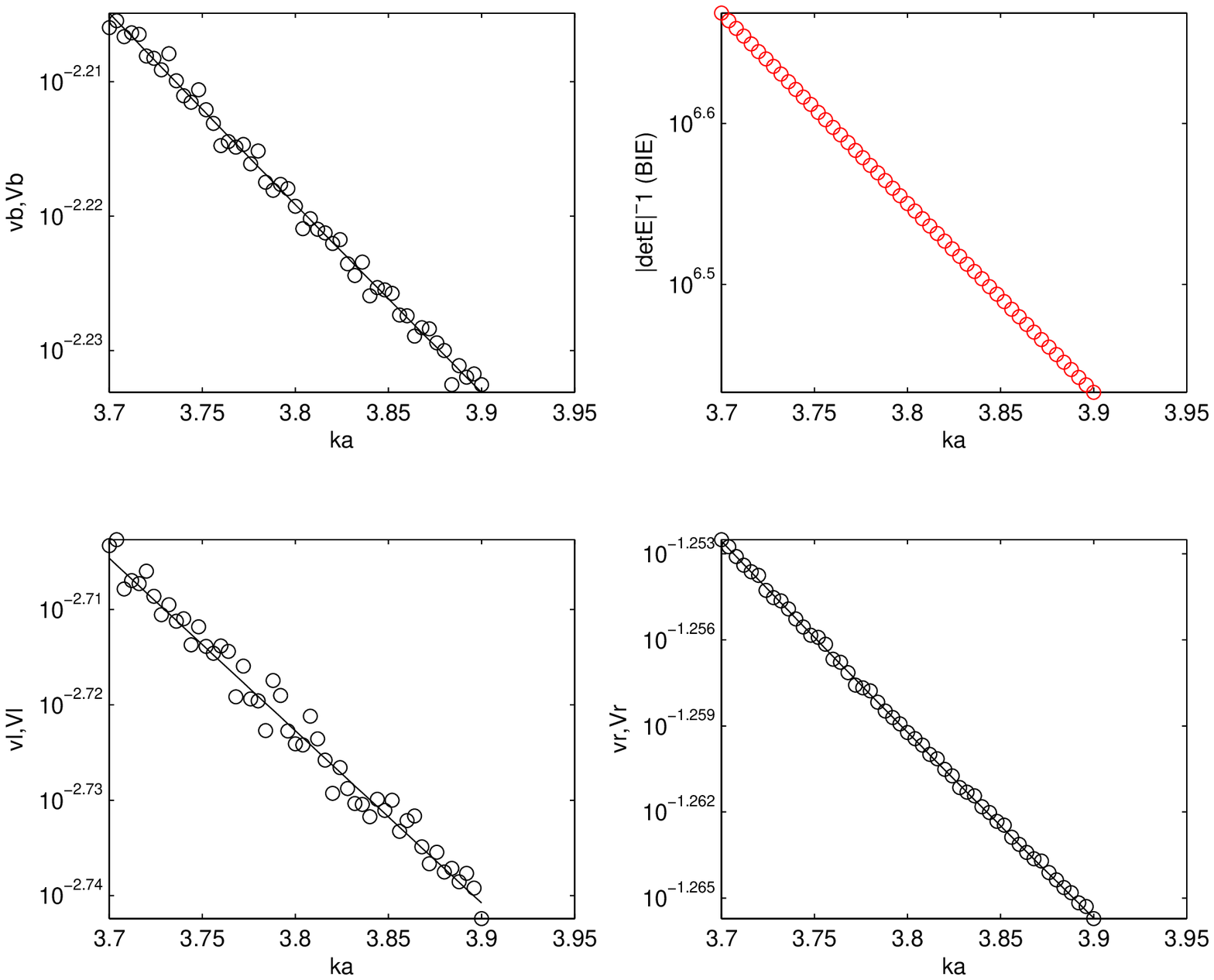}
\caption{A different zoom; otherwise the same as fig. \ref{fdcbie2-3}. Case   $N=18$ and  $\epsilon=10^{-4}$.}
\label{fdcbie2-5}
\end{center}
\end{figure}
\begin{figure}[ptb]
\begin{center}
\includegraphics[width=0.55\textwidth]{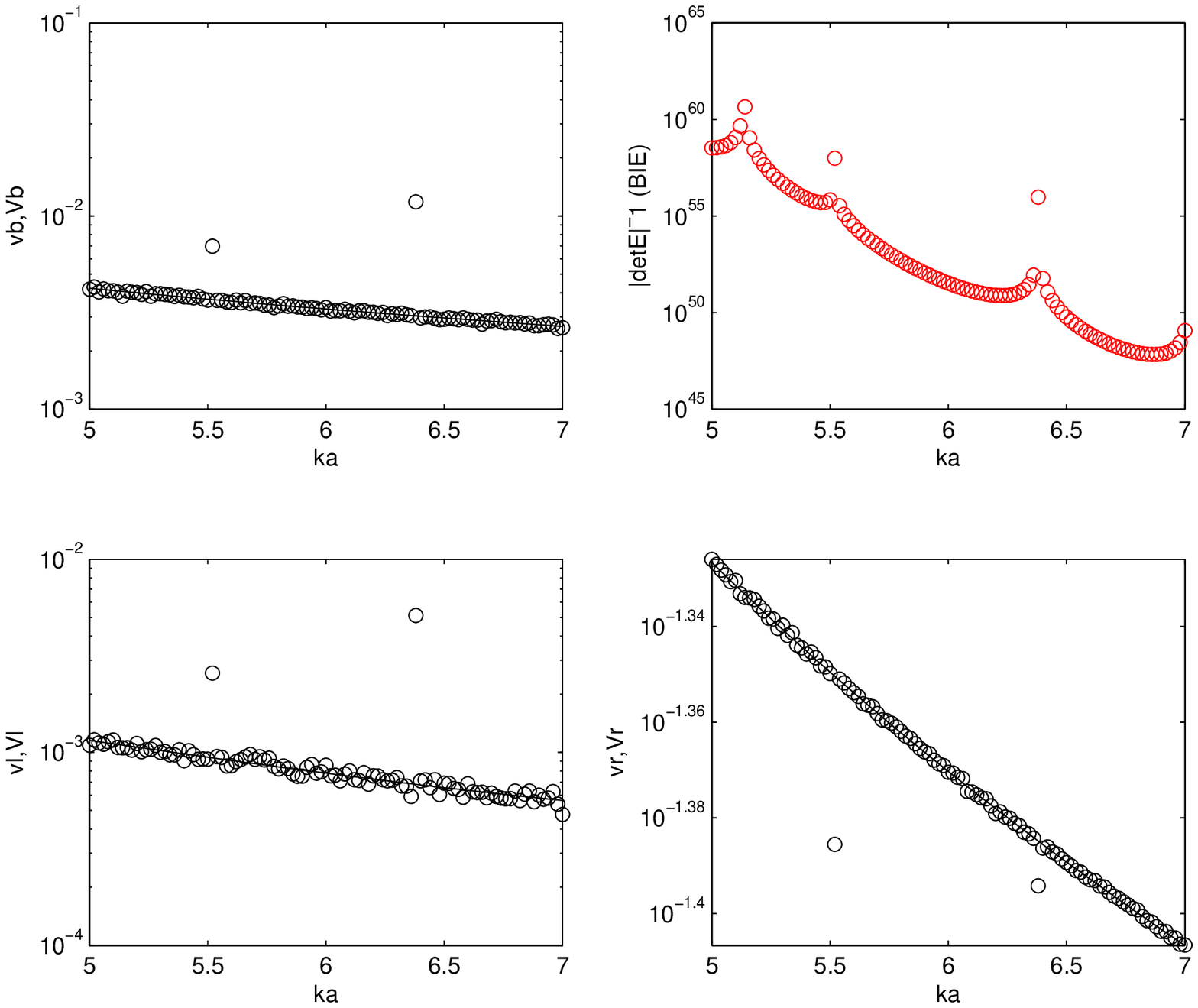}
\caption{Same as fig. \ref{fdcbie2-1} (of which the present figure is a zoom) except that  $N=38$ and  $\epsilon=10^{-4}$.}
\label{fdcbie2-6}
\end{center}
\end{figure}
\begin{figure}[ptb]
\begin{center}
\includegraphics[width=0.55\textwidth]{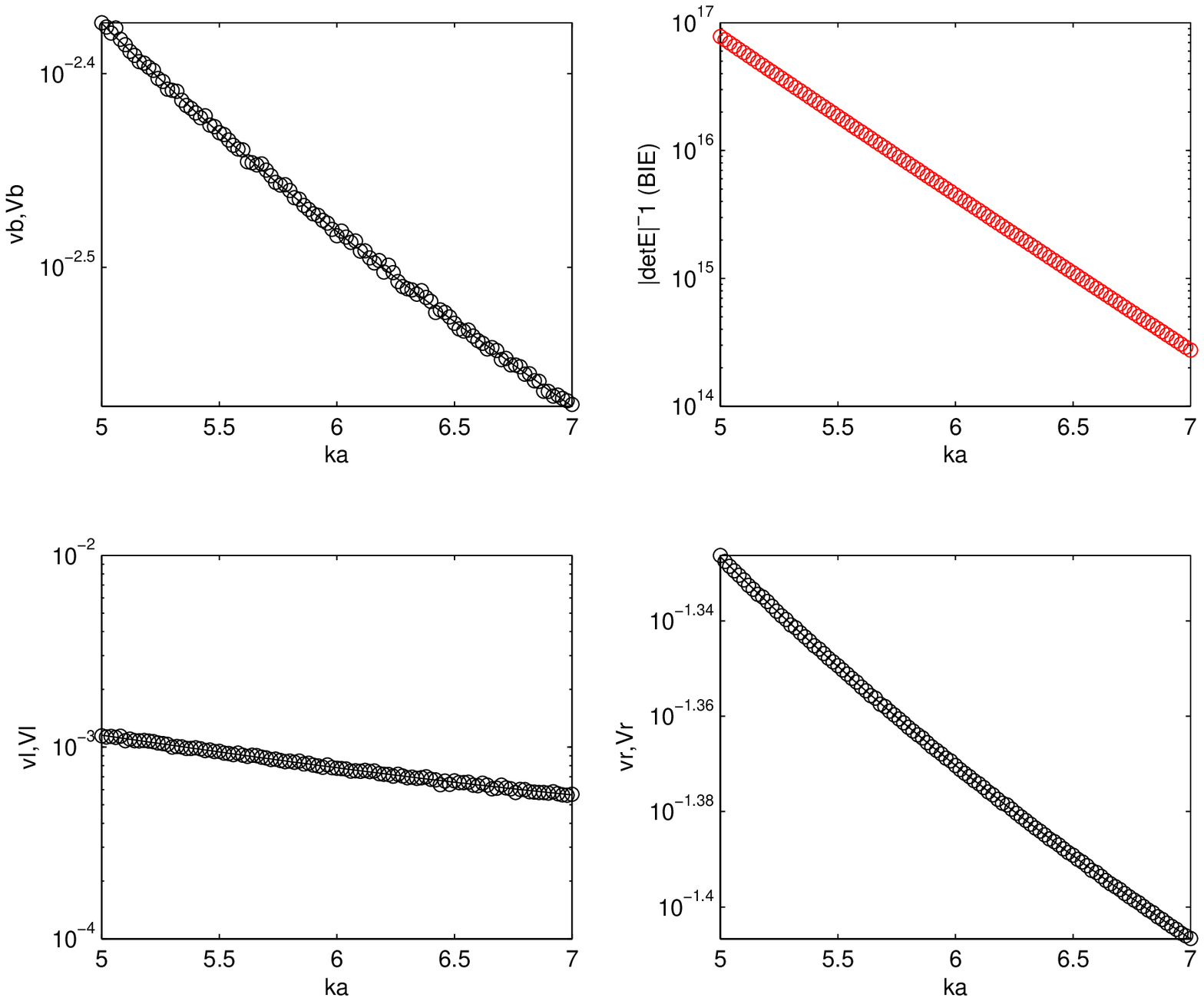}
\caption{A different zoom; otherwise the same as fig. \ref{fdcbie2-3}. Case   $N=38$ and  $\epsilon=10^{-4}$.}
\label{fdcbie2-7}
\end{center}
\end{figure}
\begin{figure}[ptb]
\begin{center}
\includegraphics[width=0.55\textwidth]{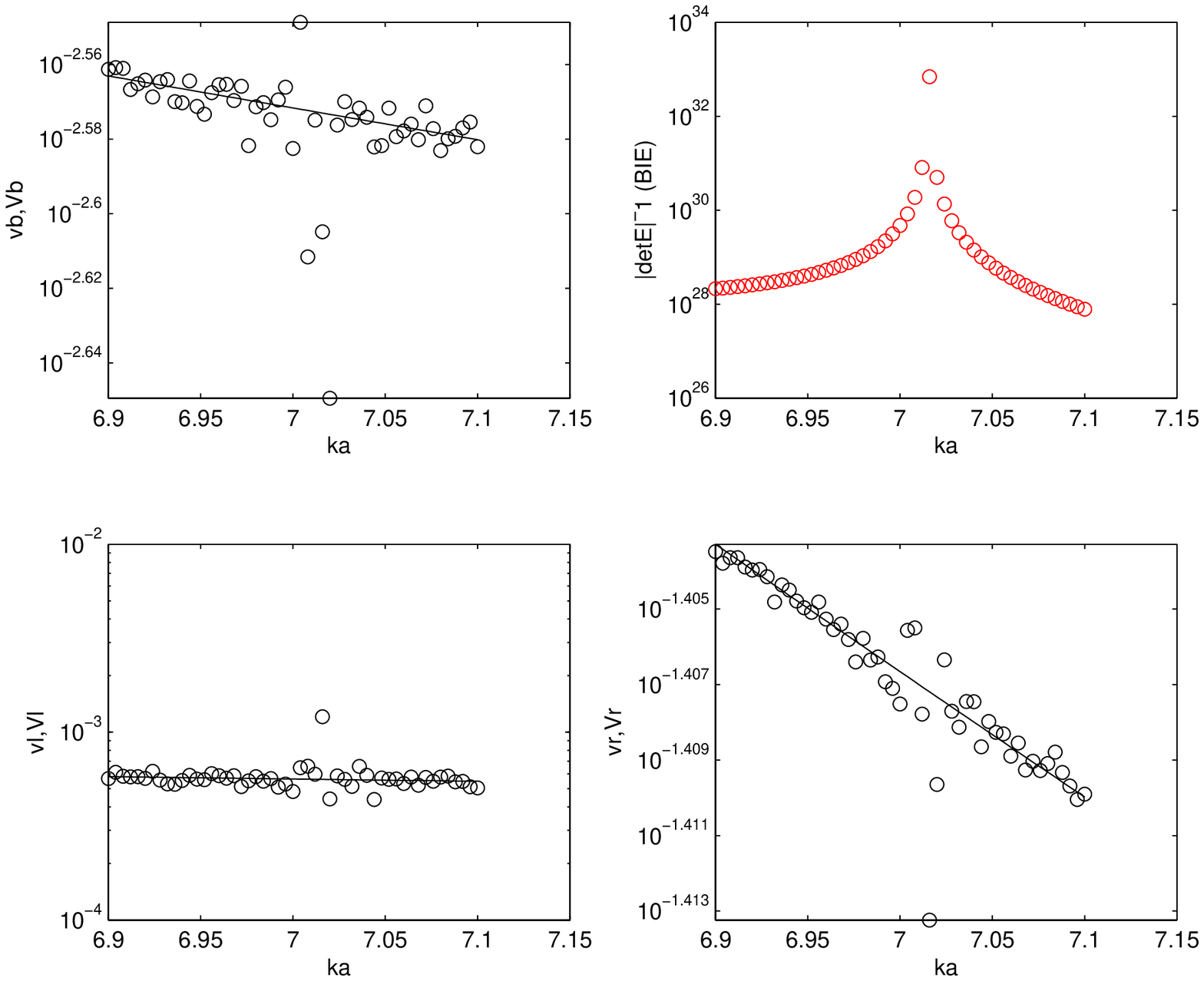}
\caption{Same as fig. \ref{fdcbie2-1} (of which the present figure is a zoom) except that  $N=28$ and  $\epsilon=10^{-4}$.}
\label{fdcbie2-8}
\end{center}
\end{figure}
\begin{figure}[ptb]
\begin{center}
\includegraphics[width=0.55\textwidth]{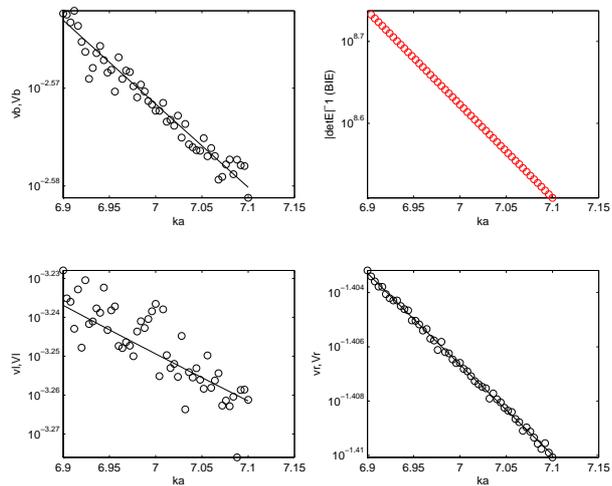}
\caption{A different zoom; otherwise the same as fig. \ref{fdcbie2-3}. Case   $N=28$ and  $\epsilon=10^{-4}$.}
\label{fdcbie2-9}
\end{center}
\end{figure}
\clearpage
\newpage
If account is taken of the scale changes in this set of figures, the latter shows convincingly that the method of cure DCBIE2 has enabled to eliminate all the resonances appearing in DBIE1. The results of other computations not appearing here here show that DCBIE2 also enables to eliminate all the resonances appearing in DBIE2 and DEBC (which, it will be recalled, occur at frequencies that  are generally-different from those at which occur the resonances appearing in DBIE1).
\section{The scattering problem in the frequency domain for the Neumann boundary condition}
%
\subsection{Governing equations}
These equations are \cite{wi19}:
\begin{equation}\label{2-010}
u(\mathbf{x})=u^{i}(\mathbf{x})+u^{s}(\mathbf{x})
~,
\end{equation}
\begin{equation}\label{2-020}
\big(\nabla\cdot\nabla+k^{2}\big)u(\mathbf{x})=-s(\mathbf{x})~;~\forall\mathbf{x}\in\Omega_{0}~,
\end{equation}
\begin{equation}\label{2-030}
u^{s}(\mathbf{x})\sim {\text{outgoing wave}};~\|\mathbf{x}\|\rightarrow\infty~,
\end{equation}
\begin{equation}\label{2-040}
\boldsymbol{\nu}\cdot\nabla u(\mathbf{x})=0~;~\mathbf{x}\in\Gamma~,
\end{equation}
wherein:\\
a) $\mathbf{x}$ is a vector in the $x-y$ (cross-section) plane  directed from the origin $0$ to an arbitrary point $(x,y)$ in cartesian coordinates or $r,\theta$ in polar coordinates,\\
b)  as concerns the displacement frequency domain fields:  $u^{i}(\mathbf{x})$ is shorthand for $u_{z}^{i}(\mathbf{x};\omega)$,  $u^{s}(\mathbf{x})$ is shorthand for $u_{z}^{s}(\mathbf{x};\omega)$,  $u(\mathbf{x})$ is shorthand for $u_{z}(\mathbf{x};\omega)$, with $\omega=2\pi f$ the angular frequency and $f$ the frequency,\\
c) $u^{i}(\mathbf{x})$ is the wave (called 'incident wave'), radiated by the source of density $s(\mathbf{x})$, that exists in the configuration in which the body is absent,\\
d) $u^{s}(\mathbf{x})$ is the scattered field,\\
e) $u(\mathbf{x})$ is the total field in the region $\Omega_{0}$ exterior to the body, the interior of the latter being denoted by $\Omega_{1}$,\\
f) $\Gamma$ is the closed curve delineating the boundary between $\Omega_{0}$ and $\Omega_{1}$, and, at present, I take this curve to be a circle of radius $a$ (note that $\Omega_{0}$, $\Omega_{1}$, and $\Gamma$ are geometric entities in the $x-y$ plane),\\
g) the frequency domain field is related to the time $(t)$ domain field by the relation $u(\mathbf{x};t)=2\Re\int_{0}^{\infty}u(\mathbf{x};\omega)\exp(-i\omega t)d\omega$,\\
h) $k=\omega/\beta$ is the (positive real) wavenumber,\\
i) $\boldsymbol{\nu}$ is the unit vector normal to $\Gamma$ and directed towards the inside of $\Omega_{1}$\\\\
Note that I am again dealing with a forward-scattering problem, i.e., $s(\mathbf{x})$ and therefore $u^{i}$, $\beta$, $a$, $\omega$ are assumed to be known and the problem is to determine $u^{s}$ and/or $u$.
\subsection{The separation of variables (SOV) solution (i.e., NSOV) for the Neumann-boundary body}
The  SOV technique again consists (for 2D problems such as mine) in assuming that the solution (actually just a representation thereof)  can be expressed as the product of two functions, each of which depends on only one of the two chosen coordinates, whereupon the partial differential (wave) equation (\ref{2-020}) separates into two independent ordinary differential equations the solution of which can be expressed in terms of elementary functions.

I choose the $r,\theta$ coordinates so that the $\theta$ differential equation turns out to have solutions $\exp(in\theta)~;~n\in \mathbb{Z}$ whereas the $r$ differential equation has solutions $J_{n}(kr)~;~n\in\mathbb{Z}$ on the one hand, and $H_{n}^{(1)}(kr)~;~n\in\mathbb{Z}$ on the other hand. The solutions in terms of the Bessel functions can be ruled out in the region $\Omega_{0}$ exterior to the scattering object because of the radiation condition (\ref{2-030}) so that the SOV representation of the scattered field in $\Omega_{0}$ becomes
\begin{equation}\label{2-170}
u^{s}(r,\theta)=\sum_{n\in\mathbb{Z}}C_{n}H_{n}^{(1)}(kr)\exp(in\theta)~~;~~\forall~\theta\in[0,2\pi[~.
\end{equation}
The actual SOV solution to the scattering problem requires the invocation of the boundary condition (\ref{2-040}) and (\ref{2-010})
\begin{equation}\label{2-180}
-\frac{\partial}{\partial r}u^{i}(a,\theta)-\frac{\partial}{\partial r}u^{s}(a,\theta)=-k\sum_{n\in\mathbb{Z}}\left[A_{n}\dot{J}_{n}(ka)+C_{n}\dot{H}_{n}^{(1)}(ka)\right]\exp(in\theta)=0
~~;~~\forall~\theta\in[0,2\pi[~.
\end{equation}
The solution for $\{C_{n}\}$ is quite obvious (recall that $\{A_{n}\}$ is known via (\ref{1-160}) and from the fact that $r^{s},\theta^{s}$ are known), but I wish to bring to the fore a feature that will be useful further on.  Thus, I choose to project (\ref{2-180}) as follows:
\begin{equation}\label{2-190}
\int_{0}^{2\pi}\sum_{n\in\mathbb{Z}}\left[A_{n}\dot{J}_{n}(ka)+C_{n}\dot{H}_{n}^{(1)}(ka)\right]\exp(in\theta)\exp(-im\theta)d\theta=0
~;\forall m\in\mathbb{Z}
~,
\end{equation}
which, after interchanging the integral and the sum, and making use of the identity (\ref{1-100}), yields
\begin{equation}\label{2-210}
\sum_{n\in\mathbb{Z}}\left[-\dot{H}_{n}^{(1)}(ka)\delta_{mn}\right]C_{n}=A_{n}\dot{J}_{n}(ka)~;\forall m\in\mathbb{Z}
~,
\end{equation}
which is an infinite-order matrix equation in which the matrix $[~]$ is diagonal and non-singular for all real frequencies $f$ due to the fact that the the derivative of the Hankel function is complex and its real and imaginary parts vanish for different values of $ka$ \cite{as68}.
It follows, by simple matrix inversion, that
\begin{equation}\label{2-220}
C_{n}=-A_{n}\frac{\dot{J}_{n}(ka)}{\dot{H}_{n}^{(1)}(ka)}~;\forall n\in\mathbb{Z}
~.
\end{equation}
Thus, on account of (\ref{2-010}) and (\ref{1-163})
\begin{multline}\label{2-230}
u(\mathbf{x})=\sum_{n\in\mathbb{Z}}\left\{H(r-r^{s})B_{n}H_{n}^{(1)}(kr)+
A_{n}\left[H(r^{s}-r)J_{n}(kr)-\frac{\dot{J}_{n}(ka)}{\dot{H}_{n}^{(1)}(ka)}H_{n}^{(1)}(kr)\right]\right\}\times\\
\exp(in\theta)~;~\forall\mathbf{x}\in\Omega_{0}
~,
\end{multline}
wherein the $A_{n}$ and $B_{n}$ are given in (\ref{1-160}) and (\ref{1-165}) respectively.
It follows that:
\begin{equation}\label{2-240}
u(a,\theta)=\sum_{n\in\mathbb{Z}}
A_{n}\left[J_{n}(kr)-\frac{\dot{J}_{n}(ka)}{\dot{H}_{n}^{(1)}(ka)}H_{n}^{(1)}(kr)\right]\exp(in\theta)=
\sum_{n\in\mathbb{Z}}\frac{2i}{ka\pi}\frac{1}{\dot{H}_{n}^{(1)}(ka)}\exp(in\theta)
~.
\end{equation}

Eq. (\ref{2-230}) can be considered as the exact solution to the scattering problem. This solution for $u$  shows no sign of resonances.
\subsection{Some consequences of Green's second identity}
I again start from Green's second identity (\ref{1-300}) which is applicable to any type of boundary conditions,  three of the consequences of which are:
\begin{equation}\label{2-310}
u(\mathbf{x})=u^{i}(\mathbf{x})+
\int_{\Gamma}\left[G(\mathbf{x};\mathbf{x}')\boldsymbol{\nu}(\mathbf{x}')\cdot\nabla (\mathbf{x}')u(\mathbf{x}')-
u(\mathbf{x}')\boldsymbol{\nu}(\mathbf{x}')\cdot\nabla (\mathbf{x}')G(\mathbf{x};\mathbf{x}')\right]d\gamma(\mathbf{x}')
~;~\forall\mathbf{x}\in\Omega_{0}~,
\end{equation}
\begin{equation}\label{2-320}
\frac{1}{2}u(\mathbf{x})=u^{i}(\mathbf{x})+
\int_{\Gamma}G(\mathbf{x};\mathbf{x}')\boldsymbol{\nu}(\mathbf{x}')\cdot\nabla (\mathbf{x}')u(\mathbf{x}')d\gamma(\mathbf{x}')-
pv\int_{\Gamma}u(\mathbf{x}')\boldsymbol{\nu}(\mathbf{x}')\cdot\nabla (\mathbf{x}')G(\mathbf{x};\mathbf{x}')d\gamma(\mathbf{x}')
~;~\forall\mathbf{x}\in\Gamma~,
\end{equation}
\begin{equation}\label{2-330}
0=u^{i}(\mathbf{x})+
\int_{\Gamma}\left[G(\mathbf{x};\mathbf{x}')\boldsymbol{\nu}(\mathbf{x}')\cdot\nabla (\mathbf{x}')u(\mathbf{x}')-
u(\mathbf{x}')\boldsymbol{\nu}(\mathbf{x}')\cdot\nabla (\mathbf{x}')G(\mathbf{x};\mathbf{x}')\right]d\gamma(\mathbf{x}')
~;~\forall\mathbf{x}\in\Omega_{1}~.
\end{equation}
The object of what follows is obviously to apply any one of these boundary integral (BI) expressions, or combinations thereof, to solve the Neumann boundary-value problem.
\subsection{The three BI expressions for the case of a Neumann boundary condition}
These are:
\begin{equation}\label{2-340}
u(\mathbf{x})=u^{i}(\mathbf{x})-
\int_{\Gamma}u(\mathbf{x}')\boldsymbol{\nu}(\mathbf{x}')\cdot\nabla (\mathbf{x}')G(\mathbf{x};\mathbf{x}')d\gamma(\mathbf{x}')
~;~\forall\mathbf{x}\in\Omega_{0}~,
\end{equation}
\begin{equation}\label{2-350}
\frac{1}{2}u(\mathbf{x})=u^{i}(\mathbf{x})-
pv\int_{\Gamma}u(\mathbf{x}')\boldsymbol{\nu}(\mathbf{x}')\cdot\nabla (\mathbf{x}')G(\mathbf{x};\mathbf{x}')d\gamma(\mathbf{x}')
~;~\forall\mathbf{x}\in\Gamma~,
\end{equation}
\begin{equation}\label{2-360}
0=u^{i}(\mathbf{x})-
\int_{\Gamma}u(\mathbf{x}')\boldsymbol{\nu}(\mathbf{x}')\cdot\nabla (\mathbf{x}')G(\mathbf{x};\mathbf{x}')d\gamma(\mathbf{x}')
~;~\forall\mathbf{x}\in\Omega_{1}~.
\end{equation}
The first of these three only enables to determine the wavefield in the outer region {\it after} determining  $u$ on $\Gamma$ either by the second or third BI equation (BIE for short), or by a combination of these two BIE. Note that (\ref{2-350}) is a second-kind BIE and (\ref{2-360}) is what is frequently called an 'extended boundary condition' (EBC).
\subsection{Solution of the second-kind BIE (i.e., NBIE2) for the case of a Neumann condition on the circular boundary}
The BIE is:
\begin{equation}\label{2-370}
\frac{1}{2}u(\mathbf{x})=u^{i}(\mathbf{x})-
pv\int_{\Gamma}u(\mathbf{x}')\boldsymbol{\nu}(\mathbf{x}')\cdot\nabla (\mathbf{x}')G(\mathbf{x};\mathbf{x}')d\gamma(\mathbf{x}')
~;~\forall\mathbf{x}\in\Gamma~.
\end{equation}
The circular nature of $\Gamma$ entails:
\begin{equation}\label{2-380}
\frac{1}{2}u(a,\theta)=u^{i}(a,\theta)+
\int_{0}^{2\pi}u(a,\theta')\frac{\partial}{\partial r'}G(a,\theta;a,\theta')ad\theta'
~;~\forall\theta\in[0,2\pi[~,
\end{equation}
and the task is henceforth to determine $u(a,\theta)$.

The $2\pi$-periodic nature (in terms of $\theta$) of $u^{i}$ and $u$ incites one to expand these functions in terms of Fourier basis functions:
\begin{equation}\label{2-390}
u^{i}(a,\theta)=\sum_{n\in\mathbb{Z}}g_{n}\exp(in\theta)~~,~~u(a,\theta)=\sum_{n\in\mathbb{Z}}f_{n}\exp(in\theta)
~;~\forall\theta\in[0,2\pi[~,
\end{equation}
and to employ a Galerkin procedure, consisting of projecting the integral equation on the same Fourier basis set of functions so as to obtain, after sum and integral exchanges and  use of (\ref{1-200}):
\begin{equation}\label{2-400}
\frac{1}{2}f_{m}=g_{m}+\sum_{n\in\mathbb{Z}}f_{n}\int_{0}^{2\pi}d\theta\exp(-im\theta)~ pv\int_{0}^{2\pi}d\theta'\frac{a}{2\pi}
\frac{\partial}{\partial r'}G(a,\theta;a,\theta')\exp(in\theta')
~;~\forall m\in\mathbb{Z}~.
\end{equation}
I now make use of (\ref{1-060}) to obtain (in the sense of its use in the $pv$ integral)
\begin{equation}\label{2-410}
\frac{\partial}{\partial r'}G(a,\theta;a,\theta')=\frac{ik}{4}\sum_{l=-\infty}^{\infty}\left[\frac{-i}{ka\pi}+\dot{H}_{l}^{(1)}(ka)J_{l}(ka)\right]\exp[il(\theta-\theta')]~,
\end{equation}
to find
\begin{multline}\label{2-420}
\frac{1}{2}f_{m}=g_{m}+\sum_{n\in\mathbb{Z}}f_{n}\sum_{l\in\mathbb{Z}}\frac{ika}{8\pi}\left[\frac{-i}{ka\pi}+
\dot{H}_{l}^{(1)}(ka)J_{l}(ka)\right]\times\\
\int_{0}^{2\pi}d\theta\exp[i(l-m)\theta]\int_{0}^{2\pi}d\theta'\exp[i(n-l)\theta']
~;~\forall m\in\mathbb{Z}~.
\end{multline}
or, on account of (\ref{1-200})
\begin{equation}\label{2-430}
\frac{1}{2}f_{m}=g_{m}+\sum_{n\in\mathbb{Z}}f_{n}\left[\frac{1}{2}\delta_{mn}+
\frac{ika\pi}{2}\dot{H}_{l}^{(1)}(ka)J_{l}(ka)\delta_{mn}\right]
~;~\forall m\in\mathbb{Z}~,
\end{equation}
which can be re-written as the  matrix equation
\begin{equation}\label{2-450}
\sum_{n\in\mathbb{Z}}E_{mn}f_{n}=g_{m}~;~\forall m\in\mathbb{Z}~,
\end{equation}
wherein
\begin{equation}\label{2-460}
E_{mn}=\frac{-ika\pi}{2}\dot{H}_{n}^{(1)}(ka)J_{n}(ka)\delta_{mn}~;~\forall m,n\in\mathbb{Z}~.
\end{equation}
Once again, I have to deal with an  infinite-order diagonal matrix, thus enabling, in theory, the obtention of a closed-form solution for ${f_{n}}$. But I forsee a major problem due to the fact that now this matrix vanishes for certain real frequencies, this being due to fact that  the derivative of the Bessel functions are equal to zero at an infinite discrete set of their real arguments \cite{as68}. Be this as it may, at real frequencies not in the neighborhood of the indicated frequencies, it is legitimate to invert $\mathbf{E}=\{E_{mn}\}$ whence
\begin{equation}\label{2-470}
\mathbf{f}=\mathbf{E}^{-1}\mathbf{g}~~\Rightarrow~~
f_{m}=\left[\frac{-ika\pi}{2}\dot{H}_{m}^{(1)}(ka)J_{m}(ka)\right]^{-1}g_{m}~;~\forall m\in\mathbb{Z}~.
\end{equation}
If I recall that for my line source
\begin{equation}\label{2-480}
u^{i}(a,\theta)=\sum_{m\in\mathbf{Z}}A_{m}J_{m}(ka)\exp(im\theta)=\sum_{m\in\mathbf{Z}}g_{m}\exp(im\theta)~,
\end{equation}
then
\begin{equation}\label{2-485}
g_{m}=A_{m}J_{m}(ka)~,
\end{equation}
whence
\begin{equation}\label{2-490}
\mathbf{f}=\mathbf{E}^{-1}\mathbf{g}~~\Rightarrow~~
f_{m}=A_{m}\left[\frac{-ika\pi}{2}\dot{H}_{m}^{(1)}(ka)\right]^{-1}~;~\forall m\in\mathbb{Z}~,
\end{equation}
which, by virtue of (\ref{2-390}), agrees with the SOV exact solution (\ref{2-240}) for $u(a,\theta)$. However, it is important to recall that this solution for $f_{m}$ is only applicable for  real frequencies that are not in the neighborhood for which $\dot{J}_{n}(ka)=0; \forall n\in \mathbb{Z}$.
\subsection{The field outside the object obtained by using the 'solution' of the second-kind BIE (i.e., NBIE2) for the case of a Neumann condition on the circular boundary}
The field outside the object is obtainable via (\ref{2-340})
\begin{equation}\label{2-500}
u(\mathbf{x})=u^{i}(\mathbf{x})-
\int_{\Gamma}u(\mathbf{x}')\boldsymbol{\nu}(\mathbf{x}')\cdot\nabla(\mathbf{x}') G(\mathbf{x};\mathbf{x}')d\gamma(\mathbf{x}')
~;~\forall\mathbf{x}\in\Omega_{0}~,
\end{equation}
Note that this is not a BIE but rather a boundary-integral representation (BIR) of the field (in the region $\Omega_{0}$). The solution for the latter field is obtained by merely introducing the previously-found $u$ into the integrand. In polar coordinates, the BIR is
\begin{equation}\label{2-510}
u(r,\theta)=u^{i}(r,\theta)+\int_{0}^{2\pi}u(a,\theta)\frac{\partial}{\partial r'}G(r,\theta;a,\theta')ad\theta
~;~ r>a~, \forall\theta\in[0,2\pi[~,
\end{equation}
I make use of
\begin{equation}\label{2-520}
\frac{\partial}{\partial r'}G(r>a,\theta;a,\theta)=\frac{ik}{4}\sum_{l\in\mathbb{Z}}H_{l}^{(1)}(kr)\dot{J}_{l}(ka)\exp[il(\theta-\theta')]
~,
\end{equation}
and previous expansions to obtain
\begin{multline}\label{2-530}
u(r,\theta)=\sum_{n\in\mathbb{Z}}\left[H(r-r^{s})B_{n}H_{n}^{(1)}(kr)+H(r^{s}-r)A_{n}J_{n}(kr)\right]\exp(in\theta)+\\
\sum_{n\in\mathbb{Z}}f_{n}\sum_{l\in\mathbb{Z}}\frac{ika}{4\pi}\sum_{l\in\mathbb{Z}}H_{l}^{(1)}(kr)\dot{J}_{l}(ka)\exp[il\theta)\int_{0}^{2\pi}\exp[i(n-l)\theta']d\theta'
~;~ r>a~, \forall\theta\in[0,2\pi[~,
\end{multline}
or
\begin{multline}\label{2-540}
u(r,\theta)=\sum_{n\in\mathbb{Z}}\left\{H(r-r^{s})B_{n}H_{n}^{(1)}(kr)+
A_{n}\left[H(r^{s}-r)J_{n}(kr)+f_{n}\frac{ika\pi}{2}H_{l}^{(1)}(kr)\dot{J}_{l}(ka)\right]\right\}\times\\
\exp[in\theta)
~;~ r>a~, \forall\theta\in[0,2\pi[~,
\end{multline}
which, after the introduction of (\ref{2-470}), becomes
\begin{multline}\label{2-550}
u(r,\theta)=\sum_{n\in\mathbb{Z}}\left\{H(r-r^{s})B_{n}H_{n}^{(1)}(kr)+A_{n}\left[H(r^{s}-r)J_{n}(kr)-
\frac{\dot{J}_{l}(ka)}{\dot{H}_{l}^{(1)}(ka)}H_{l}^{(1)}(kr)\right]\right\}\times\\
\exp[in\theta)
~;~ r>a~, \forall\theta\in[0,2\pi[~,
\end{multline}
which agrees with the exact SOV solution (\ref{2-230}). As before, I call attention to the fact that this solution relies on a 'solution' for $v$ that can only be obtained at real frequencies  that are not in the neighborhood for which $\dot{J}_{l}(ka)=0$.
\subsection{Determination of $u$ on $\Gamma$ via the extended boundary condition integral equation (i.e., NEBC) for the circular object with Neumann boundary condition}
I recall the EBC integral equation expressed in (\ref{2-360})
\begin{equation}\label{2-660}
0=u^{i}(\mathbf{x})-
\int_{\Gamma}u(\mathbf{x}')\boldsymbol{\nu}(\mathbf{x}')\cdot\nabla(\mathbf{x}')G(\mathbf{x};\mathbf{x}')d\gamma(\mathbf{x}')
~;~\forall\mathbf{x}\in\Omega_{1}~.
\end{equation}
I choose to sample this equation on $\Gamma_{in}\subset\Omega_{1}$, where $\Gamma_{in}$ is a circle, with center at the origin $O$, of radius $b<a$. Consequently, the polar coordinate expression of (\ref{2-660}) is
\begin{equation}\label{2-670}
0=u^{i}(b,\theta)+\int_{0}^{2\pi}u(a,\theta')\frac{\partial}{\partial r'}G(b,\theta;a,\theta')ad\theta'~;~\forall\theta\in[0,2\pi[
~.
\end{equation}
I employ the following expressions of the Green's function and $u^{i}(b,\theta)$ (on account of the fact that $b<a<r^{s})$
\begin{equation}\label{2-680}
\frac{\partial}{\partial r'}G(b,\theta;a,\theta')=\frac{i}{4}\sum_{l\in\mathbb{Z}}\dot{H}_{l}^{(1)}(ka)J_{l}(kb)\exp[il(\theta-\theta')]~,~
u^{i}(b,\theta)=\sum_{n\in\mathbb{Z}}A_{n}J_{n}(kb)\exp(in\theta)
~,
\end{equation}
to obtain, by the usual Galerkin procedure
\begin{equation}\label{2-690}
0=h_{m}+\sum_{n\in\mathbb{Z}}f_{n}\left[\frac{ika\pi}{2}\dot{H}_{n}^{(1)}(ka)J_{n}(kb)\right]\delta_{mn}
~;~\forall m\in\mathbb{Z}
~,
\end{equation}
wherein $h_{m}=A_{m}J_{m}(kb)$. As previously, I am confronted with a matrix equation, the matrix of which is of infinite order, diagonal, and singular at a denumerable, infinite set of frequencies for which $J{n}(kb)=0~;~\forall n\in\mathbb{Z}$ so that this matrix cannot be inverted at these frequencies. At real frequencies not in the neighborhood of these singular frequencies, the solution is, as before
\begin{equation}\label{2-695}
f_{n}=A_{n}\left[\frac{-ika\pi}{2}\dot{H}_{n}^{(1)}(ka)\right]^{-1}~;~\forall n\in\mathbb{Z}
~,
\end{equation}
which is nothing other than the exact SOV solution. It ensues, that at these frequencies the field is as previously within $\Omega_{0}$.

I waive the possibility of solving for $u$ on $\Gamma$ via a first-kind integral equation since the problems this BIE method raises are substantially the same as for the second-kind BIE and EBC. Thus I consider next a single method for curing these problems.
\subsection{A CBIE scheme (i.e., NCBIE) appealing to NBIE2 and  NEBC}
The point of departure is the two BIE's:
\begin{equation}\label{2-760}
\frac{1}{2}u(\mathbf{x})=u^{i}(\mathbf{x})-pv\int_{\Gamma} u(\mathbf{x}')\boldsymbol{\nu}(\mathbf{x}')\cdot\nabla(\mathbf{x}')d\gamma(\mathbf{x'}~;~\forall \mathbf{x}\in\Gamma
~,
\end{equation}
\begin{equation}\label{2-770}
0=u^{i}(\mathbf{x})-pv\int_{\Gamma} u(\mathbf{x}')\boldsymbol{\nu}(\mathbf{x}')\cdot\nabla(\mathbf{x}')d\gamma(\mathbf{x'}~;~\forall \mathbf{x}\in\Gamma_{in}
~,
\end{equation}
which, (assuming that $\Gamma$ is the circle $r=a$) in polar coordinates, and by choosing $\Gamma_{in}$ to ba a circle centered at the orgin $O$ with radius $b<a$, take the form:
\begin{equation}\label{2-780}
\frac{1}{2}u(a,\theta)=u^{i}(a,\theta)+\int_{0}^{2\pi}u(a,\theta')\frac{\partial}{\partial r'}G(a,\theta;a,\theta')ad\theta'~;~\forall\theta\in[0,2\pi[
~,
\end{equation}
\begin{equation}\label{2-790}
0=u^{i}(b,\theta)+\int_{0}^{2\pi}u(a,\theta')\frac{\partial}{\partial r'}G(b,\theta;a,\theta')ad\theta'~;~\forall\theta\in[0,2\pi[
~.
\end{equation}
Since both equations apply to the same $\theta$ intervals, I again form a linear combination of the two so as to obtain the single BIE
\begin{equation}\label{2-800}
\frac{1}{2}u(a,\theta)=u^{i}(a,\theta)+\eta u^{i}(b,\theta)+\int_{0}^{2\pi}au(a,\theta')\left[\frac{\partial}{\partial r'}G(a,\theta;a,\theta')+\eta\frac{\partial}{\partial  r'}G(b,\theta;a,\theta')\right]d\theta'~;~\forall\theta\in[0,2\pi[
~,
\end{equation}
wherein $\eta$ is an unspecified scalar constant for the moment and I keep in mind that the integral involving  $\frac{\partial}{\partial r}G(a,\theta;a,\theta')$ is a principal value integral.

I make the expansions:
\begin{equation}\label{2-810}
u(a,\theta)=\sum_{n\in\mathbb{Z}}f_{n}\exp(in\theta)~~,~~u^{i}(a,\theta)=\sum_{n\in\mathbb{Z}}g_{n}\exp(in\theta)~~,~~
u^{i}(b,\theta)=\sum_{n\in\mathbb{Z}}h_{n}\exp(in\theta)~~;~\forall\theta\in[0,2\pi[
~,
\end{equation}
and again invoke the Galerkin procedure to obtain
\begin{multline}\label{2-820}
\frac{1}{2}f_{m}=g_{m}+\eta h_{m}+\\
\sum_{n\in\mathbb{Z}}f_{n}\int_{0}^{2\pi}d\theta\int_{0}^{2\pi}d\theta'\frac{a}{2\pi}\left[\frac{\partial}{\partial  r'}G(a,\theta;a,\theta')+\eta \frac{\partial}{\partial  r'}G(a,\theta;a,\theta')\right]\exp[i(n-m)\theta]~;~\forall m\in\mathbb{Z}
~.
\end{multline}
By recalling previous results I find
\begin{equation}\label{2-830}
\frac{\partial}{\partial  r'}G(a,\theta;a,\theta')+\eta \frac{\partial}{\partial  r'}G(b,\theta;a,\theta')=\sum_{l\in\mathbb{Z}}\left\{\frac{1}{4\pi a}+\frac{ik}{4}\dot{H}_{l}^{(1)}(ka)[J_{l}(ka)+\eta J_{l}(kb)] \right\}\exp[i
l(\theta-\theta')]
~,
\end{equation}
so that the following matrix equation ensues
\begin{equation}\label{2-840}
\sum_{l\in\mathbb{Z}}E_{mn}f_{n}=g_{m}+\eta h_{m}=A_{m}[J_{m}(ka)+\eta J_{m}(kb)]
~,
\end{equation}
wherein
\begin{equation}\label{2-850}
E_{mn}=\frac{-ika\pi}{2}\dot{H}_{n}^{(1)}(ka)[J_{n}(ka)+\eta J_{n}(kb)]\delta_{mn}
~.
\end{equation}
Again,  $\mathbf{E}=\{E_{mn}\}$ is an infinite-order, diagonal matrix, but now it is {\it not singular at any real frequency provided $\eta$ is chosen to be an imaginary scalar constant} because the Bessel functions  are real at  real frequencies. Consequently, with this choice of $\eta$, the inverse of $\mathbf{E}$ exists at all real frequencies so that
\begin{equation}\label{2-860}
f_{n}=A_{n}\left[\frac{-ika\pi}{2}\dot{H}_{n}^{(1)}(ka)\right]^{-1}
~,
\end{equation}
which is nothing other than the exact SOV solution. Thus, this  CBIE scheme constitutes a cure for the disease that plagues traditional BIE methods (at least for scattering problems with a Neumann condition on a circular boundary).
\clearpage
\newpage
\subsection{Numerical results for the Neumann boundary circular cylinder via NBIE2 on the one hand, and the cure via NCBIE on the other hand: one source wave incidence}
The following figures, i.e., \ref{fncbie1-1}-\ref{fncbie1-8}, all apply to a traction-free circular cylinder of radius $a=1000 m$,  and outside of which  $\beta=2000 ms^{-1}$, submitted to the wave radiated by a line source situated at  $r^{s}=6000 m$, $\theta^{s}=30^{\circ}$. The responses (as a function of $f$, $f$ the frequency) are computed  by the couples (NBIE2, NCBIE). In NCBIE I choose $b=500 m$  and $\eta=0+1i$ as well as a randomization of the elements of the matrix $\mathbf{E}$ just like that of this matrix in NBIE2.
\begin{figure}[ht]
\begin{center}
\includegraphics[width=0.5\textwidth]{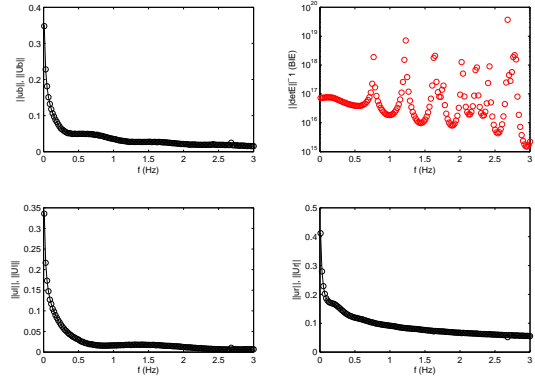}
\caption{Transfer functions of displacement at three points on the traction-free boundary. The upper left-hand, lower left-hand, lower right-hand panels are for  the transfer functions at $\theta=180^{\circ}$,  $\theta=270^{\circ}$,  $\theta=360^{\circ}$, respectively. The upper right-hand panel depicts $1/\|det(\mathbf{E}(ka))\|$. Lower-case letters and circles correspond to  NBIE2 computations, upper-case letters and continuous curves to  NSOV (exact) computations.  Case $N=28$, $\epsilon=10^{-3}$.}
\label{fncbie1-1}
\end{center}
\end{figure}
\begin{figure}[ht]
\begin{center}
\includegraphics[width=0.5\textwidth]{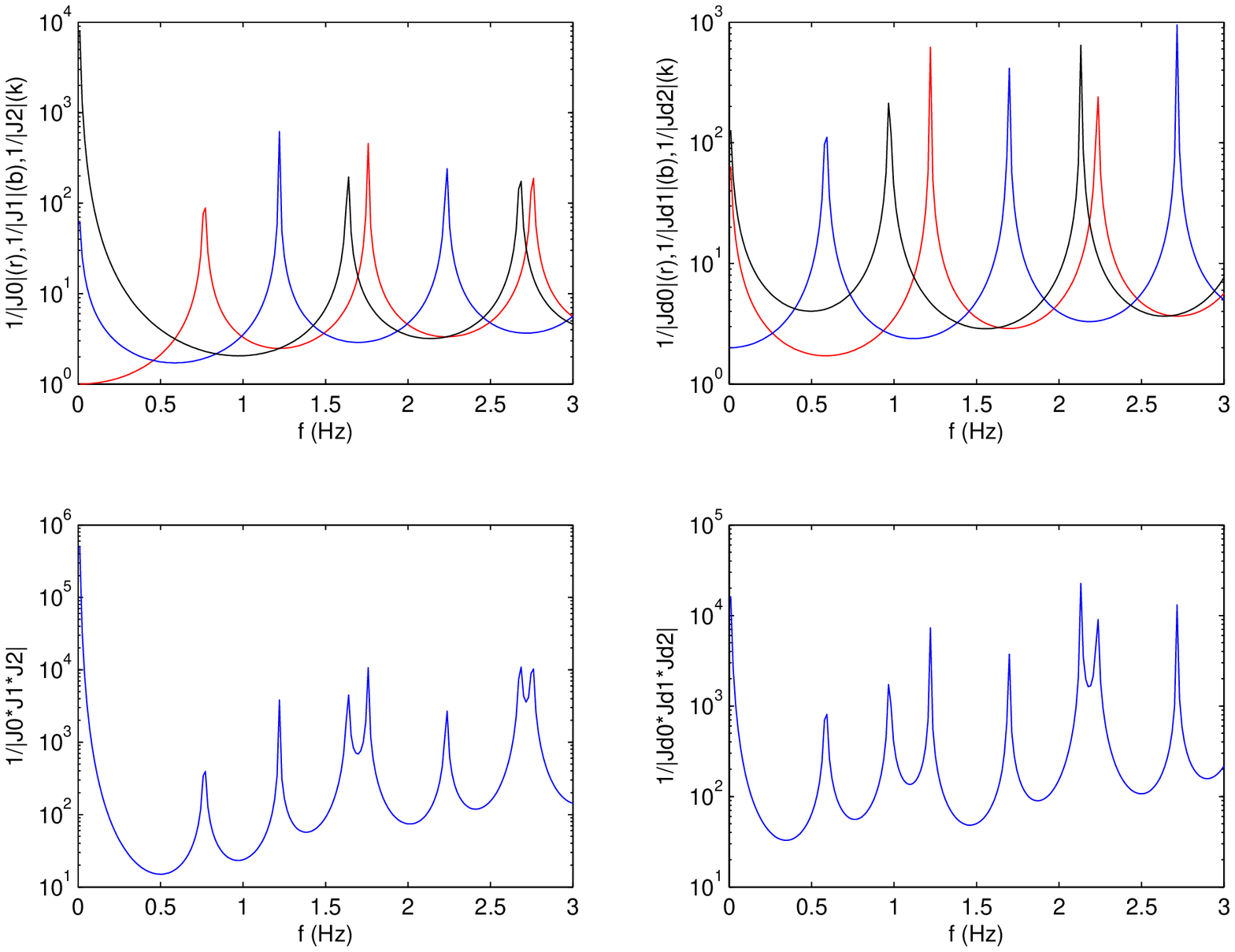}
\caption{This figure enables the connection of the observed resonance frequencies to the zeros of either $\dot{J}_{n}(ka)$ (for NBIE2) or  $\dot{J}_{n}(kb)$ (for NEBC). The upper left-hand panel is relative to $1/|J_{0}(ka)|$ (red), $1/|J_{1}(ka)|$ (blue),  $1/|J_{2}(ka)|$ (black) whereas the lower left-hand panel is relative to $1/|J_{0}(ka)J_{1}(ka)J_{0}(ka)|$. The upper right-hand panel is relative to $1/|\dot{J}_{0}(ka)|$ (red), $1/|\dot{J}_{1}(ka)|$ (blue),  $1/|\dot{J}_{2}(ka)|$ (black) whereas the lower right-hand  panel is relative to $1/|\dot{J}_{0}(ka)\dot{J}_{1}(ka)\dot{J}_{2}(ka)|$. As expected, the positions of the lower-frequency resonant features in fig. \ref{fncbie1-1} coincide with the zeros of $J_{n}(ka)~;~n=0,1,2$ and the first few maxima of $1/\|det(\mathbf{E}(ka))\|$   in fig.  \ref{fncbie1-1} are located at the same positions as those of $1/|J_{0}(ka)J_{1}(ka)J_{2}(ka)|$ herein.}
\label{fncbie1-2}
\end{center}
\end{figure}
\begin{figure}[ptb]
\begin{center}
\includegraphics[width=0.7\textwidth]{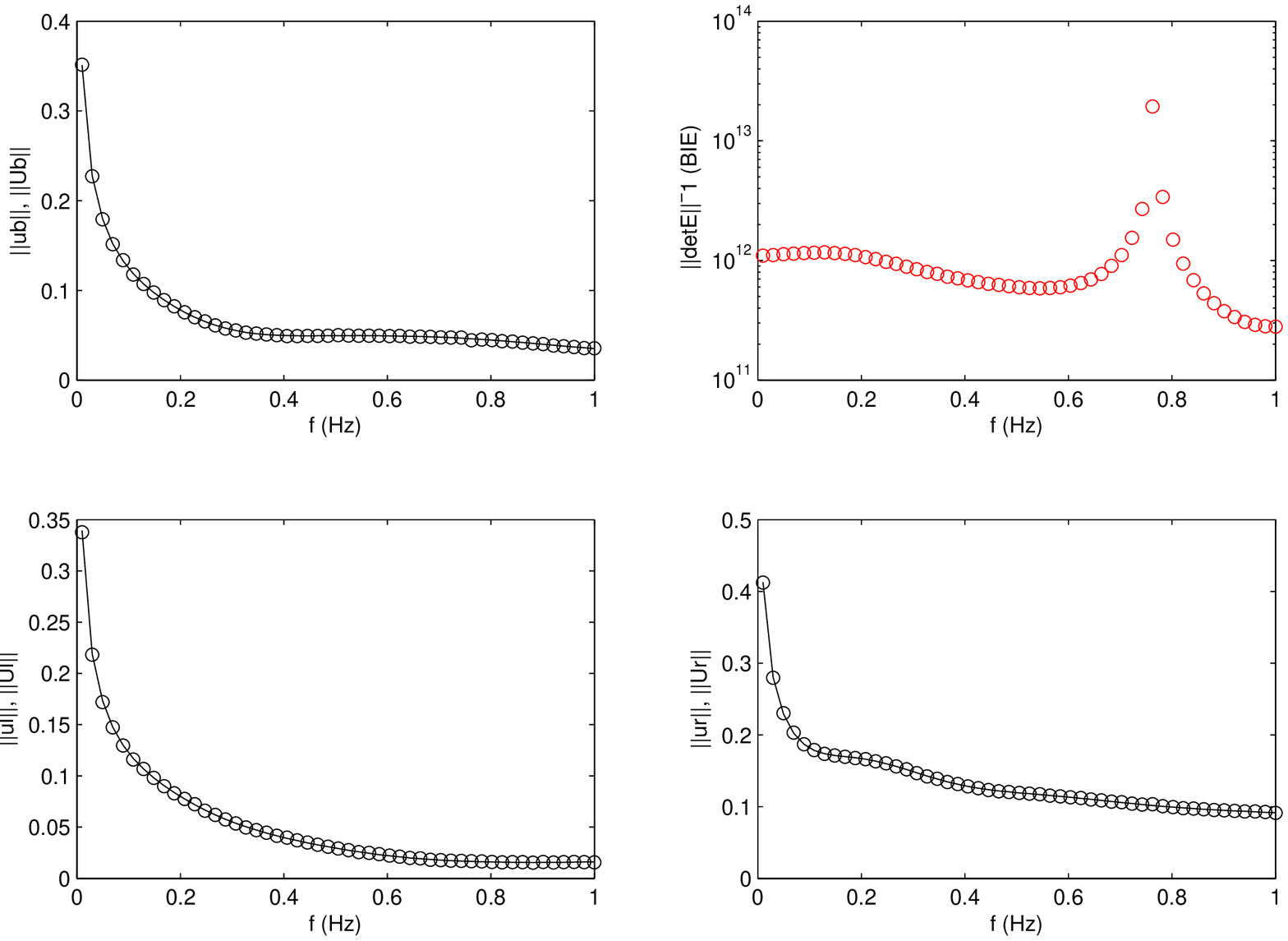}
\caption{Same as fig. \ref{fncbie1-1} of which the present figure is a zoomed version. Case $N=20$, $\epsilon=10^{-3}$.}
\label{fncbie1-3}
\end{center}
\end{figure}
\begin{figure}[ptb]
\begin{center}
\includegraphics[width=0.7\textwidth]{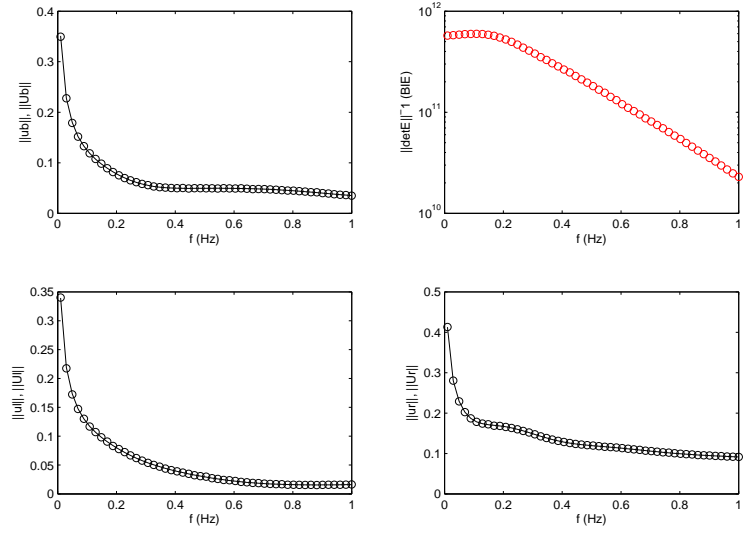}
\caption{Transfer functions of displacement at three points on the traction-free boundary for the same range of frequencies as in the previous figure. The upper left-hand, lower left-hand, lower right-hand panels are for  the transfer functions at $\theta=180^{\circ}$,  $\theta=270^{\circ}$,  $\theta=360^{\circ}$, respectively. The upper right-hand panel depicts $1/\|det(\mathbf{E}(ka))\|$. Lower-case letters and circles correspond to  NCBIE2 computations, upper-case letters and continuous curves to  NSOV (exact) computations.  Case $N=20$, $\epsilon=10^{-3}$.}
\label{fncbie1-4}
\end{center}
\end{figure}
\begin{figure}[ptb]
\begin{center}
\includegraphics[width=0.7\textwidth]{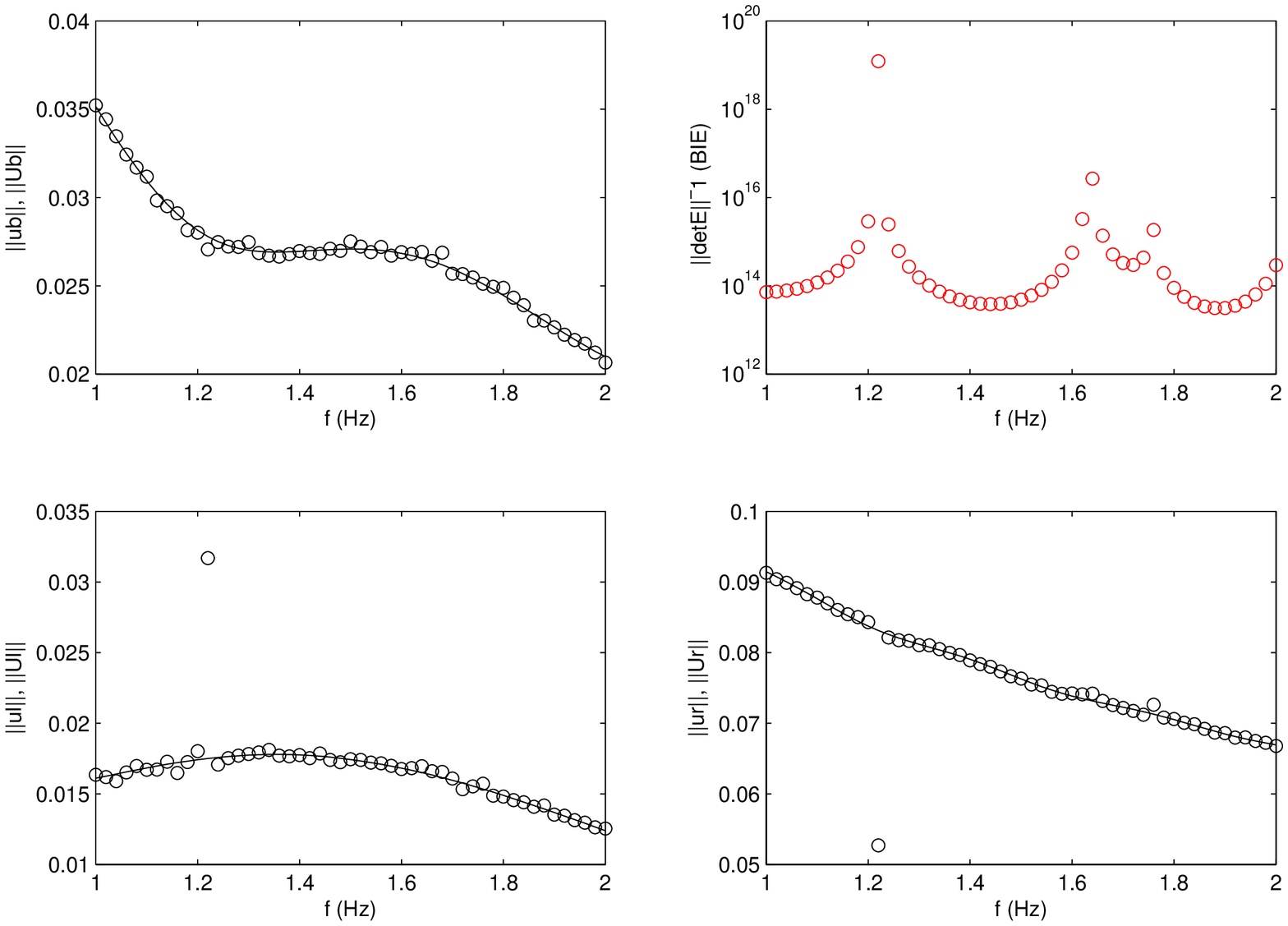}
\caption{Same as fig. \ref{fncbie1-1} of which the present figure is a zoomed version. Case $N=24$, $\epsilon=10^{-3}$.}
\label{fncbie1-5}
\end{center}
\end{figure}
\begin{figure}[ptb]
\begin{center}
\includegraphics[width=0.7\textwidth]{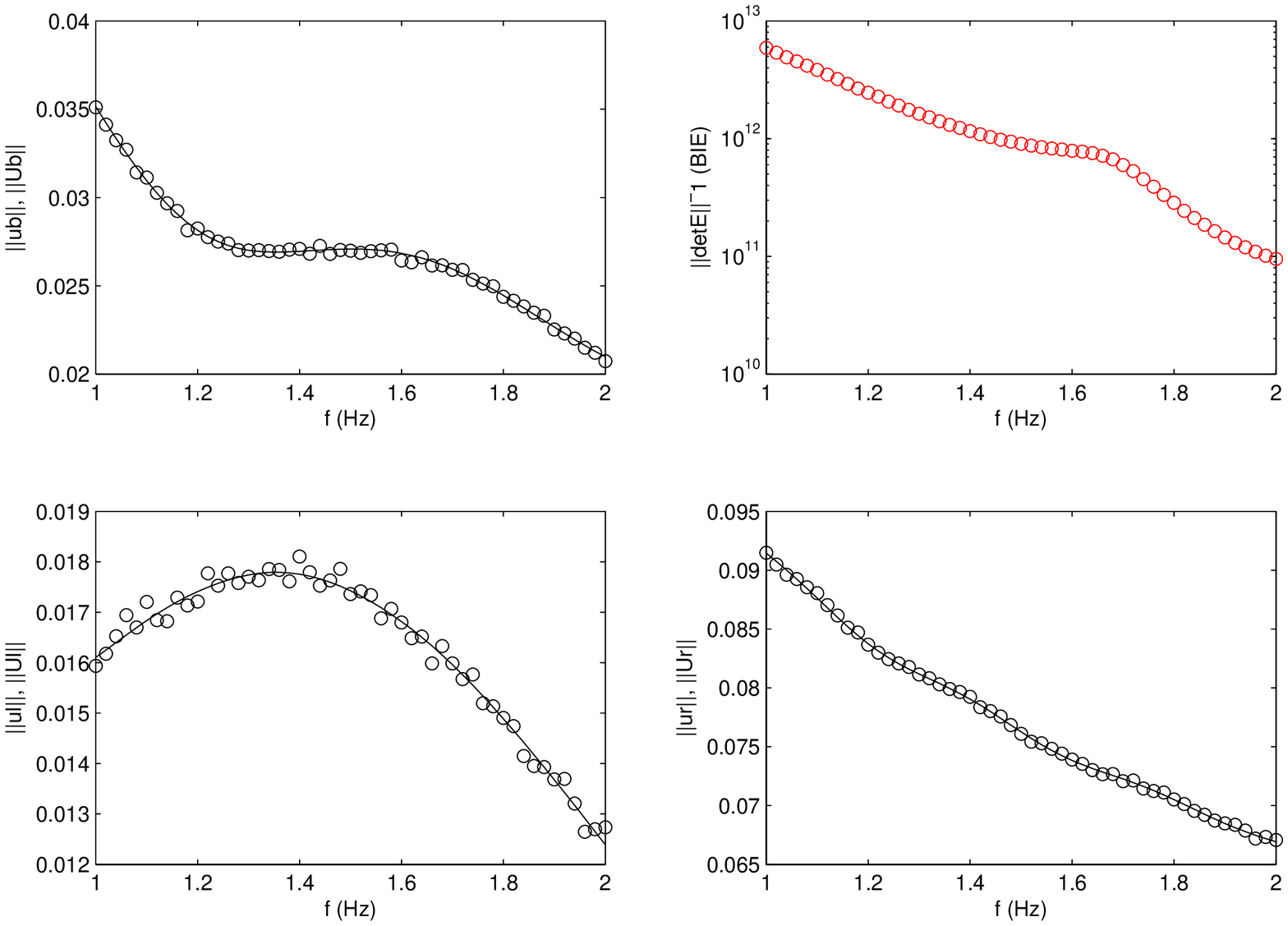}
\caption{Same as fig. \ref{fncbie1-4} of which the present figure is a different zoomed version. Case $N=24$, $\epsilon=10^{-3}$.}
\label{fncbie1-6}
\end{center}
\end{figure}
\begin{figure}[ptb]
\begin{center}
\includegraphics[width=0.7\textwidth]{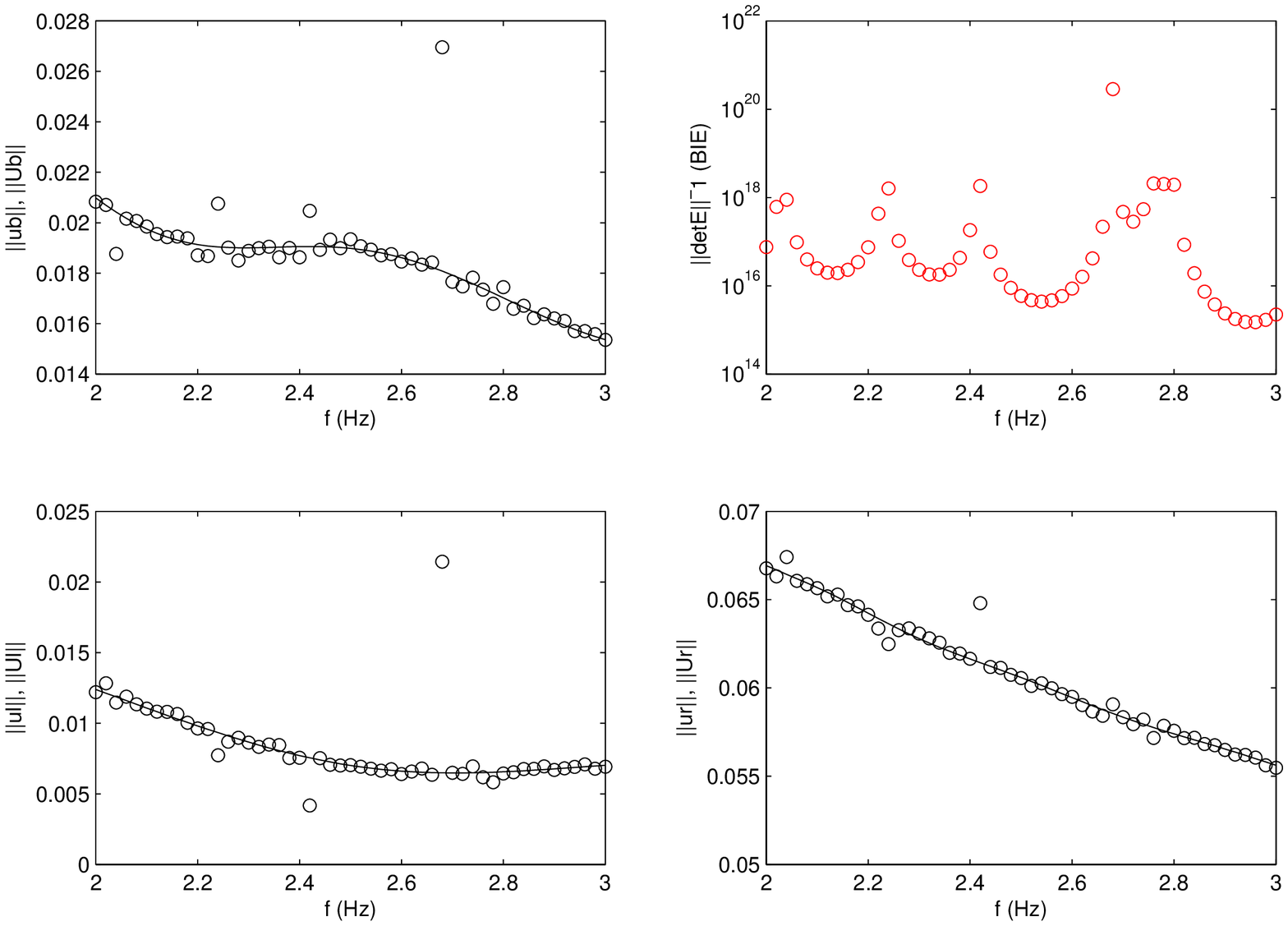}
\caption{Same as fig. \ref{fncbie1-1} of which the present figure is a zoomed version. Case $N=28$, $\epsilon=10^{-3}$.}
\label{fncbie1-7}
\end{center}
\end{figure}
\begin{figure}[ptb]
\begin{center}
\includegraphics[width=0.7\textwidth]{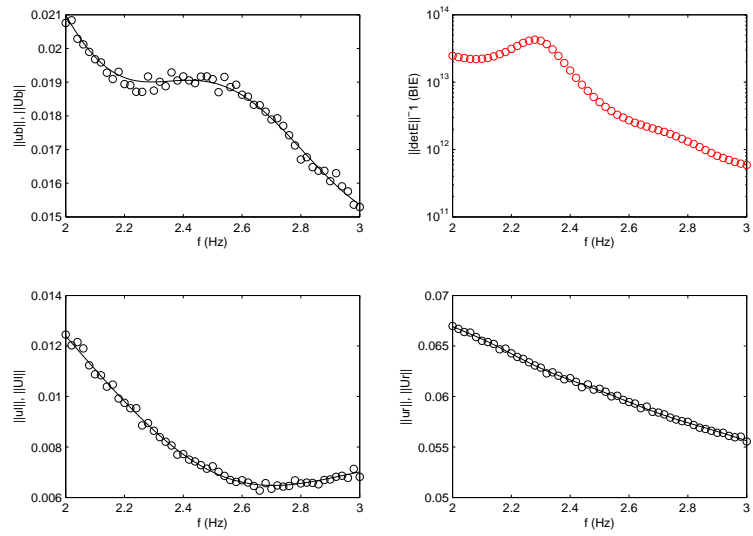}
\caption{Same as fig. \ref{fncbie1-4} of which the present figure is a different zoomed version. Case $N=28$, $\epsilon=10^{-3}$.}
\label{fncbie1-8}
\end{center}
\end{figure}
\clearpage
\newpage
If account is taken of the scale changes in this set of figures, the latter shows convincingly that the method of cure NCBIE has enabled to eliminate all the resonances appearing in NBIE2. The results of other computations not appearing here here show that NCBIE also enables to eliminate all the resonances appearing in  NEBC (which, it will be recalled, occur at frequencies that  are generally-different from those at which occur the resonances appearing in NBIE2).
\section{The problem of the frequency domain response, to the wave radiated by a line source, of a rigid body canyon}
\subsection{Equivalence of the canyon problem to the cylinder problem for a specific type of solicitation}
Here, I shall demonstrate the equivalence via the boundary integral relations (\ref{1-340})-(\ref{1-350}) relative to a Dirichlet  condition on  the boundary $\Gamma$, i.e.,
\begin{equation}\label{3-010}
u(\mathbf{x})=0~;~\forall\mathbf{x}\in\Gamma~,
\end{equation}
Recall that this boundary condition implies that the wavefield cannot penetrate within the object, i.e.,
\begin{equation}\label{3-015}
u(\mathbf{x})=0~;~\forall\mathbf{x}\in\Omega_{1}~.
\end{equation}
The BI relations were:
\begin{equation}\label{3-020}
u(\mathbf{x})=u^{i}(\mathbf{x})+
\int_{\Gamma}kG(\mathbf{x};\mathbf{x}')v(\mathbf{x}')d\gamma(\mathbf{x}')
~;~\forall\mathbf{x}'\in\Omega_{0}~,
\end{equation}
\begin{equation}\label{3-030}
0=u^{i}(\mathbf{x})+
\int_{\Gamma}kG(\mathbf{x};\mathbf{x}')v(\mathbf{x}')d\gamma(\mathbf{x}')
~;~\forall\mathbf{x}\in\Gamma~,
\end{equation}
wherein
\begin{equation}\label{3-033}
v(\mathbf{x}')=\frac{1}{k}\boldsymbol{\nu}(\mathbf{x}')\cdot\nabla (\mathbf{x}')u(\mathbf{x}')
~;~\mathbf{x}\in\Gamma~.
\end{equation}
My demonstration will be made only for the case of a circular (radius $a$) boundary, but the method (as well as the conclusion to which it leads) is easily generalized to boundaries of other shapes.
Consequently, the polar coordinate representations are appropriate, whence:
\begin{equation}\label{3-035}
u(a,\theta)=0
~;~\forall\theta\in[0,2\pi[~,
\end{equation}
\begin{equation}\label{3-040}
u(r,\theta)=u^{i}(r,\theta)+
\int_{0}^{2\pi}kaG(r,\theta;a,\theta')v(a,\theta')d\theta'
~;~\forall r>a~,~\forall\theta\in[0,2\pi[~,
\end{equation}
\begin{equation}\label{3-050}
0=u^{i}(a,\theta)+
\int_{0}^{2\pi}kaG(a,\theta;a,\theta')v(a,\theta')d\theta'
~;~\forall\theta\in[0,2\pi[~.
\end{equation}
I first pay attention to (\ref{3-050}), a consequence of which is
\begin{equation}\label{3-060}
0=u^{i}(a,\theta)+u^{i}(a,-\theta)+
\int_{0}^{2\pi}ka\left[G(a,\theta;a,\theta')+G(a,-\theta;a,\theta')\right]v(a,\theta')d\theta'
~;~\forall\in[0,\pi]~.
\end{equation}
I assume that the incident wavefield is such that:
\begin{equation}\label{3-070}
0=u^{i}(r,\theta)+u^{i}(r,-\theta)
~;~\forall r\in[0,\infty[~,~\forall\theta\in[0,\pi]~.
\end{equation}
An example of such a wavefield is the one radiated by two out-of-phase line sources located at $(r^{s},\theta^{s})$ and
 $(r^{s},-\theta^{s})$:
\begin{equation}\label{3-080}
u^{i}(r,\theta)=\frac{i}{4}H_{0}^{(1)}(k|\sqrt{r^{2}+(r^{s})^{2}-2rr^{s}\cos(\theta-\theta^{s}}|)-
\frac{i}{4}H_{0}^{(1)}(k|\sqrt{r^{2}+(r^{s})^{2}-2rr^{s}\cos(\theta+\theta^{s}}|)
~.
\end{equation}
consequently, (\ref{3-060}) tells us that
\begin{multline}\label{3-090}
0=
\int_{0}^{\pi}ka\left\{\Big[G(a,\theta;a,\theta')+G(a,-\theta;a,\theta')\right]v(a,\theta')+\\
\left[G(a,\theta;a,2\pi-\theta')+G(a,-\theta;a,2\pi-\theta')\right]v(a,2\pi-\theta')\Big\}d\theta'
~;~\forall\theta\in[0,\pi]~.
\end{multline}
However
\begin{multline}\label{3-100}
G(r,\theta;a,\theta')=\frac{i}{4}\sum_{l=-\infty}^{\infty}
\left[H(r-a)H_{l}^{(1)}(kr)J_{l}(ka)+H(a-r)J_{l}(kr)H_{l}^{(1)}(ka)\right]\exp[il(\theta-\theta')]=\\
\frac{i}{2}\sum_{l=0}^{\infty}\epsilon_{l}\left[H(r-a)H_{l}^{(1)}(kr)J_{l}(ka)+H(a-r)J_{l}^{(1)}(kr)H_{l}^{(1)}(ka)\right]\cos[l(\theta-\theta')]=\\
\sum_{l=0}^{\infty}F_{l}(r,a)\cos[l(\theta-\theta')]~,
\end{multline}
wherein
\begin{multline}\label{3-110}
F_{l}(r,a)=\frac{i}{2}\epsilon_{l}\left[H(r-a)H_{l}^{(1)}(kr)J_{l}(ka)+H(a-r)J_{l}(kr)H_{l}^{(1)}(ka)\right]~~,~~\epsilon_{0}=1~~,~~\epsilon_{l>0}=2~,
\end{multline}
so that (\ref{3-090}) becomes
\begin{multline}\label{3-120}
0=ka\sum_{l=0}^{\infty}F_{l}(a,a)
\int_{0}^{\pi}\Big\{
\left[\cos[l(\theta-\theta')]+\cos[l(-\theta-\theta')]\right]v(a,\theta')+\\
\left[\cos[l(\theta-2\pi+\theta')]+\cos[l(-\theta-2\pi+\theta')]\right]v(a,2\pi-\theta')\Big\}d\theta'
~;~\forall\theta\in[0,\pi]~.
\end{multline}
or
\begin{equation}\label{3-130}
0=ka\sum_{l=0}^{\infty}F_{l}(a,a)
\int_{0}^{\pi}
\left[\cos[l(\theta-\theta')]+\cos[l(\theta+\theta')]\right]\left[v(a,\theta')+v(a,2\pi-\theta')\right]
~;~\forall\theta\in[0,\pi]~.
\end{equation}
from which necessarily ensues
\begin{equation}\label{3-140}
v(a,\theta)+v(a,2\pi-\theta)=0
~;~\forall\theta\in[0,\pi]~,
\end{equation}
or, equivalently
\begin{equation}\label{3-150}
v(a,\theta)+v(a,-\theta)=0
~;~\forall\theta\in[0,\pi]~.
\end{equation}
Now I return to (\ref{3-040}) which, together with (\ref{3-070}), implies that
\begin{equation}\label{3-160}
u(r,\theta)+u(r,-\theta)=
\int_{0}^{2\pi}ka\left[G(r,\theta;a,\theta')+G(r,-\theta;a,\theta')\right]v(a,\theta')d\theta'
~;~\forall r>a~,~\forall\theta\in[0,\pi]~,
\end{equation}
or, on account of (\ref{3-140})
\begin{multline}\label{3-170}
u(r,\theta)+u(r,-\theta)=\\
\int_{0}^{\pi}ka\left[G(r,\theta;a,\theta')+G(r,-\theta;a,\theta')-G(r,\theta;a,2\pi-\theta')-G(r,-\theta;a,2\pi-\theta')\right]v(a,\theta')d\theta'
\\
~;~\forall r>a~,~\forall\theta\in[0,\pi]~.
\end{multline}
Proceeding as previously leads to
\begin{multline}\label{3-180}
u(r,\theta)+u(r,-\theta)=ka\sum_{l=0}^{\infty}F_{l}(r,a)
\int_{0}^{\pi}\Big\{
\left[\cos[l(\theta-\theta')]+\cos[l(-\theta-\theta')]\right]v(a,\theta')-\\
\left[\cos[l(\theta-2\pi+\theta')]-\cos[l(-\theta-2\pi+\theta')]\right]v(a,2\pi-\theta')\Big\}d\theta'
~;~\forall r>a~,~\forall\theta\in[0,\pi]~,
\end{multline}
from which finally ensues the general antisymmetry relation
\begin{equation}\label{3-190}
u(r,\theta)+u(r,-\theta)=0~;~\forall r>a~,~\forall\theta\in[0,\pi]~,
\end{equation}
and the particular relations
\begin{equation}\label{3-200}
u(r,0)=0~;~\forall r>a~,
\end{equation}
\begin{equation}\label{3-210}
u(r,\pi)=0~;~\forall r>a~.
\end{equation}
\begin{figure}[ptb]
\begin{center}
\includegraphics[width=0.85\textwidth]{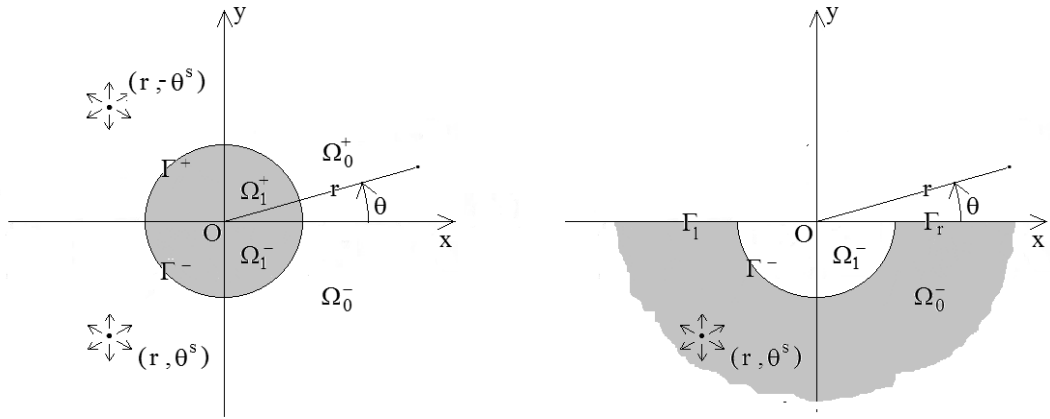}
\caption{Cross section view of two scattering configurations. The left-hand configuration is relative to a circular (radius $a$) cylinder submitted to the wavefield radiated by two line sources located at $(r^{s},\theta^{s})$ and $(r^{s},-\theta^{s})$. The right-hand configuration is relative to a semi-circular (radius $a$) canyon submitted to the wavefield radiated by one line source located at $(r^{s},\theta^{s})$. The circular ($\Gamma_{circle}=\Gamma^{-}\bigcup\Gamma^{+}$) boundary of the cylinder and semi-circular portion ($\Gamma^{+}$) portion, as well as the left-hand ($\Gamma_{l}$) and right-hand ($\Gamma_{r}$) straight portions, of the canyon boundary $\Gamma_{canyon}$ are the locus of either Dirichlet or Neumann conditions. The regions interior (impervious to the wavefield) and exterior to $\Gamma_{circ}$ are $\Omega_{1}=\Omega_{1}^{+}\bigcup\Omega_{1}^{+}$ and  $\Omega_{0}=\Omega_{0}^{+}\bigcup\Omega_{0}^{+}$ respectively. The 'exterior boundaries' of $\Omega_{0}^{+}$ and $\Omega_{0}^{-}$ can be considered to be semi-circles of infinite radius. The wavefield is non-nil only in the region $\Omega_{0}^{-}$ underneath the canyon boundary $\Gamma_{canyon}=\Gamma_{l}\bigcup\Gamma^{-}\bigcup\Gamma_{r}$.}
\label{fequivalence}
\end{center}
\end{figure}
Now refer to fig. \ref{fequivalence}. Eqs. (\ref{3-010}), (\ref{3-210}) and (\ref{3-200})  relative to the circular cylinder body submitted to the two line sources entail
\begin{equation}\label{3-220}
u(\mathbf{x})=0~;~\forall\mathbf{x}\in\Gamma^{-}~~,~~u(\mathbf{x})=0~;~\forall\mathbf{x}\in\Gamma_{l}~~,
~~u(\mathbf{x})=0~;~\forall\mathbf{x}\in\Gamma_{r}~,
\end{equation}
respectively, which amount to the composite Dirichlet boundary condition for the semi-circular cylindrical canyon submitted to the bottom line source of the previous pair
\begin{equation}\label{3-225}
u(\mathbf{x})=0~;~\forall\mathbf{x}\in\Gamma_{canyon}~.
\end{equation}
Since the only region in which the field is non-nil in the canyon configuration is $\Omega_{0}^{-}$, and the 'exterior boundary' of this region is a semi-circle of infinite radius on which a radiation condition prevails in the circular cylinder problem, the same is necessarily true for the semi-circular canyon problem. Moreover, the field in both problems satisfies the same partial differential equation  which leads to the two BI equations (\ref{3-040})-(\ref{3-050}). Thus the necessary conclusion is that {\it the two configurations in fig. \ref{fequivalence} are rigorously-equivalent} as concerns the wavefield on $\Gamma_{l}$, $\Gamma^{-}$, $\Gamma_{r}$ (which is nil as befits a Dirichlet boundary condition), and in $\Omega_{0}^{-}$, this being true only if the solicitation (due to two line sources for the cylinder and one lower line source of this pair for the semi-cylinder) is such as to satisfy (\ref{3-070}).
\subsection{Numerical results for the canyon configuration with a Dirichlet (rigid body) boundary condition}
Since this problem is of less interest in the geophysical context, I shall not pursue it any further, and prefer to henceforth concentrate my attention on the canyon configuration with a Neumann boundary condition.
\section{The problem of the  frequency domain response, to the wave radiated by a line source or to a plane wave, of a stress-free boundary canyon}
The case of plane-wave solicitation is treated in exactly the same way as for source-wave solicitation, with the $u^{i}$ in all formulae replaced by the explicit relation (for a single incident plane wave, and in polar coordinates)
\begin{equation}\label{4-003}
u^{i}(\mathbf{x})=\exp\left(ikr\cos(\theta-\theta^{i})\right)~,
\end{equation}
wherein $\theta^{i}$ is the (incident) angle between the $x$-axis and the head of the incident wavevector. The
Fourier series form of this wave is ((9.1.41) in \cite{as68})
\begin{equation}\label{4-005}
\exp\left(ikr\cos(\theta-\theta^{i})\right)=\sum_{l=-\infty}^{\infty}A_{l}J_{l}(kr)\exp(il\theta)~,
\end{equation}
in which
\begin{equation}\label{4-007}
A_{l}=i^{l}\exp(-il\theta^{i})~.
\end{equation}
\subsection{Equivalence of the canyon problem to the cylinder problem for a specific type of solicitation}
Here, I shall demonstrate the equivalence via the boundary integral relations (\ref{2-340})-(\ref{2-350}) relative to Neumann  condition on  the boundary $\Gamma$, i.e.,
\begin{equation}\label{4-010}
\boldsymbol{\nu}(\mathbf{x})\cdot\nabla(\mathbf{x})u(\mathbf{x})=0~;~\forall\mathbf{x}\in\Gamma~,
\end{equation}
Recall that this boundary condition implies that the wavefield cannot penetrate within the object, i.e.,
\begin{equation}\label{4-015}
u(\mathbf{x})=0~;~\forall\mathbf{x}\in\Omega_{1}~.
\end{equation}
The BI relations were:
\begin{equation}\label{4-020}
u(\mathbf{x})=u^{i}(\mathbf{x})-
\int_{\Gamma}u(\mathbf{x}')\boldsymbol{\nu}(\mathbf{x}')\cdot\nabla (\mathbf{x}')G(\mathbf{x};\mathbf{x}')d\gamma(\mathbf{x}')
~;~\forall\mathbf{x}\in\Omega_{0}~,
\end{equation}
\begin{equation}\label{4-030}
\frac{1}{2}u(\mathbf{x})=u^{i}(\mathbf{x})-
pv\int_{\Gamma}u(\mathbf{x}')\boldsymbol{\nu}(\mathbf{x}')\cdot\nabla (\mathbf{x}')G(\mathbf{x};\mathbf{x}')d\gamma(\mathbf{x}')
~;~\forall\mathbf{x}\in\Gamma~,
\end{equation}
My demonstration will be made only for the case of a circular (radius $a$) boundary, but the method (as well as the conclusion to which it leads) is easily generalized to boundaries of other shapes.
Consequently, the polar coordinate representations are appropriate, whence:
\begin{equation}\label{4-040}
\frac{1}{2}u(a,\theta)=u^{i}(a,\theta)+
pv\int_{0}^{2\pi}u(a,\theta')\frac{\partial}{\partial r'}G(a,\theta;a,\theta')ad\theta'
~;~\forall\theta\in[0,2\pi[~,
\end{equation}
I assume that:
\begin{equation}\label{4-050}
u^{i}(r,\theta)-u^{i}(r,\theta)=0~;~\forall r\in [0,\infty[~,~\forall\theta\in[0,\pi[~.
\end{equation}
An example of such a solicitation is that of the wavefield radiated by two in-phase line sources:
\begin{equation}\label{4-060}
u^{i}(r,\theta)=\frac{i}{4}H_{0}^{(1)}(k|\sqrt{r^{2}+(r^{s})^{2}-2rr^{s}\cos(\theta-\theta^{s}}|)+
\frac{i}{4}H_{0}^{(1)}(k|\sqrt{r^{2}+(r^{s})^{2}-2rr^{s}\cos(\theta+\theta^{s}}|)
~.
\end{equation}
Consequently, (\ref{4-040}) gives rise to:
\begin{equation}\label{4-070}
u(a,\theta)-u(a,-\theta)=
pv\int_{0}^{2\pi}2a u(a,\theta')\left[\frac{\partial}{\partial r'}G(a,\theta;a,\theta')-\frac{\partial}{\partial r'}G(a,-\theta;a,\theta')\right]d\theta'
~;~\forall\theta\in[0,\pi[~,
\end{equation}
I now make use of (\ref{1-060}) to obtain (in the sense of its use in the $pv$ integral)
\begin{multline}\label{4-080}
2a\frac{\partial}{\partial r'}G(a,\theta;a,\theta')=\frac{ika}{2}\sum_{l=-\infty}^{\infty}
\left[\frac{-i}{ka\pi}+\dot{H}_{l}^{(1)}(ka)J_{l}(ka)\right]\exp[il(\theta-\theta')]=\\
\frac{1}{2\pi}\sum_{l=-\infty}^{\infty}\exp[il(\theta-\theta')]+\sum_{l=0}^{\infty}\epsilon_{l}\frac{ika}{2}
\dot{H}_{l}^{(1)}(ka)J_{l}(ka)\cos[l(\theta-\theta')]=\\
\frac{1}{2\pi}\sum_{l=-\infty}^{\infty}\exp[il(\theta-\theta')]+\sum_{l=0}^{\infty}\mathcal{F}_{l}\cos[l(\theta-\theta')]
~,
\end{multline}
wherein
\begin{equation}\label{4-090}
\mathcal{F}_{l}(a,a)=\epsilon_{l}\frac{ika}{2}
\dot{H}_{l}^{(1)}(ka)J_{l}(ka)=\mathcal{F}_{-l}(a,a)
~.
\end{equation}
The Poisson sum formula (\cite{mf53} tells us that
\begin{equation}\label{4-100}
\frac{1}{2\pi}\sum_{l=-\infty}^{\infty}\exp[il(\theta-\theta')]=\sum_{l=-\infty}^{\infty}\delta(\theta-\theta'+2l\pi)
~,
\end{equation}
so that
\begin{multline}\label{4-110}
u(a,\theta)-u(a,-\theta)=\sum_{l=-\infty}^{\infty}\int_{0}^{2\pi}u(a,\theta')\left[\delta(\theta-\theta'+2l\pi)-\delta(-\theta-\theta'+2l\pi)\right]d\theta'+\\
\sum_{l=0}^{\infty}\mathcal{F}_{l}(a,a)\int_{0}^{2\pi}u(a,\theta')\left[\cos[l(\theta-\theta')]-\cos[l(-\theta-\theta')] \right]d\theta'
~;~\forall\theta\in[0,\pi[~,
\end{multline}
which, by the sifting property of the Dirac delta distributions, becomes
\begin{equation}\label{4-115}
0=\sum_{l=0}^{\infty}\mathcal{F}_{l}(a,a)\int_{0}^{2\pi}u(a,\theta')\left[\cos[l(\theta-\theta')]-\cos[l(-\theta-\theta')] \right]d\theta'
~;~\forall\theta\in[0,\pi[~,
\end{equation}
or,
\begin{multline}\label{4-120}
0=
\sum_{l=0}^{\infty}\mathcal{F}_{l}(a,a)\int_{0}^{\pi}\Big\{u(a,\theta')\Big[\cos[l(\theta-\theta')]-\cos[l(-\theta-\theta')]\Big]+\\
u(a,2\pi-\theta')\Big[\cos[l(\theta-2\pi+\theta')]-\cos[l(-\theta-2\pi+\theta')]\Big]\Big\}d\theta'=\\
\sum_{l=0}^{\infty}\mathcal{F}_{l}(a,a)\int_{0}^{\pi}
\Big[u(a,\theta')-u(a,2\pi-\theta')\Big]
\Big[\cos[l(\theta-\theta')]-\cos[l(-\theta-\theta')]\Big]d\theta'
~;~\forall\theta\in[0,\pi[~,
\end{multline}
from which I deduce necessarily that
\begin{equation}\label{4-130}
u(a,\theta')-u(a,2\pi-\theta')=0~;~\forall\theta\in[0,\pi[~,
\end{equation}
or equivalently
\begin{equation}\label{4-140}
u(a,\theta')-u(a,-\theta')=0~;~\forall\theta\in[0,\pi[~.
\end{equation}
Now  return to (\ref{4-020}), which, together with (\ref{4-050}), implies that
\begin{equation}\label{4-150}
u(r,\theta)-u(r,-\theta)=
\int_{0}^{2\pi}u(a,\theta')a\left[\frac{\partial}{\partial r'}G(r,\theta;a,\theta')-\frac{\partial}{\partial r'}G(r,-\theta;a,\theta')\right]d\theta'
~;~\forall r>a~,~\forall\theta\in[0,\pi[~,
\end{equation}
I make use of:
\begin{equation}\label{4-160}
a\frac{\partial}{\partial r'}G(r>a,\theta;a,\theta')=\frac{ika}{4}\sum_{l=-\infty}^{\infty}H_{l}^{(1)}(kr)\dot{J}(ka)\exp[il(\theta-\theta')]=
\sum_{l=0}^{\infty}\mathcal{G}_{l}(r,a)\cos[l(\theta-\theta')]~,
\end{equation}
wherein
\begin{equation}\label{4-170}
\mathcal{G}_{l}(r,a)=\mathcal{G}_{-l}(r,a)=\epsilon_{l}\frac{ika}{4}H_{l}^{(1)}(kr)\dot{J}(ka)~,
\end{equation}
so that
\begin{equation}\label{4-180}
u(r,\theta)-u(r,-\theta)=\sum_{l=0}^{\infty}\mathcal{G}_{l}(r,a)
\int_{0}^{2\pi}u(a,\theta')\left[\cos[l(\theta-\theta')]-\cos[l(-\theta-\theta')]\right]d\theta'
~;~\forall r>a~,~\forall\theta\in[0,\pi[~,
\end{equation}
which, after making use of (\ref{4-130}), becomes
\begin{multline}\label{4-190}
u(r,\theta)-u(r,-\theta)=\sum_{l=0}^{\infty}\mathcal{G}_{l}(r,a)
\int_{0}^{\pi}u(a,\theta')\Big[\cos[l(\theta-\theta')]-\cos[l(-\theta-\theta')]+\\
\cos[l(\theta-2\pi+\theta')]-\cos[l(-\theta-2\pi+\theta')]]\Big]d\theta'
~;~\forall r>a~,~\forall\theta\in[0,\pi[~,
\end{multline}
from which finally ensues the general symmetry relation
\begin{equation}\label{4-200}
u(r,\theta)-u(r,-\theta)=0~;~\forall r>a~,~\forall\theta\in[0,\pi]~,
\end{equation}
and the particular relations
\begin{equation}\label{4-210}
\frac{\partial}{\partial\theta}u(r,0)=0~;~\forall r>a~,
\end{equation}
\begin{equation}\label{4-220}
\frac{\partial}{\partial\theta}u(r,\pi)=0~;~\forall r>a~.
\end{equation}

 Again refer to fig. \ref{fequivalence}. Eqs. (\ref{4-010}), (\ref{4-220}) and (\ref{4-210})  relative to the circular cylinder body submitted to the two line sources entail
\begin{equation}\label{4-230}
\boldsymbol{\nu}(\mathbf{x})\cdot\nabla(\mathbf{x})u(\mathbf{x})=0~;~\forall\mathbf{x})\in\Gamma^{-}~~,~~\boldsymbol{\nu}(\mathbf{x})\cdot\nabla(\mathbf{x})u(\mathbf{x})=0~;~\forall\mathbf{x}\in\Gamma_{l}~~,
~~\boldsymbol{\nu}(\mathbf{x})\cdot\nabla(\mathbf{x})u(\mathbf{x})=0~;~\forall\mathbf{x}\in\Gamma_{r}~,
\end{equation}
respectively, which amount to the composite Dirichlet boundary condition for the semi-circular cylindrical canyon submitted to the bottom line source of the previous pair
\begin{equation}\label{4-240}
\boldsymbol{\nu}(\mathbf{x})\cdot\nabla(\mathbf{x})u(\mathbf{x})=0~;~\forall\mathbf{x}\in\Gamma_{canyon}~.
\end{equation}
Since the only region in which the field is non-nil in the canyon configuration is $\Omega_{0}^{-}$, and the 'exterior boundary' of this region is a semi-circle of infinite radius on which a radiation condition prevails in the circular cylinder problem, the same is necessarily true for the semi-circular canyon problem. Moreover, the field in both problems satisfies the same partial differential equation  which leads to the two BI equations (\ref{4-020})-(\ref{4-030}). Thus the necessary conclusion is that {\it the two configurations in fig. \ref{fequivalence} are rigorously-equivalent} as concerns the wavefield on $\Gamma_{l}$, $\Gamma^{-}$, $\Gamma_{r}$ (the normal derivative of which is nil as befits a Neumann boundary condition), and in $\Omega_{0}^{-}$, this being true only if the solicitation (due to two line sources for the cylinder and one lower line source of this pair for the semi-cylinder) is such as to satisfy (\ref{4-050}).
\clearpage
\newpage
\subsection{Numerical results for the canyon configuration with a Neumann boundary condition and one plane wave incidence: NBIE2 cured by NCBIE}
In all the following figures, i.e., figs \ref{fcanyon-1}-\ref{fcanyon-8}, the semi-circular cylindrical canyon of radius $a=1000m$ is solicited by a normally-incident plane wave (i.e., $\theta^{i}=90^{\circ}$. The incident wave propagates in the lower (i.e., negative $y$) half space in which the bulk wave velocity is $\beta=2000ms^{-1}$. In the computations involving NCBIE, $b=500m$ and $\eta=0+1i$.
\begin{figure}[ht]
\begin{center}
\includegraphics[width=0.5\textwidth]{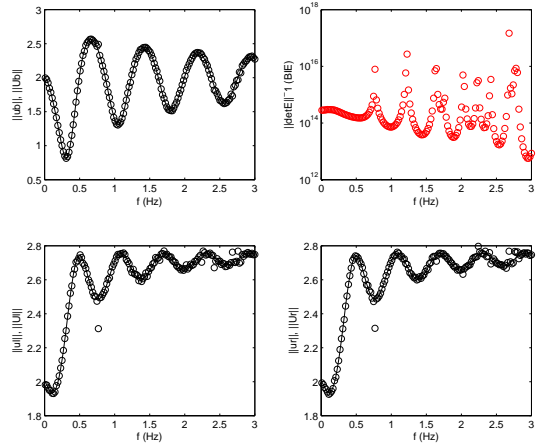}
\caption{Transfer functions of displacement at three points on the traction-free boundary of the canyon. The upper left-hand, lower left-hand, lower right-hand panels are for  the transfer functions at $\theta=180^{\circ}$,  $\theta=270^{\circ}$,  $\theta=360^{\circ}$, respectively. The upper right-hand panel depicts $1/\|det(\mathbf{E}(ka))\|$. Lower-case letters and circles correspond to  NBIE2 computations, upper-case letters and continuous curves to  NSOV (exact) computations.  Case $N=24$, $\epsilon=10^{-3}$. Compare the curve in the lower left panel to the corresponding full curve in fig. 16 of \cite{si78}. Note that Sills exhibits no 'resonance'. However, the resonances in the present figure also appear in our fig. \ref{mysills} herein. This is also true for the determinants in the upper right-hand panels of the present figure as well as of fig.  \ref{mysills} herein, which shows that the resonances  of these two figures have the same origin, i.e., that of the singularities of the matrix representing the discretized/projected  second-kind integral equation.}
\label{fcanyon-1}
\end{center}
\end{figure}
\begin{figure}[ht]
\begin{center}
\includegraphics[width=0.5\textwidth]{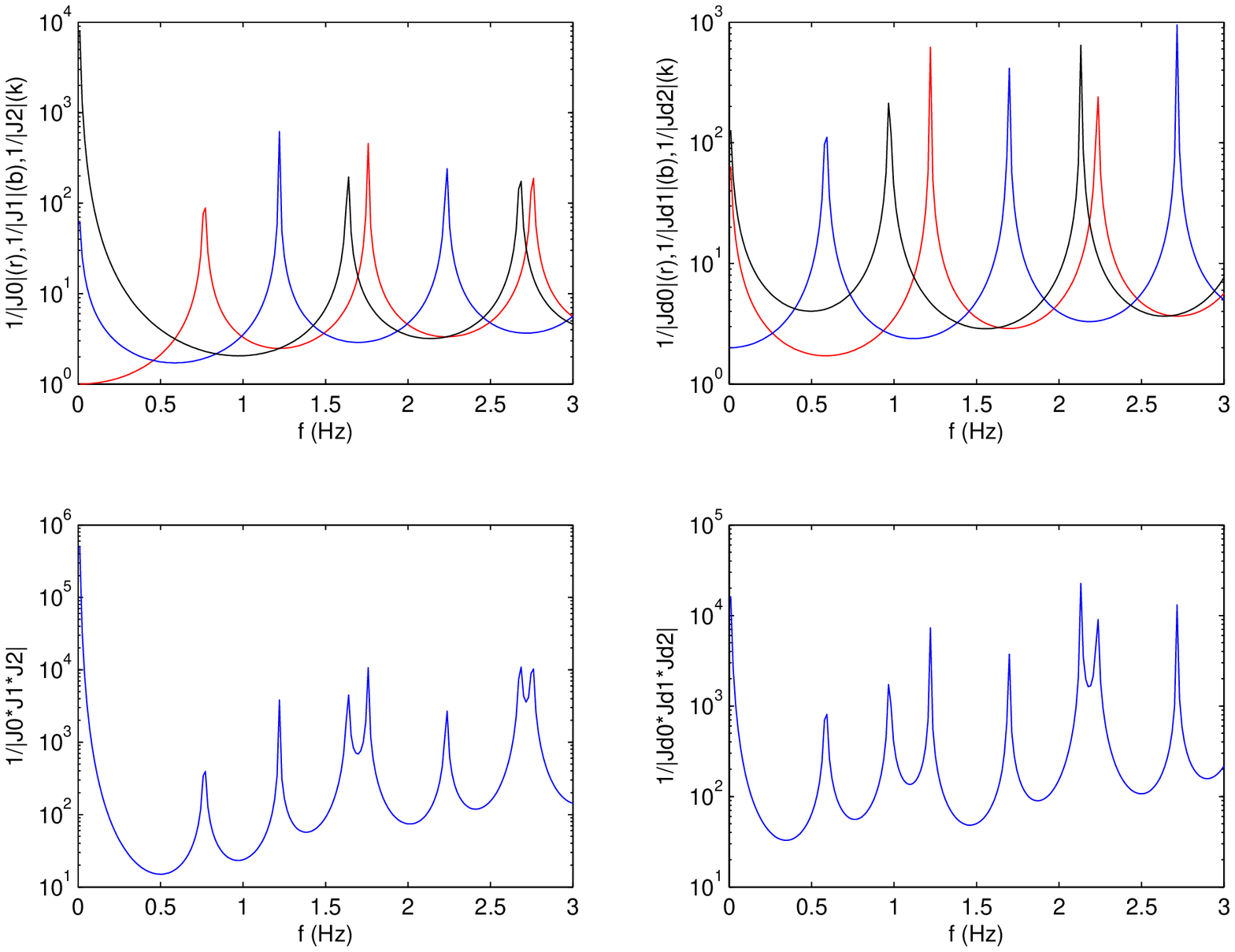}
\caption{This figure enables the connection of the observed resonance frequencies to the zeros of either $\dot{J}_{n}(ka)$ (for NBIE2) or  $\dot{J}_{n}(kb)$ (for NEBC). The upper left-hand panel is relative to $1/|J_{0}(ka)|$ (red), $1/|J_{1}(ka)|$ (blue),  $1/|J_{2}(ka)|$ (black) whereas the lower left-hand panel is relative to $1/|J_{0}(ka)J_{1}(ka)J_{0}(ka)|$. The upper right-hand panel is relative to $1/|\dot{J}_{0}(ka)|$ (red), $1/|\dot{J}_{1}(ka)|$ (blue),  $1/|\dot{J}_{2}(ka)|$ (black) whereas the lower right-hand  panel is relative to $1/|\dot{J}_{0}(ka)\dot{J}_{1}(ka)\dot{J}_{2}(ka)|$. As expected, the positions of the lower-frequency resonant features in fig. \ref{fcanyon-1} coincide with the zeros of $J_{n}(ka)~;~n=0,1,2$ and the first few maxima of $1/\|det(\mathbf{E}(ka))\|$   in fig.  \ref{fcanyon-1} are located at the same positions as those of $1/|J_{0}(ka)J_{1}(ka)J_{2}(ka)|$ herein.}
\label{fcanyon-2}
\end{center}
\end{figure}
\begin{figure}[ht]
\begin{center}
\includegraphics[width=0.7\textwidth]{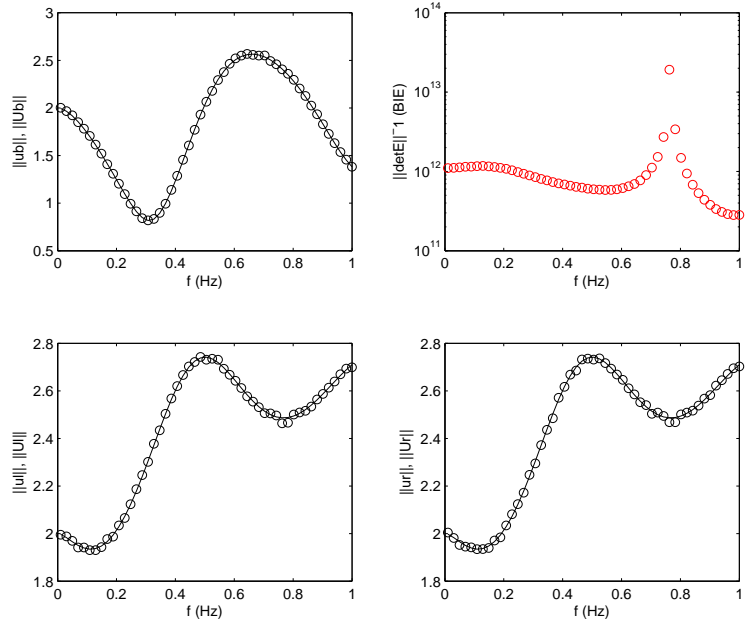}
\caption{Same as fig. \ref{fcanyon-1} of which this is a zoomed version. Case $N=20$, $\epsilon=10^{-3}$.}
\label{fcanyon-3}
\end{center}
\end{figure}
\begin{figure}[ht]
\begin{center}
\includegraphics[width=0.7\textwidth]{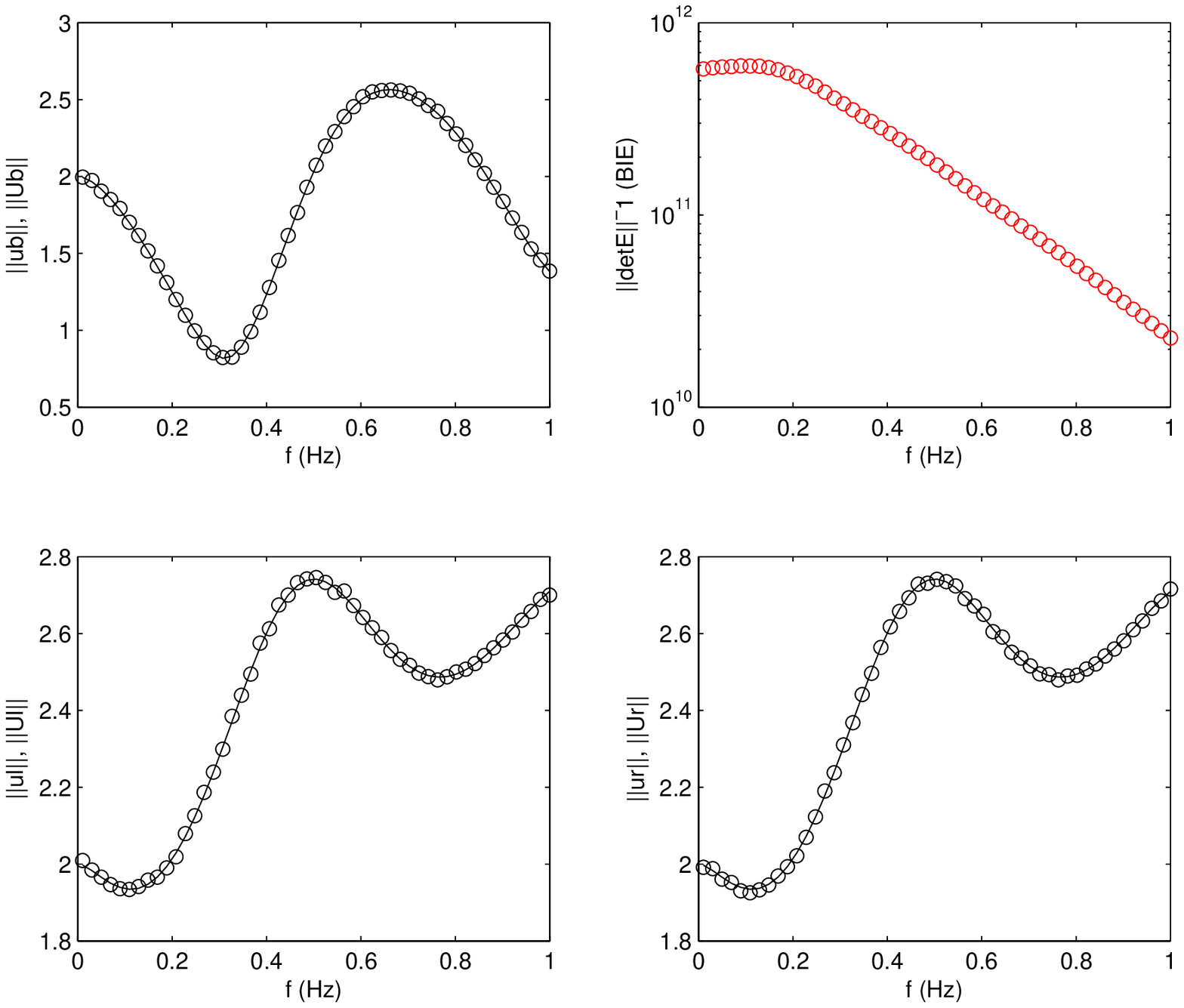}
\caption{Transfer functions of displacement at three points on the traction-free boundary of the canyon. The upper left-hand, lower left-hand, lower right-hand panels are for  the transfer functions at $\theta=180^{\circ}$,  $\theta=270^{\circ}$,  $\theta=360^{\circ}$, respectively. The upper right-hand panel depicts $1/\|det(\mathbf{E}(ka))\|$. Lower-case letters and circles correspond to  NCBIE computations, upper-case letters and continuous curves to  NSOV (exact) computations.  Case $N=20$, $\epsilon=10^{-3}$.}
\label{fcanyon-4}
\end{center}
\end{figure}
\begin{figure}[ht]
\begin{center}
\includegraphics[width=0.7\textwidth]{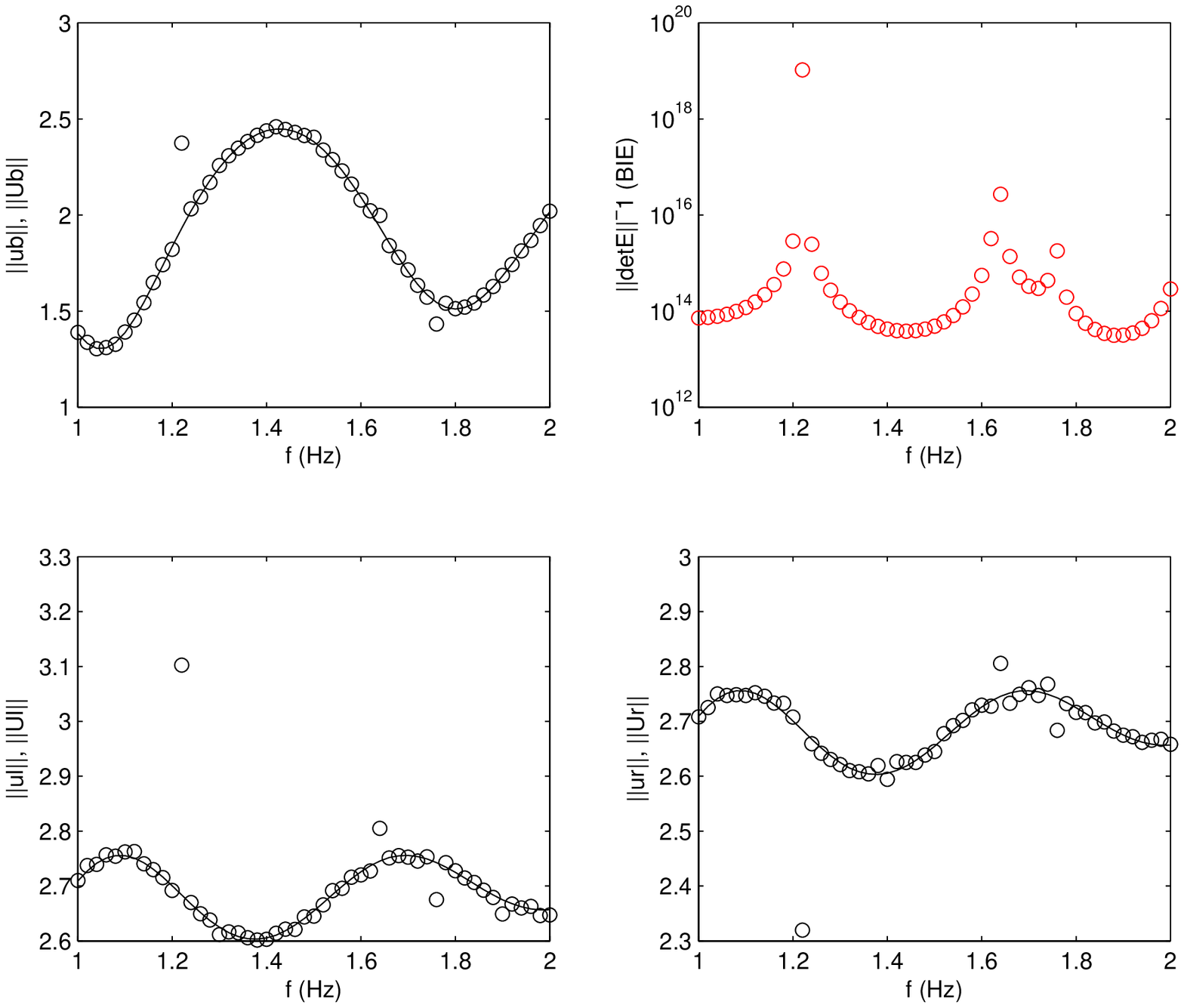}
\caption{Same as fig. \ref{fcanyon-1} of which this is another  zoomed version. Case $N=28$, $\epsilon=10^{-3}$.}
\label{fcanyon-5}
\end{center}
\end{figure}
\begin{figure}[ht]
\begin{center}
\includegraphics[width=0.7\textwidth]{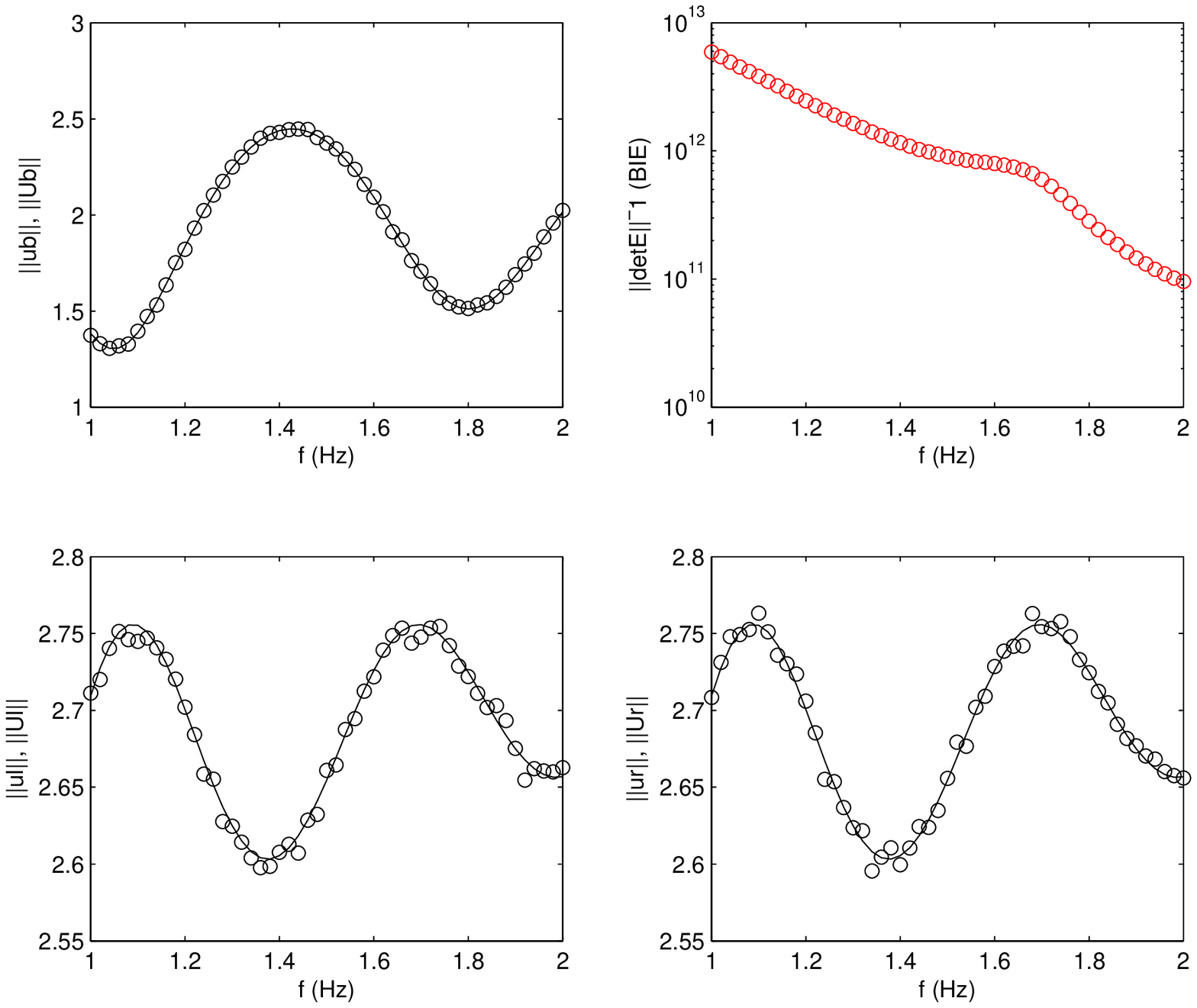}
\caption{Same as fig. \ref{fcanyon-4} for  a different zoom. Case $N=28$, $\epsilon=10^{-3}$.}
\label{fcanyon-6}
\end{center}
\end{figure}
\begin{figure}[ht]
\begin{center}
\includegraphics[width=0.7\textwidth]{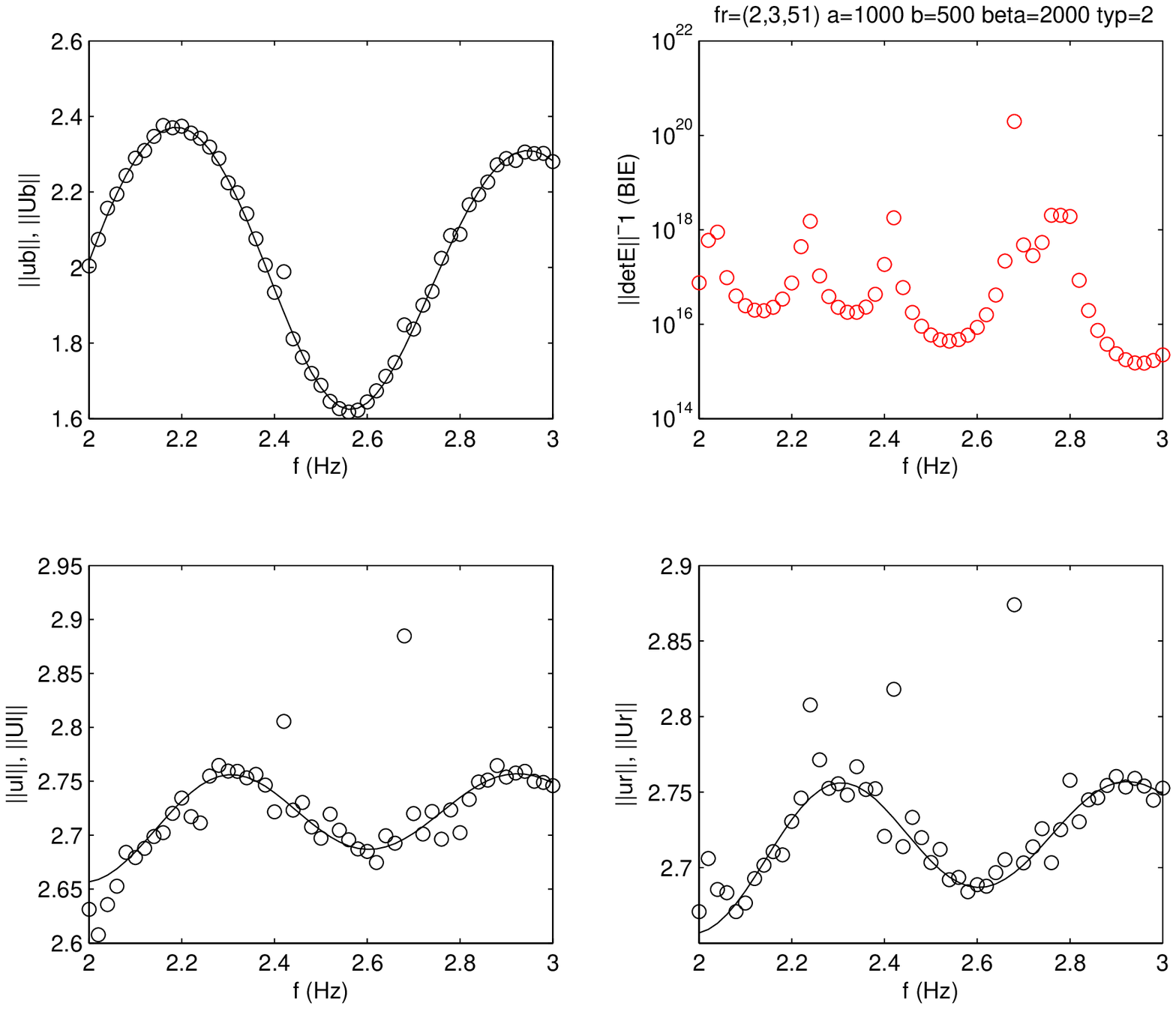}
\caption{Same as fig. \ref{fcanyon-1} of which this is another  zoomed version. Case $N=28$, $\epsilon=10^{-3}$.}
\label{fcanyon-7}
\end{center}
\end{figure}
\begin{figure}[ht]
\begin{center}
\includegraphics[width=0.7\textwidth]{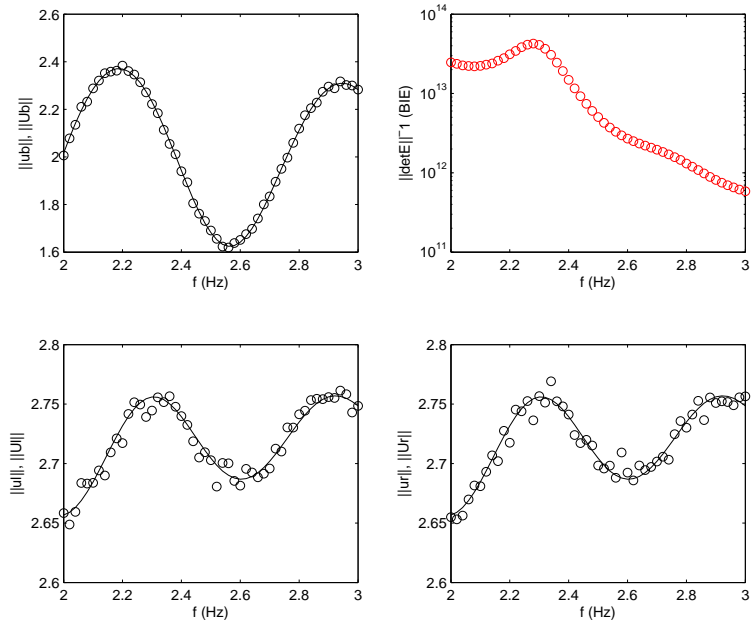}
\caption{Same as fig. \ref{fcanyon-4} for  a different zoom. Case $N=28$, $\epsilon=10^{-3}$.}
\label{fcanyon-8}
\end{center}
\end{figure}
\clearpage
\newpage
These figures show that the NCBIE cure effectively eliminates all the resonances observed as a consequence of the use of NBIE2. The same is of course true as concerns the resonances of NEBC.
\section{Conclusion}
This study has enabled to discover that resonances that we shall qualify as 'spurious' are ubiquitous  in wave scattering problems whose solution is sought by a boundary integral equation technique. Moreover, the number of such resonances increases with frequency, so that beyond the low-frequency regime their existence poses a serious problem (that we have termed 'disease'). Consequently, it is extremely important to find ways of eliminating these resonances in a rational (the first scheme of Nowak \cite{no88} is not of this nature), although convenient manner. The rationality of the technique must be such that no a priori knowledge (which is the usual case for scattering bodies of arbitrary shape) of the frequencies of occurrence of these resonances be required, . The CBIE techniques described in this study fulfill these requirements, in spite of the fact that they involve the Bessel and Hankel functions that are specific to bodies with circular or semicircular boundaries. Actually, this feature is not fundamental: what is fundamental is to combine two integral equations (such as of the first and second kind) into a single integral equation and this can be done for bodies of arbitrary shape.

Having said this, I now re-evoke the question of whether these resonances should be qualified as spurious. This question is very important because there might exist situations (e.g., for a hill instead of a canyon) in which so-called spurious (physically-unreal) and physically-real resonances co-exist; then how to decide which are unreal and which are real, and how to eliminate the former without eliminating the latter? In the problems treated herein, the situation does not appear to be that of the existence of these two types of resonances, so that the question remains of how to be sure that the resonances observed in our study are really spurious.

Suppose that we have at our disposition two methods for solving a given scattering problem  and that we have every reason to believe that both are theoretically-rigorous (this does not exclude differences at the numerical level). By proceeding to solve (usually by numerical means) our scattering problem we expect that the two solutions should be identical within the limits posed by numerical error. Since these two solutions are then not strictly identical it can be argued that they are theoretically different. To avoid this argument, suppose that we can solve our two integral equations without relying on numerical means. This was the case I chose in this study, since the problems I treated were all solved in closed-form via a Galerkin technique applied to the various integral equations. Consequently, I was able to show that the explicit solutions  of different integral equations give rise to expressions for the traction (or boundary displacement) that are radically-different. In fact, I found that one of these expressions gives rise to  resonances at one set of frequencies and the other expression to   resonances at another (different) set of frequencies. Of course, this is not an admissible situation if it is recalled that the two integral equations are solving the same physical problem. For this reason, the only reasonable conclusion is that both explicit relations resulting from these two integral equations are 'wrong' in a certain sense (this being related to the singular nature of a certain diagonal matrix $\mathbf{E}$ entering into these expressions), or what amounts to the same, the resonances, which turn out to manifest themselves not only theoretically, but also numerically, are spurious, i.e., pure consequences of the choices I made of my integral equations. To make this argument even more convincing I then showed that by combining these integral equations in a certain rigorous manner enables once again the obtention of a closed-form solution (identical to the well-known separation-of-variables solution which constitutes the reference) in which the aformentioned singularity of $\mathbf{E}$ is absent and consequently all resonances are absent.

Last, but not least, I think it useful to recall that although the issue of spurious resonances has apparently not stirred the curiosity of the elastic wave community (excepting Nowak and Hall) it is without doubt strongly-connected with themes as important to the applications-oriented elements of this community as: a) non-destructive testing of, and prediction of stress concentration, in surface-breaking cracks, b) effects of vibrations (generated by machines and vehicles or those associated with seismic waves) in underground cavities such as mine shafts, subway tunnels, etc., c) the design of open trenches to protect buildings and industrial facilities from earthquake damage,  d) the prediction of the possible effects of seismic waves on structures such as buildings, dams and bridges that are planned to be built in sites with large and/or deep topographic depressions (i.e., valleys, canyons,..). This is the main reason why my study  focused on the prototypical problem of (the seismic response of) canyons, even though Richard Ford \cite{fo01} might think this to be of no use;
\\\\
{\it "Though it was exactly, he thought, staring mutely out at the flat Brown plateau and the sheer drop straight off
the other side--how far away, you couldn't tell, since perspective was screwed up--it was exactly what he'd
expected from the pictures in High school. It was a tourist attraction. A thing to see. It was plenty big. But twenty
jillion people had already seen it, so that it felt sort of useless. A negative. Nothing like the ocean, which had a
use. Nobody needed the Grand Canyon for anything. At its most important, he guessed, it would be a terrific
impediment to somebody wanting to get to the other side.."}

\end{document}